\documentclass[english,floatfix,superscriptaddress,twocolumn,showpacs,aps,prb,reprint,groupedaddress]{revtex4-2}

\usepackage{amsmath}
\usepackage{graphicx}
\usepackage{bm}
\usepackage{dsfont}
\usepackage{xfrac}
\usepackage{siunitx}
\usepackage{rotating}
\usepackage[resetlabels]{multibib}
\newcites{sm}{References}
\usepackage{xcolor} 
\newcommand{\vect}[1]{\boldsymbol{\mathbf{#1}}}
\newcommand{\dg}{^{\dagger }}

\newcommand{\rarrow}{\rightarrow}
\usepackage{hyperref}
\hypersetup{colorlinks=true, citecolor=blue, urlcolor=blue, linkcolor=blue, breaklinks=true}

\begin{document}
\title{Unified Theory of Hastatic order and Antiferromagnetism in \texorpdfstring{URu$_2$Si$_2$}{URu2Si2}}
\author{Milan Kornja\v ca}
\author{Rebecca Flint}
\affiliation{Department of Physics and Astronomy, Iowa State University, 12 Physics Hall, Ames, Iowa 50011, USA}
\date{\today}
\begin{abstract}

The hidden order phase of URu$_2$Si$_2$ has eluded identification for over thirty-five years. A compelling proposal that explains the Ising heavy-fermion nature of the material is \emph{hastatic order}: a symmetry breaking heavy-Fermi liquid arising from a spinorial microscopic hybridization. The original hastatic proposal cannot microscopically model the pressure induced antiferromagnetic phase; and while it predicts a spinorial order parameter, it does not provide any detectable signatures of the spinorial nature. Here, we present a more realistic microscopic model of hastatic order in URu$_2$Si$_2$ based on two conducting electron bands and explore its phase diagram in detail. Our model non-trivially preserves the Ising heavy-fermion physics of the original, while allowing us to tune between the antiferromagnet and hidden order using pressure analogues and magnetic field.  Our model is also consistent with recent phenomenological predictions of not one, but two vector order parameters associated with the spinorial order that provide the first microscopic predictions for detecting the spinorial nature.  

\end{abstract}
\maketitle
\section{Introduction}

The Ising anisotropic heavy-fermion material URu$_2$Si$_2$ has been subject to intense study for over thirty-five years due to the $T_c = 17.5$K phase transition into a state known as hidden order (HO) \cite{Palstra1985, MydoshReview, Schmidt2010, Aynajian2010,Nagel2012,Trinh2015, Zhang2020}.  The main mystery is that the large entropy of $\sim 1/3 R \ln 2$ at $T_c$  \cite{Palstra1985} suggests a large order parameter, but none has been found \cite{Broholm1991,Wiebe2007}.  This mystery has prompted a plethora of theoretical proposals \cite{Amitsuka1994,Haule2009,Santini1994,Varma2006,Pepin2011,Dubi2011,Fujimoto2011,Ikeda2012,Mydosh2011,Morr2012,Kee2012,Chandra2013,Kung2015,Hsu2014, Harrison2021}, but none have really been satisfactorily reconciled with the complex and often experimentally ambiguous physics of HO.  One of the main difficulties is the proximity to a large moment antiferromagnetic phase (LMAFM) under pressure, chemical substitution and 
magnetic field \cite{Amitsuka1999, Ran2017, Knafo2020, Wolowiec2021, Kunwar2022} with closely related electronic properties and translation symmetry breaking \cite{Villaume2008,Hassinger2010,Meng2013, Frantzeskakis2021, Zhang2018}.  In addition, the Ising anisotropy, unknown valence and crystal-field ground states \cite{Jeffries2010,Sundermann2016,Booth2016}, and possible tetragonal \cite{Okazaki2011,Tonegawa2014,Riggs2015,Choi2018,Ghosh2020,Wang2020} and time-reversal symmetry breaking \cite{Schemm2015} challenge theorists to develop a unified theory of both the HO and the LMAFM in all their complexity.

One unique theoretical proposal is hastatic order \cite{Chandra2013,Chandra2015}, which naturally explains the Ising heavy-fermion physics.  It is a symmetry-breaking heavy Fermi liquid, where the primary order parameter is the hybridization associated with valence fluctuations between a ground state non-Kramers doublet and an excited state Kramers doublet.  This hybridization is therefore a consequence of the two-channel Kondo physics associated with a tetragonal U $5f^2$ $\Gamma_5$ configuration \cite{Cox1998, Amitsuka1994}, and is naturally spinorial as it mixes states that differ by half-integer angular momentum.  Hastatic order would be the first material realization of spinorial order.  Theoretically, it has been found in the two-channel Kondo model, in both infinite dimensions \cite{Hoshino2011, Hoshino2012, Hoshino2013, Hoshino2014} and in one-dimension \cite{Shauerte2005, Kornjaca2022}; and it can potentially be realized in any crystal symmetry with a non-Kramers doublet ground state \cite{Cox1998, Zhang2018, Zhang2020}.  In a model for URu$_2$Si$_2$, hastatic order captures the heavy-Fermi liquid \cite{Nagel2012,Aynajian2010,Schmidt2010,Park2012,Zhang2020,Palstra1985, Buhot2020} and Ising anisotropy \cite{Ohkuni1999,Hassinger2010,Altarawneh2011,Trinh2015,Bastien2019}, including spin-zeros in de Haas-van Alphen (dHvA) oscillations arising from an Ising form factor of the heavy quasiparticles \cite{Ohkuni1999,Altarawneh2011,Bastien2019}.  The LMAFM and HO are explained by orthogonal spinor orientations: out of the plane for the LMAFM and in-plane for the HO.

While hastatic order is an intriguing possibility, there are several weaknesses of the original model that require a more careful treatment, which we give in this paper. First, the original hastatic model predicts tetragonal symmetry breaking that is still experimentally unclear \cite{Okazaki2011,Tonegawa2014,Riggs2015,Choi2018,Bridges2020,Ghosh2020,Wang2020}; one aspect of this symmetry breaking was predicted small in-plane magnetic moments that were not found by neutrons \cite{Metoki2013,Das2013,Ross2014}.  Second, while both phases are conceptually treated with different orientations of the hastatic spinor, they could not be microscopically treated on the same footing, and only the HO was treated microscopically.  Third, the original microscopic model had a simplified structure that resulted in doubly degenerate bands throughout the Brillouin zone, which is not generic; this raises the concern that the spin-zeros may not survive in a more generic model.  Finally, there were no predictions associated with the spinorial nature of the order parameter specifically, as all predictions could be explained by an on-site vector order parameter.  We recently addressed many of these concerns by introducing a general Landau theory framework of tetragonal hastatic order \cite{Kornjaca2020}, which argued that the unusual nature of the order parameter, combined with disorder could explain the observed tetragonal symmetry breaking signatures and absence of in-plane moments. The Landau theory also predicts multiple vectorial order parameters stemming from the microscopic spinorial nature of the order, allowing for the spinorial nature to be tested directly.

This paper provides a unified microscopic hastatic theory of the HO and LMAFM phases, confirms the Landau theory predictions, and makes new predictions. The main change is that to capture both HO and LMAFM phases, two conduction electron bands are needed, while the original theory contained only one \cite{Chandra2013}.  We explore the generic phase diagram of the model, where the additional complexity allows us to consistently tune between candidate hastatic phases for HO and LMAFM both by applying pressure and magnetic field. While the phenomenological Landau theory \cite{Kornjaca2020} predicted spinorial signatures of the hastatic order, in this work we show how these signatures materialize in a realistic microscopic model.  The successes of the original microscopic theory are shown to still apply, including the spin-zeros associated with the Ising heavy Fermi liquid. Lastly, we show that hastatic order provides a framework for explaining the similarity between the electronic properties of the LMAFM and HO phases.

The structure of this paper is as follows. We begin by developing a realistic microscopic model of hastatic order in URu$_2$Si$_2$ in Sec. \ref{sec:anderson}. As the model is an interacting two-channel Anderson valence fluctuation model, we employ an $SU(N)$ large-$N$ mean-field treatment (Sec. \ref{sec:hastaticMF}) that provides multiple hastatic Ans\" atze as possible solutions. The competition between the different hastatic phases relevant for URu$_2$Si$_2$ in zero field is explored in Sec. \ref{sec:zeroBpds}, while the direct experimental signatures, including moments and susceptibilities are discussed in  Sec. \ref{sec:momsusc}. Due the the significance of the $\hat z$-axis magnetic field to the URu$_2$Si$_2$ phase diagram, the hastatic phases in-field and their properties are presented in Sec. \ref{sec:magfield}. Sec. \ref{sec:spinzeros} examines the Fermi surface properties, including the status of the dHvA spin-zeros in this more realistic model. Finally, we summarize our conclusions in Sec. \ref{sec:conclusions}.

\section{Realistic valence fluctuation model\label{sec:anderson}}

Hastatic order can arise whenever there are valence fluctuations between ground state and excited state doublets, which can described within an infinite-$U$ Anderson valence fluctuation model that contains the following basic ingredients:
\begin{equation}\label{eq:Hamsum}
    H=H_c+H_f+H_{VF}+H_{at}.
\end{equation}
These include the conduction electron kinetic energy ($H_c$), valence fluctuations ($H_{VF}$), local (atomic) interaction terms for the $f$-ions ($H_{at}$) and an effective $f$-electron hopping ($H_f$).  We are interested in developing a more realistic model for URu$_2$Si$_2$ than previously considered \cite{Chandra2013}, which we do by choosing conduction electrons arising from the Ru $d$ electrons. As there are two Ru per unit cell, our minimal model has two conduction electron bands, which we take to be $d_{z^2}$ electrons.  The previous model considered a single band of $s$-wave conduction electrons located at the U sites, a simplification that prevented the model from treating both hidden order and antiferromagnetism.  In the following subsections, we discuss each Hamiltonian component in detail.

\begin{figure}[!htb]
\includegraphics[width=1.0\columnwidth]{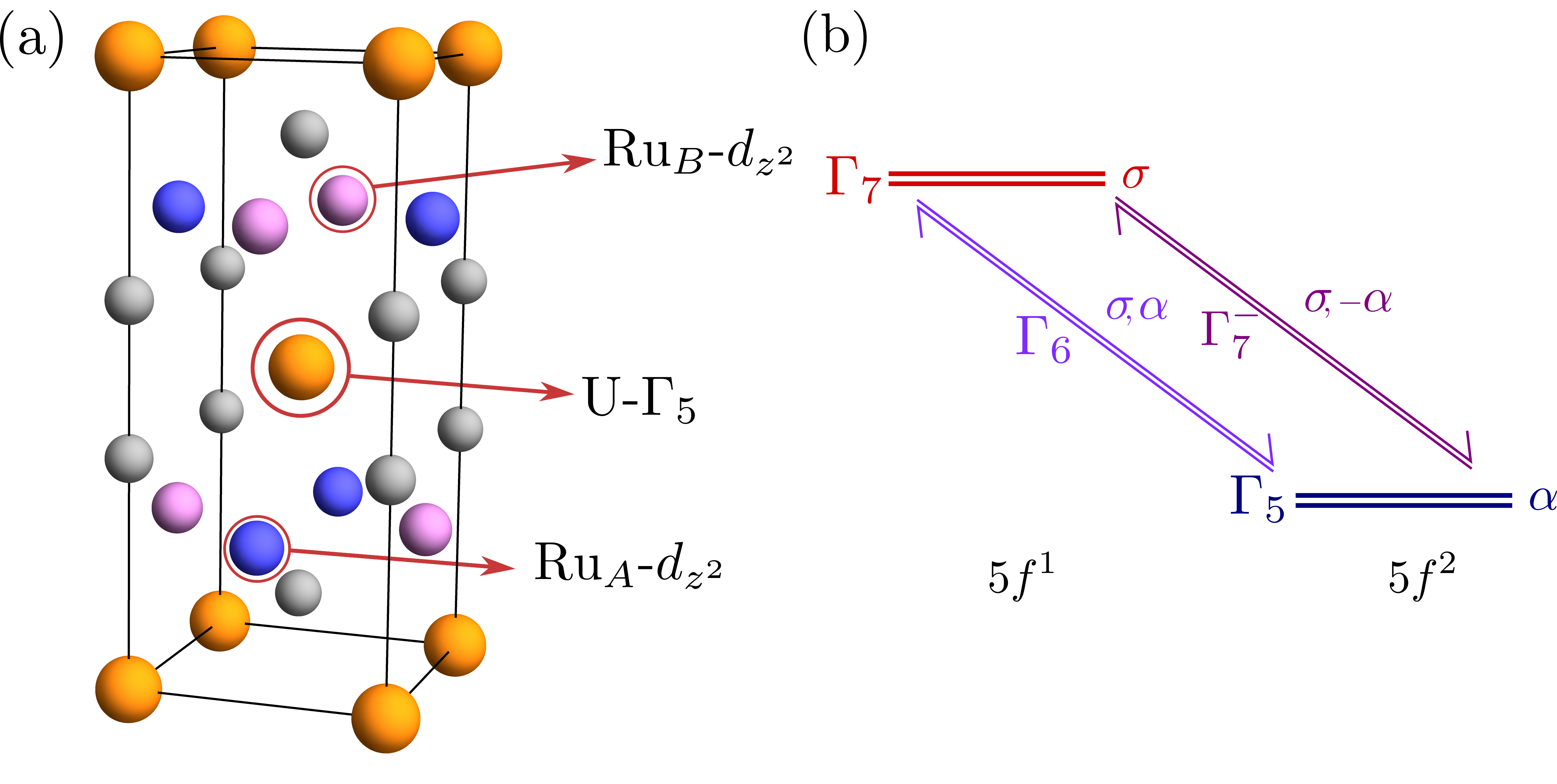}
\caption{(a) URu$_2$Si$_2$ in the conventional unit cell. U$^{4+}$ is in a tetragonal crystal field environment, with the Ising $\Gamma_5$ doublet as the ground state. Conduction electrons come from Ru and we choose the simplest case that provides generic hybridization: $d_z^2$ orbitals from the two Ru sites per unit cell. (b) Valence fluctuations between the $U^{4+}$ ground state $\Gamma_5$ non-Kramers doublet and an excited $\Gamma_7^+$ Kramers doublet lead to two-channel Kondo physics, where the on-site symmetries of the relevant conduction electrons are in $\Gamma_6 \oplus \Gamma_7^-$.  Effectively, the channel symmetry is that of the $\Gamma_7^+$ doublet. Hastatic order occurs when the excited doublet acquires a finite occupation, which allows a small excited state moment ot develop that breaks the channel symmetry. \label{fig:Fig1}}
\vspace{-0.cm}
\end{figure}

\subsection{Valence fluctuations in \texorpdfstring{URu$_2$Si$_2$}{URu2Si2}\label{sec:valfluct}}
Hastatic order requires valence fluctuations between two doublets: one ground state doublet with $f^{n}$ electrons and one excited state doublet with $f^{n\pm 1}$ electrons.  Such a situation can occur naturally for non-Kramers $f$-ions in sufficiently high symmetries, such as U$^{4+}$ in tetragonal symmetry, which can have a $5f^2$ $\Gamma_5$ ground state non-Kramers doublet. The excited ($5f^1$ or $5f^3$) states are Kramers doublets protected by time-reversal symmetry. While the ground state of URu$_2$Si$_2$ is controversial, we will assume this $5f^2$ ($J=4$) U$^{4+}$ non-Kramers ground state, where: \begin{equation}\label{eq:g5doublet}
|\Gamma_5, \pm\rangle=\cos{\xi} |\pm 3\rangle + \sin{\xi} |\mp 1\rangle.
\end{equation}
$\xi$ is a crystal field parameter that can be obtained by fitting experimental data.
Both $5f^1$ and $5f^3$ give similar excited doublet physics, so we choose an excited $5f^1$ $J=5/2$ $\Gamma_{7}^+$ doublet without loss of generality, as in the original proposal \cite{Chandra2013}:
\begin{equation}\label{eq:g7doublet}
|\Gamma_7^+, \pm\rangle=\cos{\eta} |\pm \frac{5}{2}\rangle + \sin{\eta} |\mp \frac{3}{2}\rangle,
\end{equation}
where $\eta$ is again a crystal field parameter.  Our infinite-$U$ Anderson model assumes that these two doublets are the only accessible $f$-electron configurations, with all other configurations at infinite energy.

The valence fluctuations between ground and excited doublets are shown in Fig. \ref{fig:Fig1}(b), and take the form \cite{Chandra2013}:
\begin{align}\label{eq:Valfluct}
H_{VF}(j)=&\sum_{\alpha,\sigma}\Big[V_6c\dg_{\Gamma_6\alpha}(j)\delta_{\sigma \alpha}|\Gamma_7^+\sigma\rangle\langle \Gamma_5 \alpha|\\
&+V_7c\dg_{\Gamma_7^-\alpha}(j)\delta_{\sigma \bar{\alpha}}|\Gamma_7^+\bar{\alpha}\rangle\langle \Gamma_5 \alpha|+h.c.\Big]\notag
\end{align}
within an infinite-$U$ Anderson model, where Hubbard operators like $|\Gamma_7^+\sigma\rangle\langle \Gamma_5 \alpha|$ are used to ensure only the relevant $f$ states are involved. $\alpha$ labels the pseudospin of the $\Gamma_5$ doublet, with $\bar{\alpha}=-\alpha$; $\sigma$ labels the spin of the excited state; $j$ labels the site; and $\delta_{\alpha \sigma}$ are Kronecker delta functions. The atomic Hamiltonian for the excited doublet is simply:
\begin{align}\label{eq:Hat}
H_{at}(j)=\Delta E \sum_{\sigma}|\Gamma_7^+\sigma\rangle\langle \Gamma_7^+ \sigma|,
\end{align}
with $\Delta E$ being the energy difference between ground and excited states. Two symmetry distinct channels of conduction electrons mediate the fluctuations, $\Gamma_6$ and $\Gamma_7$, stemming from $\Gamma_5\otimes\Gamma_7=\Gamma_6\oplus\Gamma_7$ \cite{Cox1998}. While the conduction electrons of required symmetry can be formed from an arbitrary band choice, we present a material realistic choice in the next section. 

\subsection{Conduction electrons \label{sec:condelhop}}

The main change in our model compared to the original model is the choice of conduction electrons.  Instead of $s$ electrons on the U sites, we choose the $d$ electrons of Ru.  This choice is motivated by density functional theory \cite{Oppeneer2010} that finds light Ru $d$ bands around the Fermi energy. We specifically choose $d_{z^2}$, as shown in Fig. \ref{fig:Fig1}(a), which simplifies the theory by restricting the number of conduction bands to two, originating from the two Ru sites per unit cell. This particular choice gives relatively generic behavior. Choosing the doublet $(d_{xz}, d_{yz})$ significantly increases the complexity of the model, while $d_{xy}$ and $d_{x^2-y^2}$ lead to non-generic simplifications in the resulting hybridization terms (see. Sec. \ref{sec:hybr}).
While treating only the $d_{z^2}$ Ru orbitals is a simplification, the generic nature of the two band hybridization allows us to explore generic features of hastatic order and, in particular to treat both HO and LMAFM phases on equal footing.

To obtain the conduction electron Hamiltonian, we use the Slater-Koster method to find symmetry allowed Ru-Ru hoppings up to the third neighbor. The method is described in detail in Appendix \ref{app:SC} and gives:
\begin{equation}
H_c=\sum_{\vect{k}\beta\beta'\sigma}\epsilon_c\left(\vect{k},\beta,\beta'\right)c\dg_{\vect{k}\beta\sigma}c_{\vect{k}\beta'\sigma},
\end{equation}
where
\begin{align}\label{eq:elbands}
\epsilon_c&\left(\vect{k},\beta,\beta'\right)=\left[-t_2\left(\cos{k_x}+\cos{k_y}\right)-\mu\right]\delta_{\beta,\beta'}\\
&+\left[-t_1\cos{\frac{k_x}{2}}\cos{\frac{k_y}{2}}-t_3\cos{\frac{k_z}{2}}\right]\left(1-\delta_{\beta,\beta'}\right)\notag.
\end{align}
$\beta, \beta'$ label the two Ru sites per unit cell (Ru$_A$ and Ru$_B$), while $\mu$ is the chemical potential, and $\sigma$ is the Ru spin; we neglect any spin-orbit effects on the Ru bandstructure.  We have set the lattice spacings $c = a = 1$. The hopping $t_1$ describes nearest neighbor Ru$_A$-Ru$_B$ hopping, $t_2$ is the second neighbor Ru$_A$-Ru$_A$ and Ru$_B$-Ru$_B$ hopping within the Ru planes, while $t_3$ introduces a $z$-dispersion as it couples Ru$_A$-Ru$_B$ in neighboring Ru planes.

\subsection{\texorpdfstring{$f$}{f} electrons \label{sec:felectrons}}

In order to capture the partially itinerant character of the U $5f$ electrons \cite{Moore2009,Oppeneer2010}, we include some $f$ electron hopping. In an infinite-$U$ Anderson model, we no longer have bare $f$ electrons hopping between sites, but rather emergent hopping arising from fluctuations.  These emergent hoppings can arise from valence fluctuations directly, as projected hopping terms, or may be treated as spin-liquid-like dispersive terms arising from inter-site RKKY interactions. To understand the range of possible $f$-dispersions, we consider both possible sources.

To treat the valence fluctuation mediated hopping in our infinite-$U$ Anderson model, the $f$-electrons have projected hoppings:
\begin{equation}\label{eq:vfFhop}
   H_{f,f}^{(1)} = \sum_{ij} t_{f,ij}^{0,\alpha\alpha'}\left(|\Gamma_7^+ \sigma, i\rangle\langle \Gamma_5 \alpha, i|\right)\left(|\Gamma_5 \alpha', j\rangle\langle \Gamma_7^+ \sigma, j|\right).
\end{equation}
This describes hopping between $\Gamma_5$ states through the partially occupied excited state doublet. Alternately, a pseudospin interaction between $\Gamma_5$ doublets can also lead to dispersion,
\begin{equation}\label{eq:HeisFhop}
H_{f,f}^{(2)} =  \sum_{<ij>,a}J_H^{(a)}\bar{\tau}^{(a)}_i\bar{\tau}^{(a)}_j,
\end{equation}
where $\bar{{\tau}}^{(a)}$ is the pseudospin of the $\Gamma_5$ doublet, with $a$ labeling the pseudospin components. This RKKY interaction is always present in materials, but must be explicitly added to be treated in the large-$N$ limit. The interactions are generically non-Heisenberg in tetragonal symmetry, with $J_H^{(\perp)}\neq J_H^{(z)}$, although we will take the Heisenberg limit here, as it is easier to treat. Both terms are discussed further, where we introduce the mean-field treatment in Sec. \ref{sec:mftheory}, and both lead to relatively simple $f$-hopping terms.

\subsection{Hybridization form factors \label{sec:hybr}}

The Ru $d_{z^2}$ conduction electrons hybridize with the $\Gamma_5$ doublet via Wannier states with $\Gamma_6$ and $\Gamma_7^-$ symmetry,
\begin{align}\label{eq:v67def}
V_6c\dg_{\Gamma_6\alpha}(j)&=\frac{1}{\sqrt{N_s}}\sum_{\vect{k}\beta\sigma}\hat{V}_{6,(\beta\sigma,\alpha)}(\vect{k})c\dg_{\vect{k}\beta\sigma}\mathrm{e}^{-i\vect{k}\cdot \bf{R}_j},\\
V_7c\dg_{\Gamma_7^-\alpha}(j)&=\frac{1}{\sqrt{N_s}}\sum_{\vect{k}\beta\sigma}\hat{V}_{7,(\beta\sigma,\alpha)}(\vect{k})c\dg_{\vect{k}\beta\sigma}\mathrm{e}^{-i\vect{k}\cdot \bf{R}_j},
\end{align}
where $N_s$ denotes the number of sites, $\beta\sigma$ indices enumerate the $(\mathrm{Ru}_A\uparrow, \mathrm{Ru}_A\downarrow,\mathrm{Ru}_B\uparrow, \mathrm{Ru}_B\downarrow)$ basis and $\hat{V}_{6/7}$ are hybridization form factor matrices that can be obtained using the Slater-Koster method described in Appendix \ref{app:SC}. Keeping only the nearest neighbor Ru sites for each U, we obtain the spin-orbit coupled hybridization matrices:
\begin{widetext}
\begin{align}\label{v67mat}
\hat{V}_6(\vect{k})=
\begin{pmatrix}
V_{6}^{(1)}\left(\mathrm{e}^{i k_z/4}\cos{\frac{k_x}{2}}-\mathrm{e}^{-i k_z/4}\cos{\frac{k_y}{2}}\right) & V_{6}^{(2)}\left(i\mathrm{e}^{i k_z/4}\sin{\frac{k_x}{2}}+\mathrm{e}^{-i k_z/4}\sin{\frac{k_y}{2}}\right)\\
V_{6}^{(2)}\left(i\mathrm{e}^{i k_z/4}\sin{\frac{k_x}{2}}-\mathrm{e}^{-i k_z/4}\sin{\frac{k_y}{2}}\right) & -V_{6}^{(1)}\left(\mathrm{e}^{i k_z/4}\cos{\frac{k_x}{2}}-\mathrm{e}^{-i k_z/4}\cos{\frac{k_y}{2}}\right)\\
V_{6}^{(1)}\left(\mathrm{e}^{i k_z/4}\cos{\frac{k_x}{2}}-\mathrm{e}^{-i k_z/4}\cos{\frac{k_y}{2}}\right) &  V_{6}^{(2)}\left(i\mathrm{e}^{-i k_z/4}\sin{\frac{k_x}{2}}+\mathrm{e}^{i k_z/4}\sin{\frac{k_y}{2}}\right)\\
V_{6}^{(2)}\left(i\mathrm{e}^{-i k_z/4}\sin{\frac{k_x}{2}}-\mathrm{e}^{i k_z/4}\sin{\frac{k_y}{2}}\right) & -V_{6}^{(1)}\left(\mathrm{e}^{i k_z/4}\cos{\frac{k_x}{2}}-\mathrm{e}^{-i k_z/4}\cos{\frac{k_y}{2}}\right),
\end{pmatrix}
\end{align}
\begin{align}
\hat{V}_7(\vect{k})=
\begin{pmatrix}
-V_{7}^{(1)}\left(\mathrm{e}^{i k_z/4}\cos{\frac{k_x}{2}}+\mathrm{e}^{-i k_z/4}\cos{\frac{k_y}{2}}\right) & V_{7}^{(2)}\left(i\mathrm{e}^{i k_z/4}\sin{\frac{k_x}{2}}-\mathrm{e}^{-i k_z/4}\sin{\frac{k_y}{2}}\right)\\
V_{7}^{(2)}\left(i\mathrm{e}^{i k_z/4}\sin{\frac{k_x}{2}}+\mathrm{e}^{-i k_z/4}\sin{\frac{k_y}{2}}\right) & V_{7}^{(1)}\left(\mathrm{e}^{i k_z/4}\cos{\frac{k_x}{2}}+\mathrm{e}^{-i k_z/4}\cos{\frac{k_y}{2}}\right)\\
V_{7}^{(1)}\left(\mathrm{e}^{i k_z/4}\cos{\frac{k_x}{2}}-\mathrm{e}^{-i k_z/4}\cos{\frac{k_y}{2}}\right) &  V_{7}^{(2)}\left(i\mathrm{e}^{-i k_z/4}\sin{\frac{k_x}{2}}-\mathrm{e}^{i k_z/4}\sin{\frac{k_y}{2}}\right)\\
V_{7}^{(2)}\left(i\mathrm{e}^{-i k_z/4}\sin{\frac{k_x}{2}}+\mathrm{e}^{i k_z/4}\sin{\frac{k_y}{2}}\right) & -V_{7}^{(1)}\left(\mathrm{e}^{i k_z/4}\cos{\frac{k_x}{2}}-\mathrm{e}^{-i k_z/4}\cos{\frac{k_y}{2}}\right)
\end{pmatrix},
\end{align}
\end{widetext}
where the $z$ position of Ru atoms is taken as $c/4$ with respect to central U atom (see Fig. \ref{fig:Fig1}). 

There are four independent parameters, $V_6^{(1)},V_6^{(2)},V_7^{(1)},V_7^{(2)}$ that are generically non-zero for $d_{z^2}$ Ru hybridization. We find that only the ratio of overall $\Gamma_7$ to $\Gamma_6$ channel strength has qualitatively significant consequences. Therefore, for simplicity, we constrain the parameters such that $V_6^{(1)}=V_6^{(2)}=V_6$ and $V_7^{(1)}=V_7^{(2)}=V_7$. We fix the overall bare hybridization, ($V_6^2+V_7^2=V^2$), such that the resulting Kondo coupling, $J_K = V^2/\Delta E$ self-consistently gives $T_c/D=1/30$ for standard parameter choices (see Fig. \ref{fig:Fig3a}). Effectively, the free parameter is the $V_7/V_6$ ratio, which is a proxy for $\sqrt{[(V_7^{(1)})^2+(V_7^{(2)})^2]/[(V_6^{(1)})^2+(V_6^{(2)})^2]}$. 
As this ratio depends on Ru-U overlaps, it should be readily tuned by applied pressure or strain.

\subsection{Coupling to magnetic field \label{sec:Bcoupling}}

URu$_2$Si$_2$ has a complicated response to magnetic fields as evidenced by Ising anisotropic signatures \cite{Ohkuni1999,Trinh2015,Bastien2019} and the complex phase diagram in strong $z$-directed field \cite{Aoki2009,Ran2017,Knafo2020, Wolowiec2021, Kunwar2022}. To capture these effects, we incorporate realistic magnetic field couplings based on a tetragonal crystal field model for a local U atom with parameters ($\xi$ and $\eta$) fit to independent thermodynamic signatures, as described in Appendix \ref{app:CEF}. The result to the first order in magnetic field is:
\begin{align}\label{eq:orBcoup}
    H_B(j)& = -\frac{g_c\mu_B}{2} \sum_{\sigma\sigma'\beta}c\dg_{j\beta\sigma}\vec{B} \cdot \vec{\tau}_{\sigma,\sigma'}c_{j\beta\sigma'}
    \cr 
    &-\frac{g_f\mu_B}{2}\sum_{\alpha\alpha'}B_z\tau^{z}_{\alpha,\alpha'}|\Gamma_5 \alpha\rangle\langle \Gamma_5 \alpha'|\cr
   &-\frac{\mu_B}{2}\sum_{\sigma\sigma'}\!\left(g_{ex}^zB_{z}\tau^{z}_{\sigma\sigma'}+g_{ex}^\perp\vec{B}_{\perp}\cdot\vec{\tau}^{\perp}_{\sigma\sigma'}\!\right)|\Gamma_7^+ \sigma\rangle\langle \Gamma_7^+ \sigma'|\cr
\end{align}
where $\vec{\tau}$ denotes the Pauli matrices in spin and the pseudospin sub-spaces. The bare $g$-factors above correspond to the conduction electrons, $g_c=2$, the $\Gamma_5$ doublet, $g_f=4.04$, and the $\Gamma_7$ doublet, both $z$, $g_{ex}^z=2.98$ and in-plane, $g_{ex}^\perp=2.28$.

The Ising nature of the $\Gamma_5$ doublet is evident in Eq. (\ref{eq:orBcoup}), as it couples linearly only to $B_z$, which generates the Ising heavy Fermi liquid signatures of the HO phase, as well as the anisotropy of the phase diagrams in magnetic field, where inaccessibly large in-plane fields are needed to affect HO/LMAFM competition. For this reason, we neglect the $\mathcal{O}(B^2)$ in-plane $\Gamma_5$ coupling. On the other hand, the excited doublet has linear splitting for all field directions, with significant anisotropy due to the tetragonal environment. This anisotropy is an important input for determining the sizes of induced magnetic moments in hastatic phases. The consequences of the magnetic field splitting of the ground and excited state doublets are discussed in detail in Sec. \ref{sec:magfield}.

\section{Hastatic order Ans\" atze and order parameters\label{sec:hastaticMF}}

Hastatic order is a channel symmetry breaking hybridization that emerges very similarly to the non-symmetry breaking hybridization in a typical single channel Kondo model.
Hastatic orders can be captured within a mean-field treatment of the above infinite-$U$ Anderson model by rewriting the Hubbard operators using auxiliary bosons representing the excited $\Gamma_7^+$ state that condense at the hastatic phase transition, $\langle b^\dagger_{j\sigma}\rangle$.  The resulting quadratic Hamiltonian is exact in the large-$N$ limit, where the $SU(2)$ ground state doublet is promoted to $SU(N)$, and can be compared to both the Landau theory of tetragonal hastatic order \cite{Kornjaca2020} and to experimental measurements.

\subsection{Mean field theory of hastatic order\label{sec:mftheory}}

Our infinite-$U$ Anderson model is written in terms of Hubbard operators. To make the model amenable to mean-field treatment, we use the standard auxiliary boson formalism \cite{Coleman1984, Chandra2013}, in which the excited state doublet is represented by auxiliary bosons, $b_{j\sigma}$, and the ground state doublet by pseudofermions, $f_{j\alpha}$:
\begin{equation}\label{eq:auxbdecoup}
|\Gamma_7^+ \sigma\rangle=b\dg_{\sigma} |\Omega\rangle, \qquad |\Gamma_5 \alpha\rangle=f\dg_{\alpha}|\Omega\rangle.
\end{equation}
Here, $\mathbf{b}$ and $\mathbf{f}$ are both spinors, although $\mathbf{b}$ represents the Kramers excited state, and $\mathbf{f}$ represents the non-Kramers ground state.

In the infinite-$U$ model, the overall occupation of ground and excited states at each site must be fixed to one, 
\begin{equation}\label{eq:constraint}
    \sum_{\alpha}f\dg_{j\alpha}f_{j\alpha}+\sum_{\sigma}b\dg_{j\sigma}b_{j\sigma}=1.
\end{equation}
We will enforce this constraint by introducing a Lagrange multiplier, $\lambda_j$.

We need to take special care with the $f$-hopping Hamiltonian, where we consider both the valence fluctuation and RKKY interaction origins.  We can rewrite the valence fluctuation mediated $f$-hopping given in Eq. (\ref{eq:vfFhop}) using auxiliary bosons to find the quartic term, $t_{f,ij}^{0,\alpha\alpha'}b_{i,\sigma}b\dg_{j,\sigma}f\dg_{i,\alpha}f_{j,\alpha'}$. The condensation of auxiliary bosons will then lead to simple $f$-hopping terms that reflect the symmetry of the hastatic ansatz.  The pseudospin RKKY interaction of Eq. (\ref{eq:HeisFhop}) between $\Gamma_5$ doublets can also be rewritten in terms of pseudofermions, using $\bar{\tau}^{a}_i=\frac{1}{2}f\dg_{i\alpha}\tau^{(a)}_{\alpha\alpha'}f_{i\alpha'}$, again leading to a quartic interaction.  This quartic interaction can be decoupled into $f$-hopping terms using mean-field theory justified in the large-$N$ limit \cite{Arovas1988,Zhang2018}. Regardless of the origin of the hopping, the resulting term in the Hamiltonian takes the form:
\begin{equation}\label{eq:fhopdec}
    H_{f,f} = \sum_{ij,\alpha \alpha'}t_{ij}^{\alpha \alpha'} f^\dagger_{i\alpha} f_{j\alpha'}.
\end{equation}
Most of the time, we will take the $f$-dispersion to be consistent with tetragonal symmetry, but this is not required, and projected hopping generically \emph{breaks} symmetries, as we will discuss in section \ref{sec:dautheory}.
A simple form can be obtained by taking the nearest neighbor overlaps of U $\Gamma_5$ orbitals using the Slater-Koster method (see Appendix \ref{app:SC}). Overlaps between $\Gamma_5$ doublets on neighboring U give the hopping,
\begin{equation}\label{eq:fbands}
\epsilon_f\left(\vect{k},\alpha\right)=-t_f\cos{\frac{k_x}{2}}\cos{\frac{k_y}{2}}\cos{\frac{k_z}{2}}.
\end{equation}
These $f$-hopping terms are required to realize the full spinorial nature of the order, as they allow the interference of spinors at neighboring sites, as found in Landau theory \cite{Kornjaca2020} and discussed below.

The full Hamiltonian in $B_z$ now takes the form:
\begin{widetext}
\begin{align}\label{eq:bHam}
    &H=\sum_{\vect{k}\beta\beta'\sigma}\epsilon_c\left(\vect{k},\beta,\beta',\sigma\right)c\dg_{\vect{k}\beta\sigma}c_{\vect{k}\beta'\sigma}+\sum_{\vect{k}\alpha}\epsilon_f\left(\vect{k},\alpha\right) f\dg_{\vect{k}\alpha}f_{\vect{k}\alpha}+\Delta E \sum_{j\sigma}b\dg_{j\sigma}b_{j\sigma}+\sum_{j}\lambda_j\left(\sum_{\alpha}f\dg_{j\alpha}f_{j\alpha}+\sum_{\sigma}b\dg_{j\sigma}b_{j\sigma}-1\right)\cr
    &+\frac{1}{\sqrt{N_s}}\sum_{\vect{k}j\alpha\beta\sigma}\left[\hat{V}_{6,(\beta\sigma,\alpha)}(\vect{k})c\dg_{\vect{k}\beta\sigma}b\dg_{j\sigma}f_{j\alpha}\mathrm{e}^{-i\vect{k}\cdot \bf{R}_j}+\hat{V}_{7,(\beta\sigma,\alpha)}(\vect{k})c\dg_{\vect{k}\beta\sigma}b\dg_{j\bar{\sigma}}f_{j\alpha}\mathrm{e}^{-i\vect{k}\cdot \bf{R}_j}+h.c.\right]+\sum_j\mu\left(n_c-\sum_{\sigma}b\dg_{j\sigma}b_{j\sigma}\right)\cr&
    -\frac{1}{2} g_c\mu_B B_z\sum_{\vect{k}\sigma\beta}c\dg_{\vect{k}\beta\sigma}{\tau^{z}}_{\sigma\sigma}c_{\vect{k}\beta\sigma}
    -\frac{1}{2}g_fB_z\mu_B\sum_{\vect{k}\alpha}f\dg_{\vect{k}\alpha}\tau^{z}_{\alpha,\alpha}f_{\vect{k}\alpha}-\frac{1}{2}\mu_Bg_{ex}^zB_z\sum_{j\sigma} b\dg_{j\sigma} \tau^{z}_{\sigma\sigma} b_{j\sigma}.
\end{align}
\end{widetext}
We have added a chemical potential term for the conduction electrons, as we switch to the canonical ensemble to preserve the conduction electron filling ($n_c$) as temperature/parameters are varied. The occupation of the excited doublet is subtracted from the usual constant $\mu n_c$ term in order to respect global charge conservation \cite{VanDyke2019}.

This model can be solved exactly in a large-$N$ $SU(N)$ treatment where the ground state multiplet  has $N$ components, $\alpha=\pm \frac{1}{2}, ..., \pm \frac{N}{2}$; the conduction electron spin index $\beta$ also has $N$ components. Effectively, the large-$N$ limit of this solution leads to a mean-field theory where the auxiliary boson spinors condense at low temperatures.  We take the \emph{amplitude} of the spinor to be uniform, but allow the $SU(2)$ spinor's \emph{direction} to retain spatial dependence:
\begin{equation}\label{eq:bmf}
   \langle b_j \rangle = |b|e^{i\chi_j}\left( \begin{array}{c} \cos \frac{\theta_j}{2} \mathrm{e}^{-i\phi_j/2} \\ \sin \frac{\theta_j}{2} \mathrm{e}^{i\phi_j/2}\end{array}\right), \qquad \lambda_j=\lambda.
\end{equation}
Here $|b|$ denotes the auxiliary boson amplitude, $\chi_j$ an overall phase, while $\theta_j$ and $\phi_j$ determine the spinor direction. We also assume that the constraint is enforced on average with $\lambda_j$ replaced by the uniform expectation value, $\lambda$. In contrast to the single-channel Kondo case, the condensation of these bosons leads to symmetry breaking, as the auxiliary boson occupation breaks channel, time reversal, and spatial symmetries [see Fig. \ref{fig:Fig2} (a)]. The resulting ordered phase is the symmetry breaking heavy-Fermi liquid that we call hastatic order \cite{Chandra2013}.

The possible phase space of hastatic orders is quite rich, as it contains not only uniform (ferrohastatic), but also staggered (antiferrohastatic) orders, potentially with a wide range of ordering wave-vectors and moment orientations. In addition to spatially varying orientations ($\theta_j$ and $\phi_j$), the overall phase of the spinor, $\chi_j$ can vary between sites, leading to unique spinorial orders with distinct broken symmetries \cite{Kornjaca2020}. 

There is a $U(1)$ gauge transformation inherent in the auxiliary boson representation:
\begin{align}\label{eq:gaugeT}
    &b_{j\sigma} \rarrow b_{j\sigma} \mathrm{e}^{-i\xi_j}, \qquad f_{j\alpha} \rarrow f_{j\alpha} \mathrm{e}^{-i \xi_j}, \\
    &t_{f,ij} \rarrow t_{f,ij} \mathrm{e}^{i(\xi_i-\xi_j)}.\notag
\end{align}
As the spinor itself is not gauge invariant, it is expected to be washed out via gauge fluctuations for any finite $N$, similar to $\langle b\rangle$ in the single channel Anderson model \cite{coleman2015book}.  However, as real symmetries are also broken by this spinorial order parameter, there must be gauge invariant order parameters that can be constructed from bilinears of the spinorial order parameter. One of these, the composite order parameter, $\vec{\Psi}_j = \langle b_j^\dagger \vec{\sigma}b_j\rangle$ is well known \cite{Shauerte2005,Hoshino2011,Zhang2018} and describes magnetic moments in the excited state $\Gamma_7^+$ doublet on each site.  This order parameter is blind to the spinorial nature (e.g. - $\chi_j$) of hastatic order, but we recently showed that signatures of the spinorial nature can survive and are carried by an additional inter-site vector order parameter \cite{Kornjaca2020}, $\vec{\Phi}_{ij}$ that picks up the phase differences between spinors on different sites.

For a given composite moment arrangement, $\vec{\Psi}_j$, there are actually multiple spinorial orders with distinct phase, $\chi_j$ arrangements that break different symmetries. These phases are essential because the spinor is only invariant under four operations of time-reversal, unlike vectorial moments that are invariant under double time-reversal symmetry.  If we consider a one dimensional arrangement, we have two distinct spinor possibilities that have the same composite moments. We can introduce $b_A$ as the spinor on sublattice A, and $\hat \theta = K i\sigma_2$ as the operation of time-reversal, where $K$ is complex conjugation.  If we have two sublattices (2SL), where $b_B = \hat \theta b_A$, the resulting staggered hastatic order will not be invariant under time-reversal followed by a lattice symmetry.  Only a \emph{four} sublattice (4SL) order, with $b_C = \hat \theta^2 b_A = -b_A$ and $b_D = \hat \theta^3 b_A = -b_B$ preserves a time-reversal like anti-unitary symmetry (here, time-reversal followed by lattice translation).  This scenario is sketched out in Fig. \ref{fig:Fig2}(c,d).  Note that only the phases, $\chi_j$ vary between the 2SL and 4SL orders, while the moments $\vec{\Psi}_j$ are identical.  Nevertheless, these two phases break different symmetries, which are detected by an order parameter capturing the intersite interference between different sublattices; this order parameter requires $f$-electron hopping, $\vec{\Phi}_{ij} = t_{f,ij}\langle b\dg_i\vec{\sigma}b_j\rangle$, as sketched out in Fig. \ref{fig:Fig2}.  The Landau-Ginzburg theory of these order parameters was explored in Ref. [\onlinecite{Kornjaca2020}], and we summarize it briefly here.

\subsection{Vector order parameters and Landau theory of tetragonal hastatic order\label{sec:dautheory}}

The nature of the hastatic order depends on the spatial ordering.
For simplicity, we will limit ourselves to a few possible ordering wave-vectors $\bf{Q}$ associated with the staggering of the composite moments. A simple ($Q = \pi$) antiferrohastatic order is found at quarter-filling in the simple two channel Kondo model in one dimension from density matrix renormalization group (DMRG) studies \cite{Shauerte2005}, as well as in a range of fillings for infinite dimensions in dynamical mean-field theory (DMFT) studies \cite{Hoshino2012, Hoshino2014}.  Ferrohastatic order has been found as a metastable state near half-filling in DMFT \cite{Hoshino2013}.  All of these results are for simple $SU(2)$ two-channel Kondo models, which may be derived from cubic Anderson models in the Kondo limit.  Our tetragonal model has a more complicated Kondo limit \cite{Cox1998} and we expect a more complicated phase diagram.  As we are primarily interested in hastatic order as a candidate for HO in URu$_2$Si$_2$, we will focus on a single staggering pattern, with $\bf{Q} = [001]$ as found for the HO \cite{Wiebe2007,Villaume2008,Bareille2014}, as well as the competition with $\bf{Q} = 0$ ferrohastatic order.

Generically, there are three gauge invariant bilinear quantities of physical interest that can develop independently, in principle. In the mean-field ($N\rarrow \infty$) limit used in this paper, all three onset at the same transition, but could onset at three different temperatures in more realistic situations. 

The first of these is the local excited state occupation:
\begin{equation}\label{eq:nb}
    n_{b,i} = \langle b\dg_i b_i\rangle,
\end{equation}
which does not break any symmetries. It is responsible for the development of heavy-Fermi liquid signatures above the hastatic transition in a regime that might be described as ``parahastatic''; the hastatic spinors gain a small magnitude, but do not order.

The second quantity is the composite vectorial order parameter:
\begin{equation}\label{eq:psi}
    \vec{\Psi}_{i} = \langle b_i\dg\vec{\sigma}b_i\rangle.
\end{equation}
It represents the moments of the local excited doublet, which form an $SO(3)$ order parameter in the cubic case, but are decomposed into out of plane, $\Psi_z$ (Ising) and in-plane, $\vec{\Psi}_\perp$ (XY) components in tetragonal symmetry. This order parameter is associated with the diagonal composite order reported in DMFT \cite{Hoshino2013}, and carries the main thermodynamic signatures of the hastatic transition. It is also responsible for moments in the direction of the hastatic spinor (see Tab. \ref{tab:table1} for staggered phases) and for in-plane hastatic phases, $\vec{\Psi}_\perp$ also leads to tetragonal symmetry breaking. Effectively, this order parameter is a magnetic order parameter, although its magnitude is quite small, suppressed by $T_c/D$ \cite{Chandra2013}, where $D$ is the conduction electron bandwidth.  

Finally, the spinorial nature is captured by the complex quantity,
\begin{equation}\label{eq:phi}
    \vec{\Phi}_{ij} = t_{f,ij}\langle b\dg_i\vec{\sigma}b_j\rangle,
\end{equation}
which is the result of interference between neighboring sublattices mediated by $f$-hopping and only exists for phases with inequivalent sublattices and $f$-electron hopping. Note that this order parameter is found in the two-channel Kondo model in one-dimension \cite{Kornjaca2022}. This order parameter captures a difference between staggered hastatic phases that has no magnetic equivalent, and it breaks additional symmetries (see the additional moments in Table \ref{tab:table1}). We will usually choose a uniform $t_{f,ij}$ between nearest-neighbors, in which case all the symmetry breaking arises from the spinor phases.  However, symmetry-breaking $t_f$'s are also possible, and it is even possible for the symmetry-breaking of $t_f$ and $\langle b\dg_i\vec{\sigma}b_j\rangle$ to cancel, rendering $\vec{\Phi}_j$ a non-symmetry breaking singlet. This scenario is actually the one considered in the original hastatic proposal \cite{Chandra2013}, and naturally occurs for projected $f$-hopping with staggered in-plane $\vec{\Psi}_j$.  However, there is no a priori reason to favor this $\vec{\Psi}$-only phase over phases with both $\vec{\Psi}$ and  $\vec{\Phi}$ present and we will treat all of them on the same footing. In the mean-field picture explored here, $\vec{\Phi}_j$ has two vectorial components, Re$\vec{\Phi}$ and Im$\vec{\Phi}$, which are strictly orthogonal to $\vec{\Psi}$ in the large-$N$ limit. These two components break additional symmetries, with Im$\vec{\Phi}$ breaking additional symmetries associated only with $\vec{\Phi}$ order, and Re$\vec{\Phi}$ contained in $\vec{\Psi} \otimes \mathrm{Re}\vec{\Phi}$. Note that the presence of $f$-hopping is necessary for the gauge invariance, and thus all effects dependent on $\vec{\Phi}$ are further suppressed by $D_f/D$, the ratio of $f$ to conduction electron bandwidth, for a total magnitude on the order of $D_fT_c/D^2$. 

\begin{figure}[!htb]
\includegraphics[width=0.95\columnwidth]{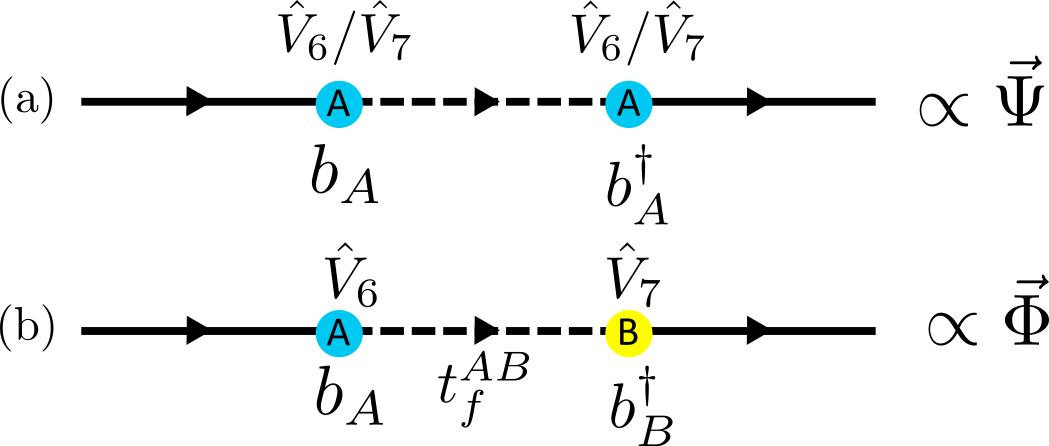}
\caption{Hastatic order parameters and their microscopic origin. While most of the hastatic order physics is captured by the composite order parameter, $\vec{\Psi}= \langle b\dg\vec{\sigma}b\rangle$, the spinorial nature of the order results in an additional vectorial order parameter, $\vec{\Phi}=t_f\langle b\dg\vec{\sigma}b\rangle$. The microscopic origins of these shown in this cartoon, where we see their effect on the conduction electron propagator; here, solid lines are the bare $c$ propagators, while the dashed lines represent $f$ propagators, and the circles are hybridizations at sites in either the $A$ or $B$ sublattices (blue, yellow, respectively).   $\vec{\Psi}$ (a) is the result of intra-sublattice scattering, while $\vec{\Phi}$ originates via interference between scattering channels on different sublattices, mediated by $f$-hopping. This scattering requires the existence of inequivalent sublattices, and thus at least antiferrohastatic order. \label{fig:Fig4}}
\vspace{-0.cm}
\end{figure}

A Landau theory analysis identified the broken symmetries associated with the different order parameters, which are summarized in the below table, where we indicate which moments are found in the 2SL and 4SL orders with out-of-plane ($\Psi_z$) or in-plane ($\vec{\Psi}_\perp$) composite moments.  The composite moments by themselves are staggered magnetic dipoles, while the two aspects of $\vec{\Phi}$ give uniform and staggered electric, magnetic and toroidal multipole moments of various ranks. For a given $\vec{\Psi}_j$, there are generally multiple 4SL phases, and we choose the one shown in Fig. \ref{fig:Fig2}, where the ABCD sublattices are stacked vertically.

\begin{table}[!ht]
\centering
\begin{tabular}{l|cc}
& $\Psi_z$ & $\vec{\Psi}_\perp$ \\
& staggered $m_z$ & staggered $\vec{m}_\perp$\\
\hline
\rule{0pt}{2.7ex}
2SL [$\mathrm{Im}\vec{\Phi}$] & $\mathrm{Re}\vec{\Phi}$ &$\mathrm{Re}\vec{\Phi}$\\
\; staggered ($Q_{xz}, Q_{yz}$) & uniform $\vec{m}_\perp$ & uniform $m_z$ \\
\hline
\rule{0pt}{2.7ex}
4SL [$\mathrm{Im}\vec{\Phi}$] & $\mathrm{Re}\vec{\Phi}$& $\mathrm{Re}\vec{\Phi}$ \\
\; uniform $\vec{p}_\perp$ & staggered $\vec{\Omega}_\perp$ & staggered $\Omega_z$ \\
\hline
\end{tabular}
\caption{\label{tab:table1} Possible antiferrohastatic (AFH) phases and associated moments, reproduced from \cite{Kornjaca2020}. For $t_f=0$ or other $\Psi$-only phases, $\vec{\Psi}$ moments are the only ones present, while for generic AFH phases, $\mathrm{Re}\vec{\Phi}$ or $\mathrm{Im}\vec{\Phi}$ moments are also nonzero. $m, p, \Omega$ refer to magnetic, electric and toroidal dipoles, respectively, while $Q$ indicates electric quadrupoles.}
\end{table}
The above moments are predicted by the Landau theory, and confirmed below within our realistic microscopic calculation.

\subsection{Mean-field equations \label{sec:MFequations}}

In the rest of the paper, we will self-consistently solve the mean-field equations provided by the realistic model constructed, for specific ansatze corresponding to different hastatic orders. In general, the mean-field decoupled Hamiltonian can be obtained from Eq. (\ref{eq:bHam}) by replacing $b_{j\sigma} \rarrow \langle b_{j\sigma}\rangle$ everywhere, leading to a quadratic fermionic Hamiltonian that can be conveniently separated into three parts:
\begin{equation}\label{eq:hamfandb}
    H=\sum_{\vect{k}}\mathcal{H}(\vect{k})+\mathcal{H}_0+\mathcal{H}_{ex}(B),
\end{equation}
where $\mathcal{H}(\vect{k})$ represents the $\vect{k}$-dependent quadratic contribution; $\mathcal{H}_0$ is a field independent constant given by:
\begin{align}\label{eq:H0}
\mathcal{H}_0=N_s\left[\Delta E |b|^2+\lambda\left(|b|^2-1\right)+\mu\left(n_c-|b|^2\right)\right];
\end{align}
and $\mathcal{H}_{ex}(B)$ is a field-dependent constant:
\begin{equation}\label{eq:Hexc}
    H_{ex}(B)=-\frac{1}{2}\mu_Bg_{ex}^zB_z\sum_{j\sigma}\langle b\dg_{j\sigma}\rangle \tau^{z}_{\sigma\sigma}\langle b_{j\sigma}\rangle.
\end{equation}
After this mean-field decoupling, all the dynamics is contained in $\mathcal{H}(\vect{k})$, which can be diagonalized to obtain a set of hybridized bands, $E_{\vect{k}\gamma}$. It is then straightforward to calculate the free energy functional:
\begin{equation}\label{eq:fe}
    \mathcal{F}=-k_BT\sum_{\vect{k}\gamma}\log{\left(1+\mathrm{e}^{-\frac{E_{\vect{k}\gamma}}{k_BT}}\right)}+\mathcal{H}_0+\mathcal{H}_{ex}(B),
\end{equation}
and obtain self-consistency equations for the different mean-field parameters:
\begin{equation}\label{eq:mfeq}
    \left(\frac{\partial F}{\partial |b|},\frac{\partial F}{\partial \theta},\frac{\partial F}{\partial \phi},\frac{\partial F}{\partial \lambda},\frac{\partial F}{\partial \mu}\right)=0,
\end{equation}
which are solved numerically for a given Ansatz. 

\subsection{Hastatic order Ans\" atze\label{sec:ansatze}}

\begin{figure}[!htb]
\includegraphics[width=1.0\columnwidth]{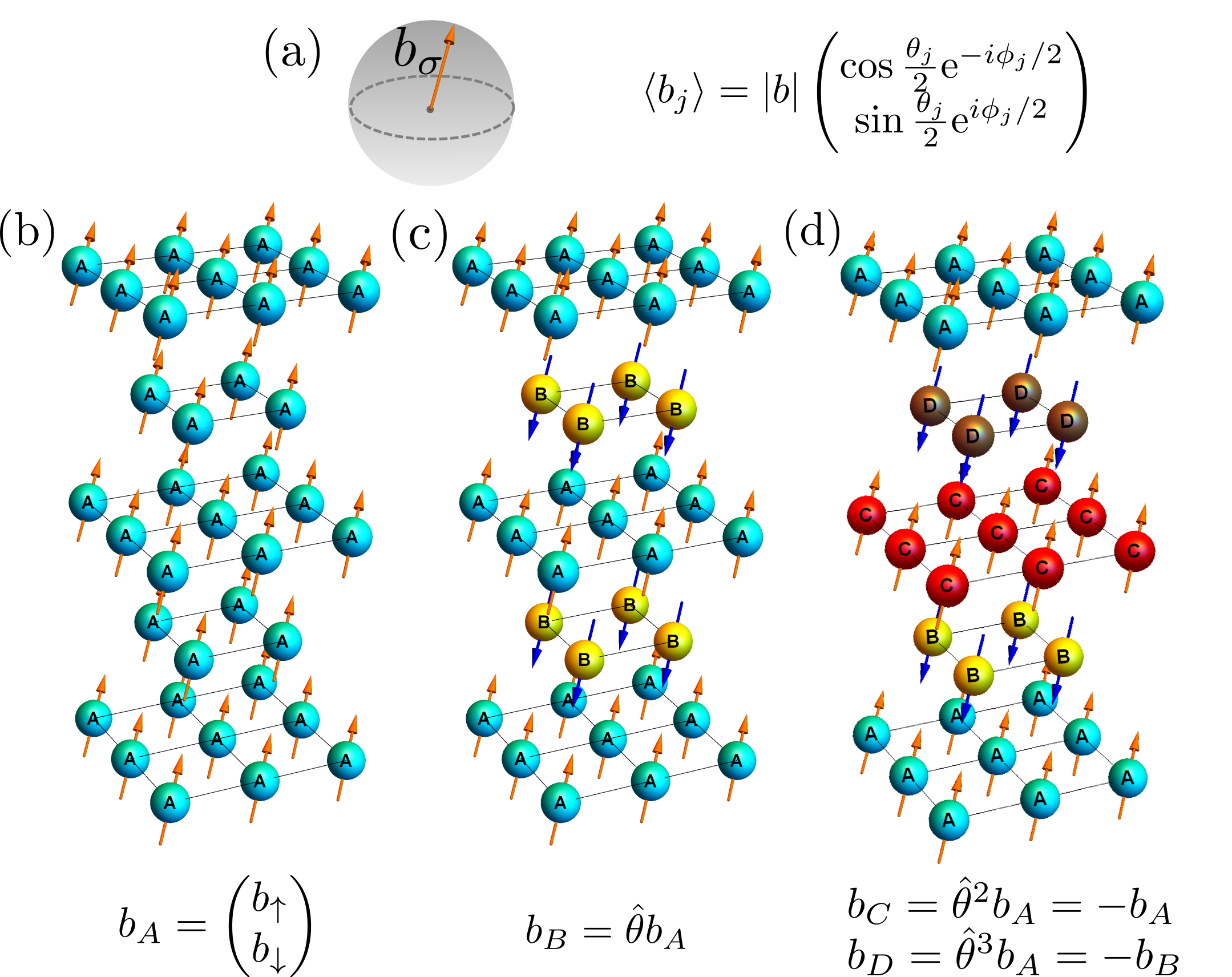}
\caption{Hastatic Ans\" atze on the U sublattice. (a) The microscopic hastatic order parameter is fundamentally a spinor describing the excited doublet occupation. (b)-(d) Distinct spinor arrangements (color) break different symmetries, even as the moment structure (determined by $\langle b\dg \vec{\sigma} b \rangle$) stays the same (arrows). The simplest hastatic Ansatz is uniform (ferrohastatic, FH), but this allows us to consider distinct staggered (antiferrohastatic, AFH) Ans\" atze.  The spinorial nature of $b_\sigma$ means the square of the time-reversal operator  ($\hat{\theta}^2 = -1$, and thus four-sublattices are needed to obtain a staggered phase with a time-reversal-like symmetry. Indeed, two and four sublattice antiferrohastatic phases have the same staggered moment arrangement, but break either time-reversal and inversion symmetries, respectively. \label{fig:Fig2}}
\vspace{-0.cm}
\end{figure}

We now describe the different hastatic Ans\" atze that we explore.  As stated above, we restrict ourselves to either uniform (ferrohastatic, FH) or staggered (antiferrohastatic, AFH) phases motivated by URu$_2$Si$_2$ ordering vector of $\vect{Q}=(0,0,1)$ found in the LMAFM, and likely relevant for hidden order \cite{Hassinger2008,Hassinger2010,Meng2013,Bareille2014, Zhang2022}. In addition, we consider possible canted phases that combine the two, although we will show that these are not energetically favored in tetragonal symmetry. 
\begin{itemize}
    \item Ferrohastatic (FH), uniform Ansatz, as shown in Fig. \ref{fig:Fig2}(b). This phase is fully described by $\vec{\Psi}$ \cite{Kornjaca2020}. There are two distinct possibilities,
        \begin{itemize}
        \item $FH_\perp$, with the spinor angles $\theta = \pi/2$, and $\phi$ with either $\phi=0$ and $\phi=\pi/4$ energetically favored (e.g. - in-plane moments with a $Z_4$ symmetry).
        \item $FH_z$, with $\theta = 0$, where $\phi$ is irrelevant.
        \end{itemize}
    \item Antiferrohastatic (AFH), staggered Ans\" atze.  The two-sublattice (2SL) Ansatz necessarily breaks time-reversal symmetry, while the four-sublattice (4SL) Ansatz breaks inversion, as shown in Fig. \ref{fig:Fig2}(c)-(d), and the singlet $\vec{\Phi}$ case, found when the $f$-electron hopping symmetry breaking cancels out the $\chi_j$ symmetry breaking. All three AFH Ans\" atze are equivalent for vanishing $f$-hoppings, where $\vec{\Phi}$ vanishes. The possibilities are thus:
    \begin{itemize}
        \item $2SL_\perp$, with $\theta = \pi/2$ and similar $\phi$ behavior to FH$_\perp$.
        \item $2SL_z$, with $\theta = 0$. Surprisingly, there is still $\phi$ dependence, which enters into $\vec{\Phi}$ and leads to four-fold symmetry breaking.
        \item $4SL_\perp$, with $\theta = \pi/2$ and similar $\phi$ behavior to FH$_\perp$.
        \item $4SL_z$, with $\theta = 0$. Again, there is still $\phi$ dependence, which enters into $\vec{\Phi}$ and leads to four-fold symmetry breaking.
        \item AFH$_{\Psi\perp}$, with $\theta = \pi/2$ and similar $\phi$ behavior to FH$_\perp$.
        \end{itemize}
    \item Canted Ans\" atze, which are a linear combination of FH$_z$ and AFH$_\perp$ spinors that one might expect to be favored in longitudinal magnetic field ($B_z$).  These can be constructed for any of the AFH Ans\" atze (2SL, 4SL, AFH$_\Psi$), although we never find these to be energetically favored.
\end{itemize}

Below, we give further details of the different Ans\"atze before deriving the mean-field phase diagram.

\subsubsection{Ferrohastatic phases}

Ferrohastatic (FH) order has a uniform hastatic spinor, and the only  associated order parameter is $\vec{\Psi}$, making the FH symmetry equivalent to a ferromagnet, although there are additional consequences due to the hybridization.  The moments turn out to be quite small ($O(T_c/D)$) and the bandstructure contains complicated hybridization gaps.

The field-independent constant term, $H_0$ is as in Eq. (\ref{eq:H0}), while the field-dependent constant term, $H_{ex}(B)$, Eq. \ref{eq:Hexc} simplifies to:
\begin{equation}\label{HexcFH}
    H_{ex}^{FH}(B)=-\frac{1}{2}N_s\mu_Bg_{ex}^zB_z|b|^2\cos{\theta},
\end{equation}
while the fermionic part can be written in the $(c_{\vect{k}\beta\sigma},f_{\vect{k}\alpha})$ basis as a 6x6 matrix:
\begin{widetext}
\begin{equation}\label{eq:FHham}
   \mathcal{H}^{FH}(\vect{k})=\begin{pmatrix}
   \epsilon_c\left(\vect{k}\right)\mathds{1}_\beta \otimes \mathds{1}_\sigma-\frac{1}{2}g_c\mu_B B_z\mathds{1}_\beta \otimes \sigma_z &\hat{V}_{6}(\vect{k})a_1(\theta,\phi)+\hat{V}_{7}(\vect{k})a_1(\theta,\phi)\tau_x\\
    \left[\hat{V}_{6}(\vect{k})a_1(\theta,\phi)+\hat{V}_{7}(\vect{k})a_1(\theta,\phi)\tau_x\right]\dg& \left(\epsilon_f\left(\vect{k}\right)+\lambda\right)\mathds{1}_\alpha-\frac{1}{2}g_f\mu_B B_z \tau_z
    \end{pmatrix}.
\end{equation}
\end{widetext}
The conduction electron part (upper left) is a 4x4 matrix, where the matrix $\epsilon_c(\vect{k})$ is given by Eq. (\ref{eq:elbands}), while the $f$ electron part (lower right) is a 2x2 matrix and the hybridization term in the upper right is a 4x2 matrix obtained from Eq.\ref{eq:v67def}. $\vec{\tau}$ are the Pauli matrices in $\alpha$ pseudospin space, and $\mathds{1}_{\alpha,\beta,\sigma}$ are two by two identity matrices in the indicated spaces, and we define $a_1(\theta,\phi)$ as a matrix in $\alpha$ pseudospin space,
\begin{equation}\label{eq:a1}
    a_1(\theta,\phi)=|b|\begin{pmatrix}
    \cos{\frac{\theta}{2}}\mathrm{e}^{i\phi/2}&0\\
    0&\sin{\frac{\theta}{2}}\mathrm{e}^{-i\phi/2}
    \end{pmatrix}.
\end{equation}

\begin{figure}[htb]
\includegraphics[width=1.0\columnwidth]{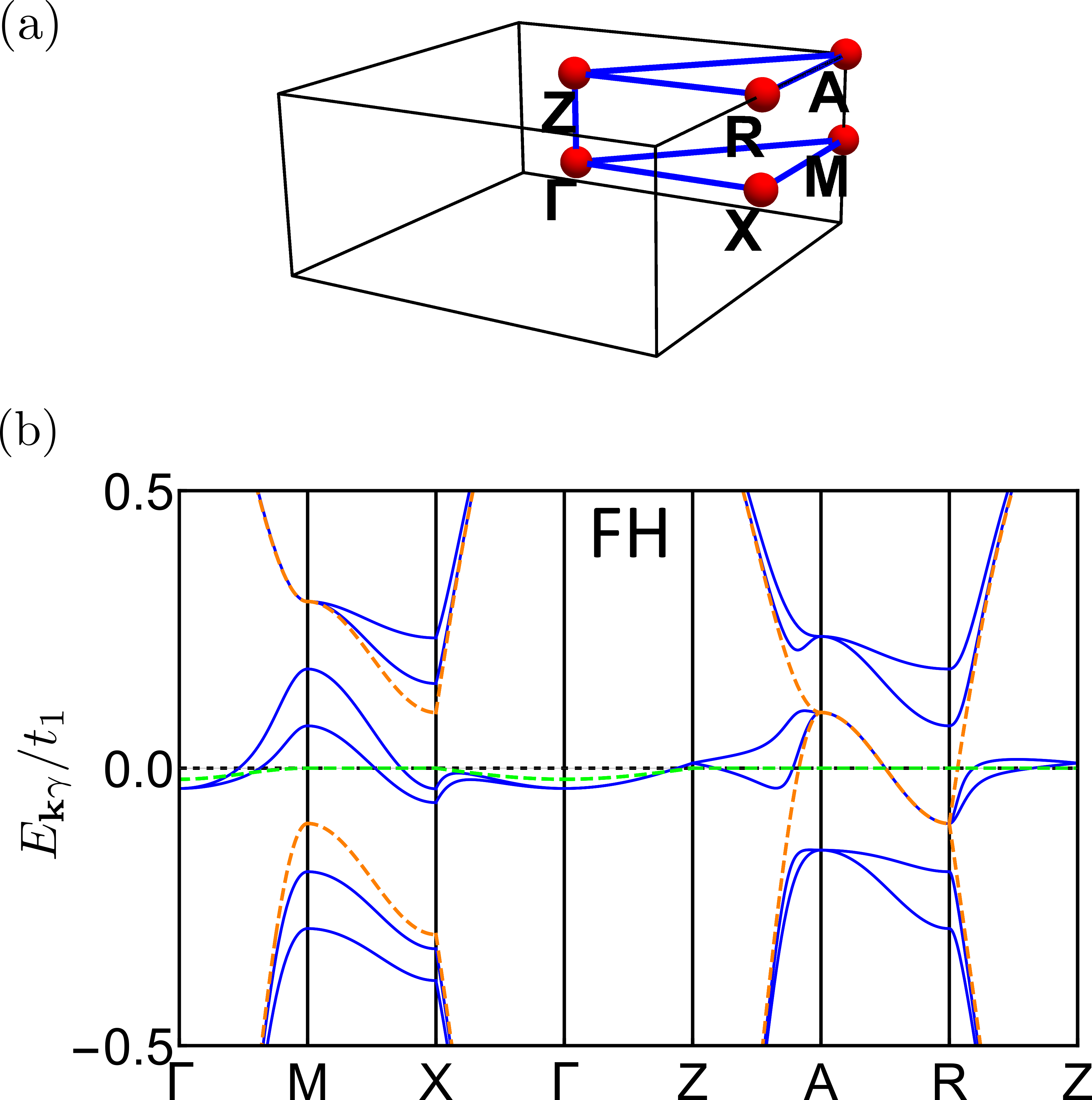}
\caption{(a) The simple tetragonal Brillouin zone used for plotting the band-structures, which is convenient for comparing ferrohastatic (FH) and antiferrohastatic (AFH) phases. (b) The FH bandstructure near the heavy bands, obtained self-consistently for $t_1=1,\; t_2=0.1,\; t_3=0.2,\; t_f=0.02,\; V_6=1,\; V_7=0.35, \;\Delta E=7.5,\; n_c=1/2$, which is a set of parameters leading to $\theta=\pi/2$ $\phi=\pi/4$ FH order. Dashed orange lines denote bare conduction bands, dashed green lines bare $f$ bands, and the hybridized bands are blue. \label{fig:Fig3a}}
\vspace{-0.0cm}
\end{figure}

FH order breaks the channel symmetry maximally. In the simpler cubic models previously explored \cite{VanDyke2019,Zhang2018}, it was possible to find a basis in which one conduction electron band is completely decoupled while the other hybridizes with the $f$-band. This is no longer possible in our realistic tetragonal model due to the complicated form of the hybridization. An example FH bandstructure obtained by diagonalizing $\mathcal{H}(\vect{k})$ is shown in Fig. \ref{fig:Fig3a}. While all FH bands are hybridized, some are heavy and others are mainly light.  Bands that hybridize with the $f$ bands roughly linearly in $b$ behave like typical heavy bands: they are mainly $f$ in character at the Fermi surface and quite flat.  Other bands are mainly light, with the hybridization at most quadratic in $b$, and the bands are mainly $c$ in character at the Fermi energy. As FH order breaks time-reversal, the original three doubly degenerate bands generically split into six non-degenerate bands.  There will be small magnetic moments aligned with the FH spinor, both in the excited state doublet and in the conduction band.

\subsubsection{Antiferrohastatic phases\label{sec:AFHph}}

\begin{figure}[!htb]
\includegraphics[width=0.97\columnwidth]{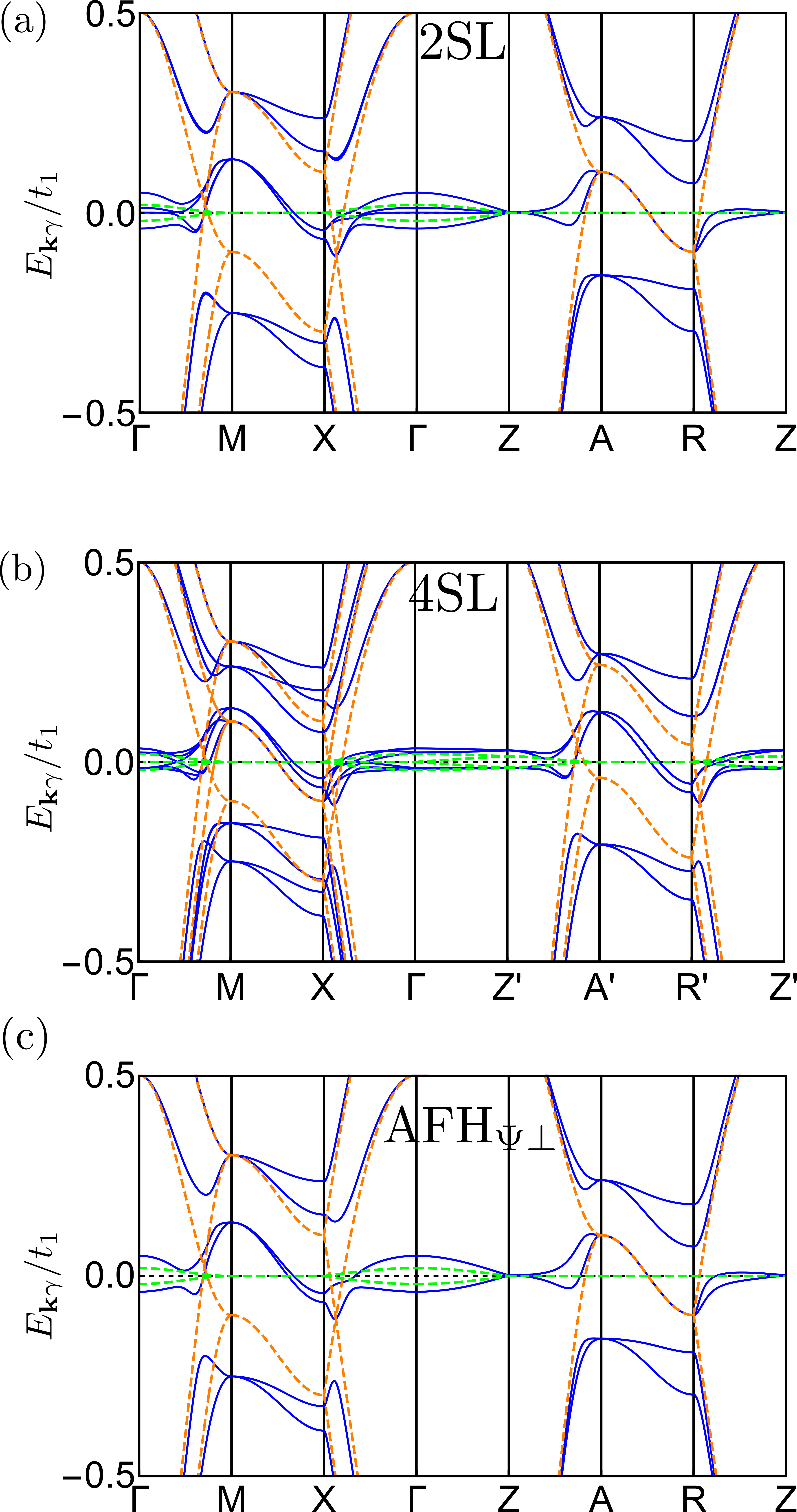}
\caption{The bandstructure for three different AFH phases.  These were obtained self-consistently for $t_1=1,\; t_2=0.1,\; t_3=0.2,\; t_f=0.02,\; V_6=1,\; V_7=0.35, \;\Delta E=7.5,\; n_c=1/2$ set of parameters, leading to $\theta=\pi/2, \phi=\pi/4$ in all the cases. Dashed orange lines denote bare conduction bands, dashed green lines denote bare $f$ bands, and the hybridized bands are blue. (a) Two-sublattice (2SL) AFH bandstructure. All bands are split at $\Gamma$ point as time-reversal symmetry is broken for the 2SL phase. (b) Four-sublattice (4SL) AFH bands. The 4SL phase preserves a time-reversal-like symmetry, allowing the bands to remain doubly degenerate at $\Gamma$ point. The bands are plotted in reduced simple tetragonal BZ ($Z'=Z/2$ and similar, see Fig. \ref{fig:Fig3a}) due to band doubling between 4SL and 2SL cases. (c) AFH$_{\Psi\perp}$ bandstructure. This is a two-sublattice phase where all bands are Kramers degenerate due to the time-reversal breaking form of the induced $f$-hopping that cancels out any symmetry breaking in $\vec{\Phi}$. \label{fig:Fig3b}}
\vspace{-0.2cm}
\end{figure}

We are most interested in the AFH phases with $\vect{Q}=(0,0,1)$, as we believe these are good candidates for the HO and LMAFM phases in URu$_2$Si$_2$. We consider the singlet $\vec{\Phi}$ phase and both two and four sublattice phases (2SL, 4SL), as depicted in Fig. \ref{fig:Fig2}, which double  and quadruple the unit cell, respectively.  We leave the detailed AFH Hamiltonians for Appendix \ref{app:AFH}, due to their complexity.  For all three Ans\"atze, $\vec{\Psi}$ is identical, and all differences emerge from $\vec{\Phi}$, whose effects are suppressed by $D_f/D$.  Example bandstructures for the different AFH phases are shown in Fig. \ref{fig:Fig3b}, and below we discuss the key features of each of these phases.

\paragraph{Two-sublattice antiferrohastatic phases\label{sec:2slsub}}

The 2SL Ansatz is constructed similarly to a typical antiferromagnetic Ansatz, as the hastatic spinors have two distinct sublattices, A and B related by time-reversal:
\begin{equation}\label{eq:2slab}
    b_B=\hat{\theta}b_{A}.
\end{equation}
For antiferromagnets, the vector order parameter returns to itself under double time-reversal, and there is an anti-unitary symmetry consisting of time-reversal followed by sublattice exchange; antiferromagnets are generally said to 'preserve' time-reversal for this reason.  As spinors pick up an additional sign under double time-reversal, the 2SL phase has no anti-unitary time-reversal-like symmetry, and really does breaks time-reversal symmetry. In the Landau theory, this is captured in the $\vec{\Phi}$ order parameter \cite{Kornjaca2020}, which generically has a uniform magnetic dipolar contribution.  This symmetry breaking can also be seen in the bandstructure, shown in Fig. \ref{fig:Fig4}(a), where all bands split, even at the $\Gamma$ point, as there is no longer any Kramers symmetry.  

The number of bands doubles, which can be understood by examining the momentum space structure of the 2SL order parameters\cite{Kornjaca2020}, where 
\begin{align}
    \vec{\Psi} &=\mathrm{Re}\langle b_{\mathbf{Q}}\dg\vec{\sigma}b_{\mathbf{0}}\rangle;\;
    \mathrm{Re}\vec{\Phi} =t_f\left(\langle b_{\mathbf{0}}\dg\vec{\sigma}b_{\mathbf{0}}\rangle-\langle b_{\mathbf{Q}}\dg\vec{\sigma}b_{\mathbf{Q}}\rangle\right)\cr
    \mathrm{Im}\vec{\Phi}& =t_f\mathrm{Im}\langle b_{\mathbf{Q}}\dg\vec{\sigma}b_{\mathbf{0}}\rangle
\end{align}  
with $b_{\mathbf{Q}}=\frac{1}{\sqrt{N_s}}\sum_{i}e^{i\mathbf{Q}\cdot\mathbf{R}_i}b_{i}$ representing the Fourier transform of hastatic spinor.

The 2SL$_\perp$ phase has staggered in-plane moments parallel $\vec{\Psi}$, while $\vec{\Phi}$ has two components, both orthogonal to $\vec{\Psi}$ that lead to uniform moments along $z$. The 2SL$_z$ phase with staggered moments along $z$ due to $\vec{\Psi}$, and the components of $\vec{\Phi}$ lead to uniform in-plane moments, as well as broken tetragonal symmetry due to the $\mathds{Z}_4$ pinning of the moments, in \emph{both} 2SL phases \cite{Kornjaca2020}.

\paragraph{Four-sublattice antiferrohastatic phases\label{sec:4slsub}}

The 4SL AFH phase preserves time-reversal symmetry, where we consider the sublattices to be stacked along $\hat z$, as shown in Fig. \ref{fig:Fig2}(d), with
\begin{equation}\label{eq:4slabcd}
    b_B=\hat{\theta}b_{A}, \; b_C=\hat{\theta}^2b_{A}=-b_A, \; b_D=\hat{\theta}^3 b_{A}=-b_B.
\end{equation}
The number of bands doubles once more compared to the 2SL phase, which can again be understood from the momentum space order parameters, which are constructed from $\pm \frac{\mathbf{Q}}{2}$ spinors \cite{Kornjaca2020}, 
\begin{align}
\vec{\Psi} & =\mathrm{Re}\langle \mathbf{b}_{-\frac{\mathbf{Q}}{2}}\dg\vec{\sigma}\mathbf{b}_{\frac{\mathbf{Q}}{2}}\rangle;\;\mathrm{Re}\vec{\Phi} =t_f\mathrm{Im}{\langle \mathbf{b}_{\frac{\mathbf{Q}}{2}}\dg\vec{\sigma}\mathbf{b}_{\frac{\mathbf{Q}}{2}}\rangle}\cr
\mathrm{Im}\vec{\Phi} &=t_f\mathrm{Im}{\langle \mathbf{b}_{-\frac{\mathbf{Q}}{2}}\dg\vec{\sigma}\mathbf{b}_{\frac{\mathbf{Q}}{2}}\rangle}.
\end{align}
As we have an anti-unitary time-reversal-like symmetry, Kramers degeneracy is guaranteed, although here the Kramers pairs are not at the same position in momentum space, with doubly degenerate bands guaranteed only at the time-reversal invariant $\mathbf{k}$-points, which include the $\Gamma$ point, as seen in Fig. \ref{fig:Fig4}(b). As the double degeneracy is generically lost for AFH phases, one might worry that the $g$-factor of the heavy bands is poorly defined and that the Ising-like $g$-factor found experimentally in the HO phase of URu$_2$Si$_2$ cannot be reproduced. We revisit the original spin-zero argument in Sec. \ref{sec:spinzeros}, and find that our AFH Ans\"atze will still generically exhibit these spin-zeros.

This 4SL phase also breaks inversion symmetry, which results in odd-parity multipole moments.  In principle, $\vec{\Phi}$ results in uniform in-plane electric dipole moments, for either $\vec{\Psi}$ orientation, although these will be screened in the metallic cases found away from quarter-filling.  In addition, there are staggered toroidal dipole moments perpendicular to $\vec{\Psi}$.  These moments are in-plane for $\Psi_z$ and along $\hat z$ for $\vec{\Psi}_\perp$, and the associated broken inversion symmetry could be measured via second harmonic generation \cite{Fiebig05, Zhao2016}.  Any in-plane moments will be pinned to the high symmetry directions, breaking tetragonal symmetry in \emph{both} 4SL phases.

\paragraph{\texorpdfstring{$\Psi$}{Psi}-only antiferrohastatic phases \label{sec:afhperp}}

The final AFH phase that we consider is the unusual case where the $\vec{\Phi}$ order parameter is a scalar.  In this case, $\vec{\Psi}$ is the only order parameter. This phase was actually considered in the original hastatic proposal \cite{Chandra2013}; there a unitary transformation $U$ was performed to absorb the angular dependence of the condensed bosonic spinors, [Eq. (\ref{eq:bHam}) and Eq. (\ref{eq:bmf})] into redefined $\chi$ fermions:
\begin{equation}\label{eq:afhpunitdef}
\chi_{j\alpha}=U_{\alpha,\alpha'}f_{j\alpha'}, \qquad \hat{U}_j=\frac{1}{|b|}\begin{pmatrix}
   \langle b\dg_{j\uparrow}\rangle &0\\
    0&\langle b\dg_{j\downarrow}\rangle
    \end{pmatrix}.
\end{equation}
This transformation should be accompanied by a transformed hopping term for the new $\chi$ fermions, 
\begin{equation}\label{eq:tftilde}
    \tilde{t}_{f,ij}=t_{f,ij}\hat{U}_i\hat{U}\dg_j=t_{f,ij}\frac{1}{|b|^2}\langle b_{j\sigma}\rangle\langle b_{i\sigma}\rangle\dg.
\end{equation}
The original hastatic proposal choose $\tilde{t}_f$ to be uniform, and treated the $\chi$ fermions as having the dispersion given by Eq. (\ref{eq:fbands}). This treatment assumes a hidden symmetry breaking $f$ electron hopping, $t_f$, which is plausible for the in-plane AFH phase, but is not generic. In general, to obtain a uniform $\chi$ hopping, $t_{f,ij}$ would have to absorb $\langle b_{j\sigma}\rangle\langle b_{i\sigma}\rangle\dg$. This symmetry breaking $t_{f,ij}$ allows the resulting $\vec{\Phi}_{ij} = t_{f,ij}\langle b\dg_i\vec{\sigma}b_j\rangle$ to be a scalar, thus giving rise to a $\Psi$-only phase.

This symmetry breaking $f$-electron hopping is plausible for the projected hopping mediated by valence fluctuations, $t_{f,ij}^{0, \alpha\alpha'}b_{i,\sigma}b\dg_{j,\sigma}f\dg_{i,\alpha}f_{j,\alpha'}$, if we assume that $\langle b_{i,\sigma}b\dg_{j,\sigma}\rangle$ above $T_c$ has the same form as $\langle b_{i,\sigma}\rangle \langle b\dg_{j,\sigma}\rangle$ does below.  For $\Psi_z$ phases, this projected hopping will vanish between sublattices as $\langle b_A\rangle \propto (1,0)$ and $\langle b_B\rangle \propto (0,1)$; this is why we only consider this Ansatz for $\vec{\Psi}_\perp$.  For in-plane spinors, this $f$-hopping does not vanish, and the effective $\chi$ hopping is uniform. 

In the Kondo limit, where $f$-hopping likely emerges from an RKKY interaction of the $\Gamma_5$ moments, this phase is still possible, as $t_{f,ij}$ is a mean-field value that could spontaneously break symmetry, but it is probably less likely than the uniform $t_f$ we consider for the other Ans\"atze.  Nevertheless, we include the AFH$_{\Psi\perp}$ as a possible AFH in-plane phase and treat it on the same footing as 2SL and 4SL phases. To do this, we assume that the resulting $\chi$ hopping after unitary transformation is of the form given by Eq. (\ref{eq:fbands}) and we fix $t_f$. The full form of the fermionic, $\mathcal{H}(\vec{k})$ part of the Hamiltonian after the unitary transformation is given in Appendix \ref{app:AFH}, and an example bandstructure is shown in Fig. \ref{fig:Fig3b}(c). Taking into account the most probable valence fluctuation origin of the fluctuations we check that both fixing $t_f$ itself or $t_f^{0}$ such that effective $t_f^{0}|b|^2\approx t_f$ at $T=0$, gives essentially equivalent zero-temperature phase diagrams.

The symmetry-breaking properties of the AFH$_{\Psi\perp}$ are completely determined by $\vec{\Psi}$, which, as for other in-plane AFH phases, leads to staggered magnetic moments parallel to itself. At the special point, $t_f=0$, all AFH phases coincide as $\vec{\Phi}$ vanishes. Both at this special point, and for AFH$_{\Psi\perp}$ itself, the bands are doubly degenerate everywhere, as seen in Fig. \ref{fig:Fig4}(c) and found in the original proposal. 

\subsubsection{Canted phases}

As we are interested in possible hastatic phases in magnetic field, we also consider a class of ``canted'' hastatic phases that represent the coexistence of FH and AFH order. One might imagine these phases to be competitive if the zero field state is AFH and applied field favors FH order, as was found in the cubic case \cite{Zhang2018}.  We will actually find that these phases are never the ground state for longitudinal fields (see Sec. \ref{sec:pBTphasediag}), although they are likely present to some degree for in-plane fields by analogy to the cubic case. The canted phases can be described as a linear combination of AFH and FH spinors:
\begin{equation}\label{eq:bcanted}
    b_{j\sigma}^{canted}=\lambda_c b_{j\sigma}^{FH}+(1-\lambda_c)b_{j\sigma}^{AFH},
\end{equation}
where $0\leq\lambda_c\leq1$ is an additional ``canting'' mean-field parameter. 

We construct canted phases for 2SL, 4SL and AFH$_{\Psi\perp}$ phases. The fermionic parts of the Hamiltonian are at most linear in $b$, which means that we can write
\begin{equation}\label{eq:cantedf}
    \mathcal{H}^{can}(\vect{k})=\lambda_c\mathcal{H}^{FH}(\vect{k})+(1-\lambda_c)\mathcal{H}^{AFH}(\vect{k}),
\end{equation}
while the bosonic constant term is quadratic:
\begin{align}\label{eq:cantedb}
    \mathcal{H}_0^{can}&=\lambda_c^2\mathcal{H}_0^{FH}+(1-\lambda_c)^2\mathcal{H}_0^{AFH}+\Delta E \mathcal{A}^{can},\cr
    \mathcal{H}_{exc}^{can}(B)&=-\frac{1}{2}N_s\mu_Bg_{exc}^zB^{(z)}|b|^2\left(\lambda_c^2+\mathcal{A}^{can}\right).
\end{align}
$\mathcal{A}^{can}$ denotes the following ``interference'' term:
\begin{equation}\label{eq:cantedif}
        \mathcal{A}^{can}=2\lambda_c(1-\lambda_c)\left(\cos{\frac{\Delta \theta}{2}}-\sin{\frac{\Delta \theta}{2}}\right)\cos{\frac{\Delta \phi}{2}},
\end{equation}
where $\Delta \theta = \theta_{AFH}-\theta_{FH}$ and $\Delta \phi = \phi_{AFH}-\phi_{FH}$, and both AFH and FH hastatic spinor angles are mean-field parameters.  

\section{Phase diagrams in zero magnetic field \label{sec:zeroBpds}}

In this section, we explore the zero-field phase diagrams and show how different FH and AFH Ans\"atze compete as the conduction electron filling, $n_c$; hybridization ratio, $V_6/V_7$; and temperature, $T$ are varied.  The main result is that all of the non-canted Ans\"atze discussed in the previous section are found for some reasonable parameter choice. In addition, the URu$_2$Si$_2$ $p-T$ phase diagram can be reproduced within the AFH region, if $V_6/V_7$ is taken as a proxy for pressure.  We find that there are relatively large energy differences between FH and AFH phases, as well as between the $z$ and in-plane $\vec{\Psi}$ orientations, with significantly smaller energy differences between the different AFH phases, and extremely weak $\mathds{Z}_4$ pinning of $\phi$ within the indicated phases.

Our microscopic model has a number of parameters that we fix for most of our calculations, as the main qualitative features do not depend on these.  We fix the hopping parameters, $t_1=1,\; t_2=0.1,\; t_3=0.2,\; t_f=0.02$, such that the unhybridized Fermi surface features roughly resemble those obtained in DFT calculations and ARPES measurements \cite{Oppeneer2010,Meng2013,Bareille2014,Denlinger2021}.  The excited state energy, $\Delta E = 7.5$ mainly controls the hastatic transition temperature ($T_c$) through the Kondo coupling, $J_K\approx V^2/\Delta E$. It was chosen such that, for standard parameter choices, $T_c/D \approx 1/30$, a reasonable value for the mixed valent URu$_2$Si$_2$ \cite{Chandra2015}. The remaining parameters ($V_7/V_6$, $n_c$, $T$, $B$) are then varied to explore potential phase diagrams.

\subsection{Competition between different hastatic Ans\" atze \label{sec:ancomp}}

\begin{figure}[!htb]
\includegraphics[width=0.99\columnwidth]{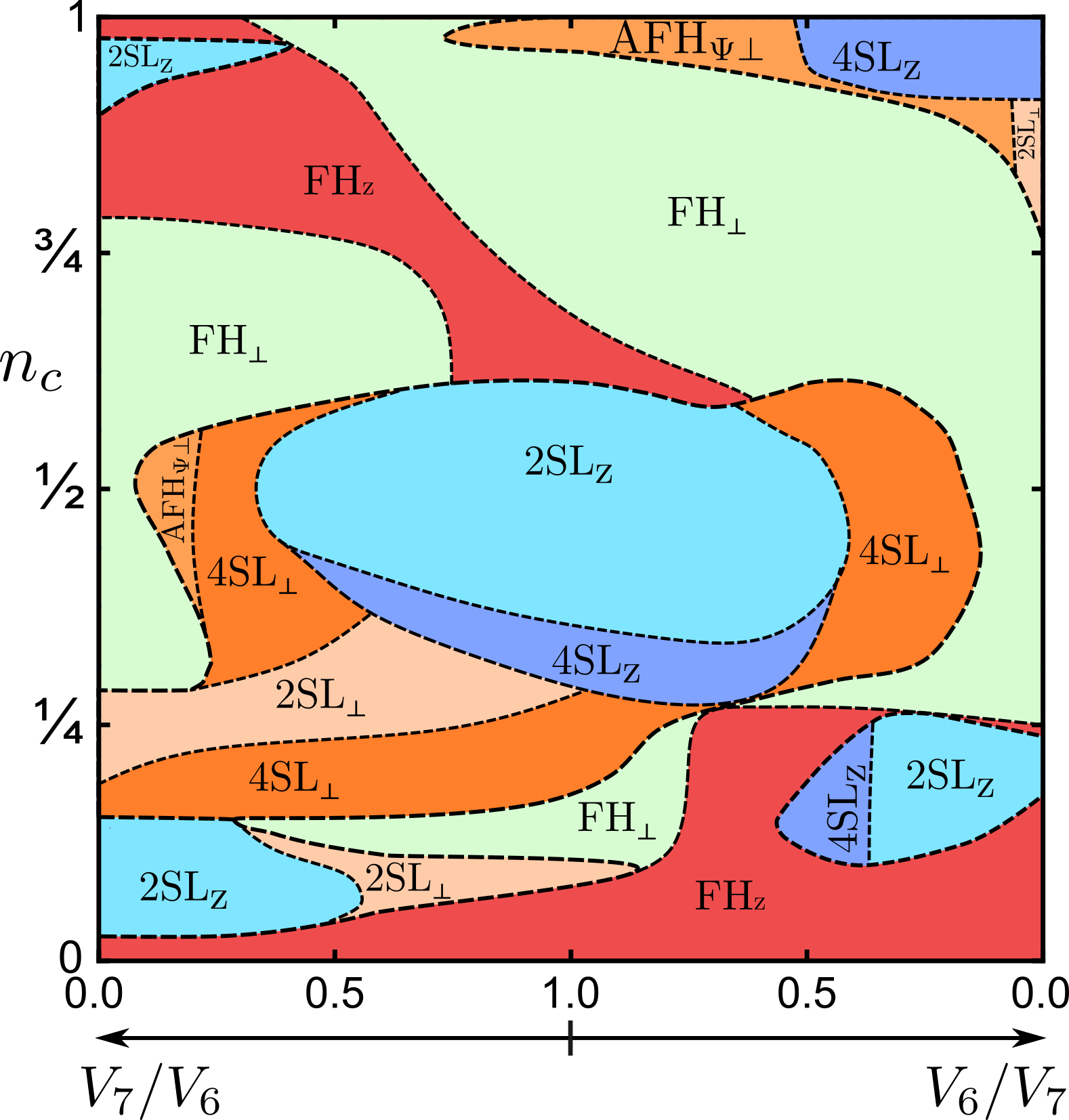}
\caption{The zero temperature phase diagram of tetragonal hastatic order in $n_c$ vs. hybridization ratio .  Here, the conduction electron filling varies from $n_c = 0$ to $n_c = 1$ with $n_c = 1$ denoting four conduction electrons per site (two bands, two spins). The hybridization is tuned such that the $V_7/V_6$ ratio changes while the overall magnitude, $V_6^2+V_7^2$ is fixed. The phase diagram was obtained by self-consistent calculations with the parameters given in the text. For each Ans\" atze, $z$ and in plane ($\perp$) directed phases are differentiated, but we do not distinguish different $\phi$ orientations here. Phases that differ only due to effects of the second, $\vec{\Phi}$ order parameter are represented with similar colors (e.g. - shades of blue for 2SL$_z$ and 4SL$_z$, and shades of orange for 2SL$_\perp$, 4SL$_\perp$ and AFH$_{\Psi\perp}$) . All phase transitions shown are first-order, although the energy differences between similar phases are generically small.  Tuning the hybridization ratio should be possible with pressure, and transitions between $\Psi_z$ and $\vec{\Psi}_\perp$ are easily obtained. \label{fig:Fig5}}
\vspace{-0.cm}
\end{figure}

We find the low temperature phase diagram for our realistic microscopic model by numerically solving the mean-field self-consistent equations given in Eq. (\ref{eq:mfeq}). The phases are plotted as the conduction electron filling, $n_c$ and hybridization ratio, $V_7/V_6$ are varied, as shown in Fig. \ref{fig:Fig5}.  Most of the phase space is occupied by the AFH phases that are expected to be relevant for URu$_2$Si$_2$. AFH order is particularly stable near quarter and half-filling, although this phase diagram is significantly more complicated than the cubic case \cite{Zhang2018}.  The complications arise from the particle-hole asymmetric nature of the model in tetragonal symmetry, as well as the additional $V_7/V_6$ tuning parameter.  Cuts along $V_6 = 0$ or $V_7 = 0$ more closely resemble the cubic case, where FH order was stabilized for half-filling and very low fillings, while AFH order was stabilized near quarter-filling. 

\begin{figure}[htbp]
\includegraphics[width=1.0\columnwidth]{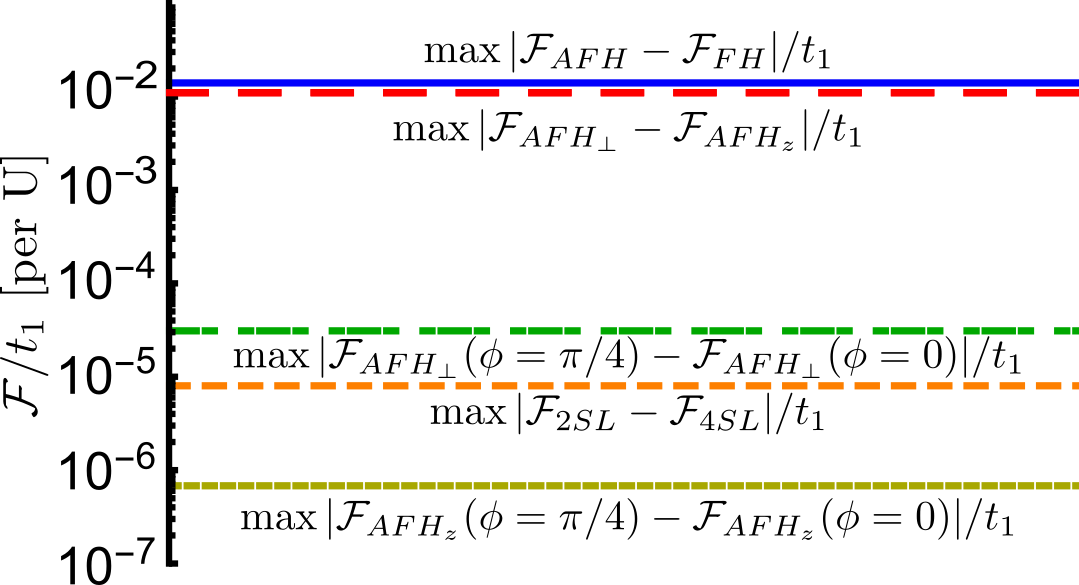}
\caption{Maximum free energy density differences found between different hastatic phases along the $n_c=1/2$ cut of the phase diagram in Fig. \ref{fig:Fig5}. While the AFH/FH and $\theta$ free energy differences are large, and comparable to one another, the $\phi$ pinning in $\theta=\pi/2$ phases is suppressed by $(T_c/D)^2$ comparatively; the 2SL-4SL energy differences stemming from $\vec{\Phi}$ are suppressed by $(D_f/D)^2$; and the $\phi$ pinning in $\theta=0$ phases is suppressed by $T_c^2D_f/D^3$. \label{fig:Fig6_fescales}}
\vspace{-0.cm}
\end{figure}

In a tetragonal system, it is expected that the $\Psi_z$ and $\vec{\Psi}_\perp$ phases will have substantially different properties. We will show in later sections that the $\Psi_z$ phases generically have relatively large 5f$^2$ moments oriented along $\hat z$, resembling local moment ferro- or antiferromagnets, although they also open up hybridization gaps and otherwise exhibit heavy Fermi liquid physics.  The in-plane phases have no large moments, and in fact only have excited state ($5f^1$) and conduction electron moments whose magnitude is suppressed by $T_c/D$; there are no $5f^2$ moments at all. Additional $\vec{\Phi}$ related signatures will generally be small, and for most physical aspects, the differences between different $\Psi_z$ or $\vec{\Psi}_\perp$ AFH phases can be neglected, which can be seen from free energy scale comparison in Fig. \ref{fig:Fig6_fescales}. All transitions here are first order, and it is straightforward to reproduce the pressure ($V_7/V_6$) induced transition between in-plane and out-of-plane orders found in URu$_2$Si$_2$, which we explore in detail in Sec. \ref{sec:pTPD}.

One important caveat is that we can only find phases that we look for.  We examined FH, $\bf{Q} = [001]$ AFH and canted orders here. We did not include Ans\"atze for other $\bf{Q}$ AFH phases; magnetic or quadrupolar orders; or superconductivity, and so this phase diagram merely suggests the possibilities for real materials captured by this model.

\subsection{Pinning of hastatic order to the lattice \label{sec:pinning}}

\begin{figure}[htbp]
\includegraphics[width=1.0\columnwidth]{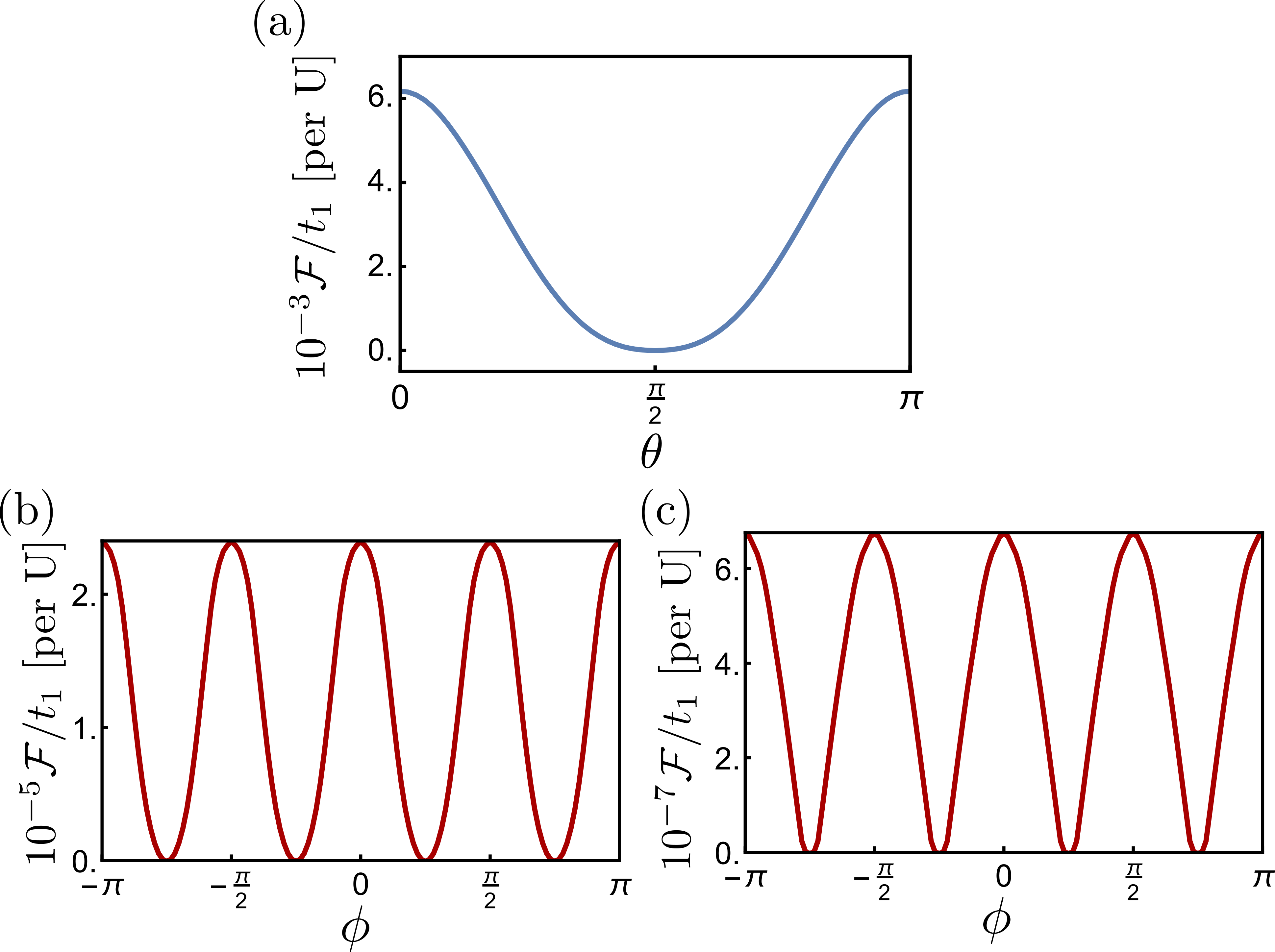}
\caption{Free energy density dependence on the spinor angles within the 4SL phase. (a) $\mathcal{F}(\theta)$ for $\phi=0$, (b) $\mathcal{F}(\phi)$ for $\theta=\pi/2$ (in-plane), (c)  $\mathcal{F}(\phi)$ for $\theta=0$ ($\hat z$). While the energy barriers in $\theta$ are large, the $\phi$ pinning is suppressed by a factor of $(T_{c}/D)^2$ or $T_{c}^2D_f/D^3$ for the $\vec{\Psi}_\perp$ and $\Psi_z$ phases respectively. The plots are obtained for the parameters used in Fig. \ref{fig:Fig5}, at the point $n_c=1/2$, $V_7/V_6=0.35$, where the 4SL$_\perp$ phase has the lowest energy. Note that for these parameters, $D_f/D\approx 1/60$ and $T_{c}/D\approx 1/30$, and the free energy density is scaled by $t_1$. \label{fig:Fig6}} 
\vspace{-0.cm}
\end{figure}

Previous work \cite{Chandra2015, Kornjaca2020} suggests weak $\phi$ pinning, which we confirm by plotting the angular dependence of the free energy density, $\mathcal{F}$ in Fig. \ref{fig:Fig6} and comparing energy scales in Fig. \ref{fig:Fig6_fescales}. for a point within the 4SL$_\perp$ phase at half-filling; the results generalize well to the rest of the phase diagram.  While the barriers in $\theta$ are large, as expected for tetragonal symmetry, the in-plane ($\theta=\pi/2$) $\phi$ barriers are suppressed by $\mathcal{O}[(T_{c}/D)^2]$, which is consistent with weak in-plane pinning of main $\vec{\Psi}$ order parameter \cite{Kornjaca2020}.  The metastable $4SL_z$ phase ($\theta = 0$) also has $\phi$ dependence, due to the $\vec{\Phi}$ order parameter, but the pinning is even weaker, $\mathcal{O}[T_{c}^2D_f/D^3]$. 

The weak $\phi$ pinning means that the $\vec{\Psi}_\perp$ phases are nearly XY-like, where the in-plane symmetry breaking can be washed out by random strain disorder \cite{Kornjaca2020}. For the rest of the paper, we depict different $\phi$ phases as a single phase, only distinguishing for $\phi$ when in-plane symmetry breaking signatures are discussed.

\subsection{Hastatic order and p-T phase diagram of \texorpdfstring{URu$_2$Si$_2$}{URu2Si2}\label{sec:pTPD}}

\begin{figure}[!htb]
\includegraphics[width=0.95\columnwidth]{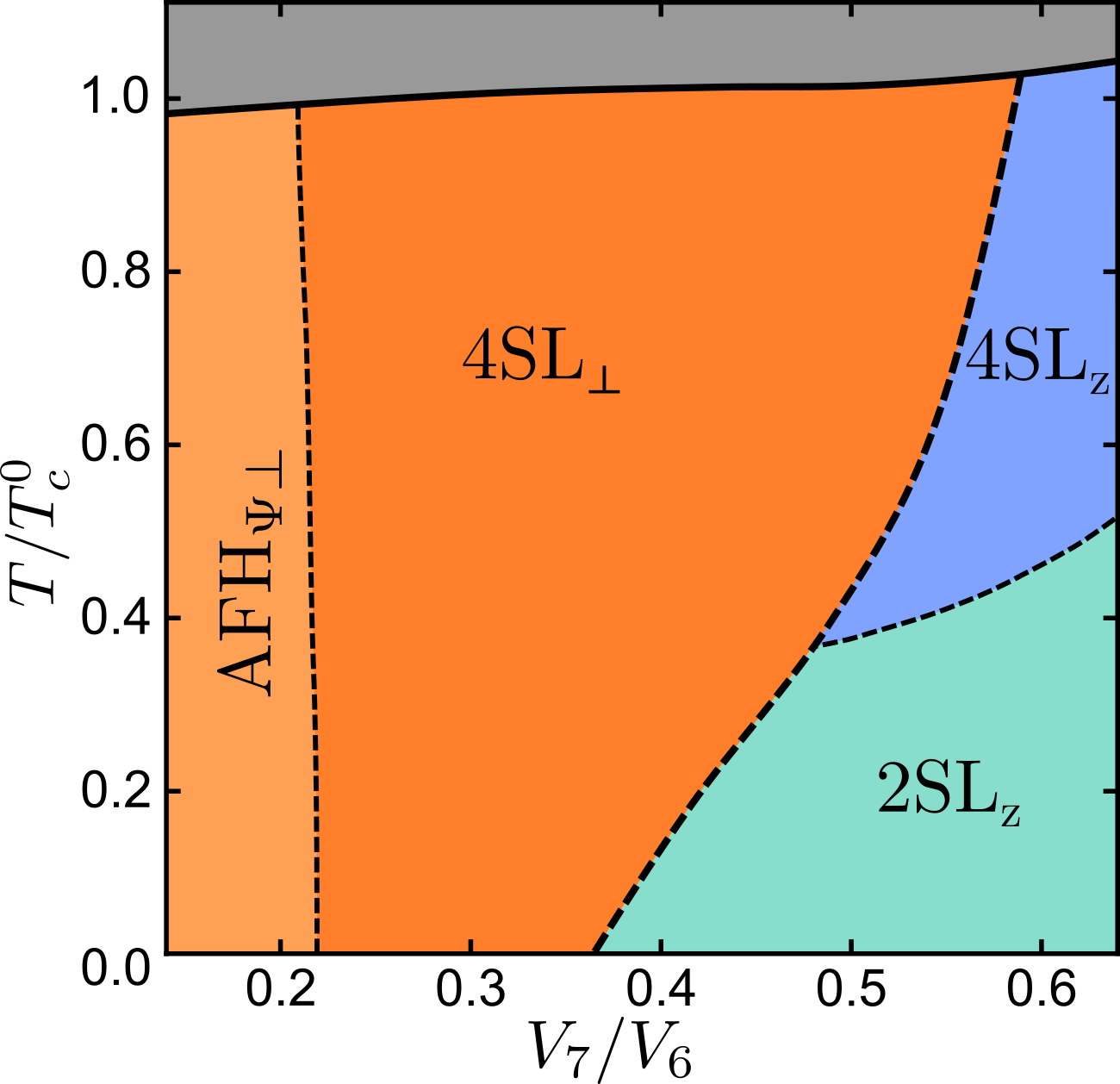}
\caption{Temperature-hybridization ratio phase diagram of hastatic order for $n_c=1/2$. Temperature is measured in the units of critical temperature for the parameters given in the text ($V_7/V_6=0.35$), $T_c^0$. Solid lines denote second order transitions, while dashed lines denote first order transitions. Phases that differ only by the $\vec{\Phi}$ order parameter are represented with similar colors: orange for AFH$_\perp$ and blue for AFH$_z$. Above $T_c$, there is a light Fermi liquid with free $\Gamma_5$ moments. The hybridization ratio tunes between in-plane and $z$ AFH phases that are separated by a first order phase transition, with a slightly increasing $T_c$. In URu$_2$Si$_2$, pressure or isoelectronic chemical substitution (Ru with Fe or Os) can plausibly tune the hybridization ratio. Temperature is seen to suppress the $z$-order in favour of in-plane order. The phase diagram is in qualitative agreement with experimental URu$_2$Si$_2$ phase diagram with a natural identification of large moment $z$ phases as LMAFM and small moment in-plane phases as hidden order. \label{fig:Fig7}}
\vspace{-0.cm}
\end{figure}

Now we explore the tuning between AFH phases in more detail, fixing $n_c = 1/2$ and exploring how the AFH phase changes from an in-plane, HO-like phase to an out-of-plane, LMAFM-like phase as a function of temperature and $V_7/V_6$, which acts as a proxy for pressure or isoelectronic substitution.  We show this phase diagram in Fig. \ref{fig:Fig7}. At first sight, this phase diagram is significantly more complicated than the experimental URu$_2$Si$_2$ phase diagram \cite{MydoshReview, Ran2017}, however, the differences are solely due to $\vec{\Phi}$.  Both AFH$_{\Psi\perp}$ and 4SL$_\perp$ have the same $\vec{\Psi}_\perp$, and resemble the HO phase in many regards, while both $4SL_z$ and $2SL_z$ have the same $\Psi_z$ and same large ($\sim .4\mu_B$) staggered 5f$^2$ moments.  The energy differences between phases distinguished only by $\vec{\Phi}$ are very small, suppressed by $(D_f/D)^2 \sim (1/60)^2 \sim 10^{-4}$ (see Fig. \ref{fig:Fig6_fescales}) compared to the energy differences between the disordered and hastatic phases; this energy difference is one order of magnitude smaller than the difference between $\Psi_z$ and $\vec{\Psi}_\perp$ states, for our choice of parameters. These additional first order transitions will have latent heats suppressed by $(D_f/D)^2$, and any discontinuities in $\vec{\Psi}$ are similarly suppressed, making these transitions extremely difficult to distinguished experimentally, except through $\vec{\Phi}$ specific measurements, like Kerr effect or second harmonic generation. Since the energy differences are so small, there may be domains of different $\vec{\Phi}$ phases, or the particular $\vec{\Phi}$ phase may vary between samples. The transition between $z$ and $\perp$ AFH phases is generically first-order, and no counter examples are found in the $n_c-V_6/V_7-T$ phase diagrams shown. However, it is possible, when considering the full parameter set of $V_6^{(1)}, V_6^{(2)}, V_7^{(1)}, V_7^{(2)}$, to find rare cases where the hastatic spinor rotates smoothly between the basal plane and the $z$-axis, resulting in two second-order transitions replacing the first order transition; we do not believe this scenario is experimentally relevant, as it requires fixing two of the four hybridization parameters to be zero. 

\section{Moments and susceptibilities\label{sec:momsusc}}

In this section, we reexamine the symmetry breaking signatures of AFH order within our more realistic model.  We are particularly interested in the magnetic moments and susceptibilities, as well as the elastic and nematic responses.  In addition to the previously found transverse magnetic moments in AFH$_\perp$, we can examine the moments in the AFH$_z$ phases, as well as additional moments due to the new $\vec{\Phi}$ order parameter.  One particularly interesting feature arises from the tetragonal  symmetry breaking, which can manifest within the magnetic moments; as a lattice distortion; and as an electronic nematicity.  Within our microscopic theory, we find that these occur in a hierarchy, with a large electronic nematicity and very small lattice and magnetic responses, as used in the Landau-Ginzburg theory to resolve apparently inconsistent tetragonal symmetry breaking experimental data\cite{Kornjaca2020}.

\subsection{Conventional magnetic dipole moments \label{sec:stgmom}}

The magnetic dipole moments arising from $\vec{\Psi}$  are the most natural consequences of the hastatic symmetry breaking.  FH order gives uniform moments, while AFH order gives staggered moments in the direction of $\vec{\Psi}\approx \langle b^{\dagger}\vec{\sigma}b\rangle$.  There are multiple possible microscopic origins for these moments: the ground state $\Gamma_5$ doublet, the excited state $\Gamma_7^+$ doublet and the conduction electrons.  We can calculate these moments separately, but typically calculate them together by adding a single conjugate staggered magnetic field, $\vec{B}_s$ to our Hamiltonian, with the realistic couplings ($g_c,g_f,g_{ex}$) obtained for uniform field in Sec. \ref{sec:Bcoupling}:
\begin{align}\label{eq:momcoupling}
    H_s & =-\frac{g_c\mu_B}{2}\! \sum_{\vect{k}\sigma\sigma'\beta}c\dg_{\vect{k}\beta\sigma}\vec{B}_s \cdot \vec{\tau}_{\sigma,\sigma'}c_{\vect{k}+\vect{Q}\beta\sigma'}\\
    &-\frac{g_f\mu_B}{2}\!\sum_{\vect{k}\alpha\alpha'}f\dg_{\vect{k}\alpha}B_{s,z}\tau^{z}_{\alpha\alpha'}f_{\vect{k}+\vect{Q}\alpha'}\cr
    &-\frac{\mu_B}{2}\!\sum_{j\sigma\sigma'}\langle b\dg_{j\sigma}\rangle\left(g_{ex}^zB_{s,z}\tau^{z}+g_{ex}^\perp\vec{B}_{s,\perp}\cdot\vec{\tau}^{\perp}\right)_{\sigma\sigma'}\!\!\langle b_{j\sigma'}\rangle\notag.
\end{align}
The three terms come from the coupling to conduction electrons (isotropic), the ground state 5$f^2$ $\Gamma_5$ doublet (Ising), and the excited 5$f^1$ $\Gamma_7$ doublet (anisotropic), respectively. We obtain the staggered moments by taking the numerical derivative of the free energy with respect to the staggered field:
\begin{equation}\label{eq:momder}
    \vec{m}_s=-\frac{\partial\mathcal{F}}{\partial\vec{B}_s }\Bigg|_{B_s\rightarrow 0}.
\end{equation}

\begin{figure}[!htb]
\includegraphics[width=1.0\columnwidth]{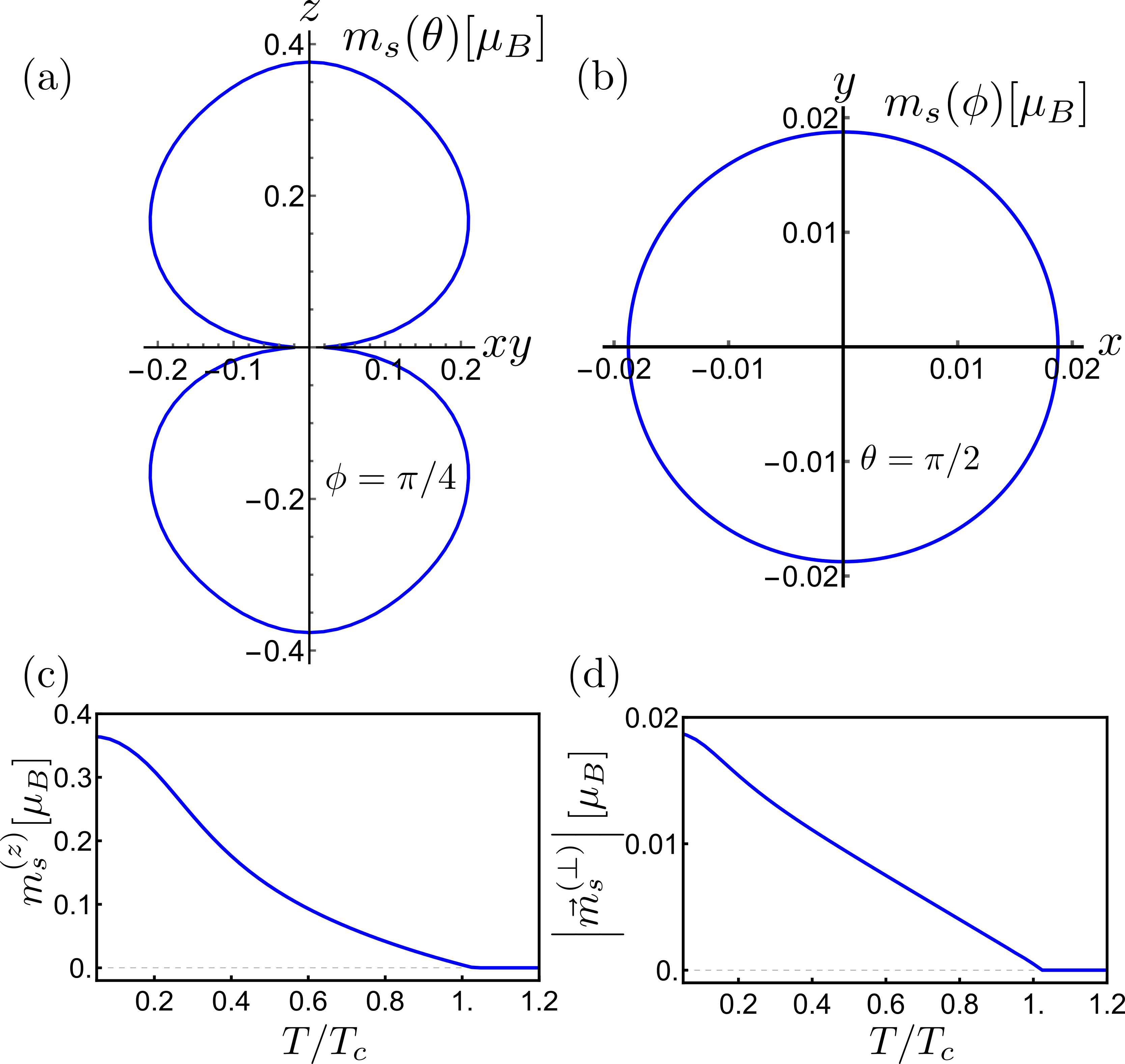}
\caption{Staggered magnetic moments within the 4SL AFH phase as a function of: (a) the $\theta$ angle of the hastatic spinor for $\phi=\pi/4$, at zero temperature, (b) the $\phi$ angle of the hastatic spinor for $\theta=\pi/2$, at zero temperature, (c) temperature for the $\Psi_z$ phase (with $\phi=\pi/4$), (d) temperature for the in-plane $\vec{\Psi}_\perp$ phase (with $\phi=\pi/4$). All plots are obtained using the self-consistent solutions for $n_c=1/2$, $V_7/V_6=0.35$, with the usual set of parameters. While the $\Psi_z$ phase has large magnetic moments of primarily 5$f^2$ origin, the moments in the in-plane phase stem from the 5$f^1$ excited doublet and conduction electrons, and are suppressed by $T_c/D\sim 1/30$; their in-plane anisotropy is unobservably weak. All of these moments are the consequence of $\vec{\Psi}$ order parameter alone and behave similarly for all AFH phases. \label{fig:Fig8}}
\vspace{-0.cm}
\end{figure}

We show the angular and temperature dependence of these staggered moments in a 4SL phase in Fig. \ref{fig:Fig8}, although results are similar for 2SL, 4SL and AFH$_{\Psi\perp}$ phases. The moments in the $z$ phase are large ($\sim 0.4 \mu_B$), allowing us to identify the phase as a candidate for the LMAFM phase. They are predominantly 5$f^2$ $\Gamma_5$ in character, and the $\theta$ angular dependence shows the characteristic Ising anisotropy. In the in-plane phase, however, the moments are suppressed by $T_c/D$, due to their origin from the hybridization induced polarization of conduction electrons and the occupancy of  5$f^1$ excited doublet; these two components to the moments are approximately identical in size and parallel to one another. Not only are these in-plane moments small, $\sim .01 \mu_B$, but they are also very susceptible to disordering by random strain. In our calculation, the temperature dependence of the moments is mean-field-like, growing linearly with $T_c-T$ close to $T_c$, reflecting the quadratic dependence on the fundamental $\langle b_j\rangle$ order parameter.

\subsection{Unconventional magnetic dipole moments\label{sec:unmom}}

We now turn to the moments stemming from $\vec{\Phi}$. In our microscopic mean-field theory, $\vec{\Phi}$ and $\vec{\Psi}$ both turn on at $T_c$, and thus we always have the signatures of both, although beyond mean-field, one might find $\vec{\Phi}$ only, as well as $\vec{\Psi}$ only, phases. To show the generic behavior of these $\vec{\Phi}$ associated moments, we treat the 2SL phases, where there are uniform magnetic dipole moments associated with the time-reversal symmetry breaking.  The 4SL phase also has unconventional moments, mainly uniform in-plane electric dipoles that will be screened, and staggered toroidal dipole moments; we expect these moments will behave similarly to the uniform 2SL moments in zero magnetic field.

\begin{figure}[!htb]
\includegraphics[width=1.0\columnwidth]{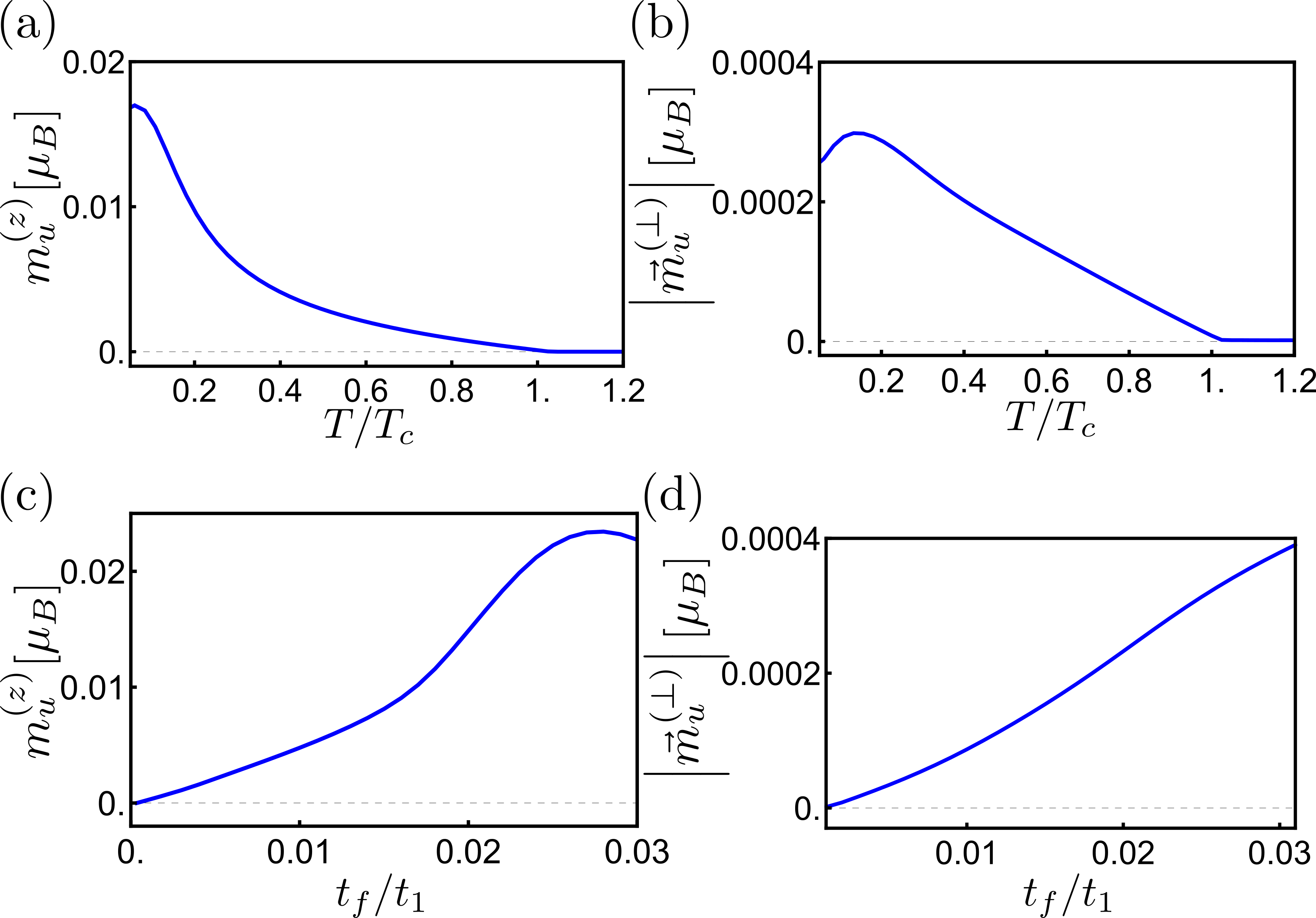}
\caption{Uniform magnetic moments found in the 2SL phases as a function of: (a) temperature for 2SL$_\perp$, with $\phi=\pi/4$, where the $m_u^{(z)}$ moments are non-zero ($t_f=0.02$); (b) temperature for the 2SL$_z$ phase, also with $\phi=\pi/4$, where in-plane moments, $m_u^{(\perp)}$ are non-zero ($t_f=0.02$); (c) $t_f/t_1$ at zero temperature for the 2SL$_\perp$ phase with $\phi=\pi/4$; (d) $t_f/t_1$ at zero temperature for the 2SL$_z$ phase with $\phi=\pi/4$. All plots are obtained self-consistently for the usual set of parameters at $n_c=1/2$, $V_7/V_6=0.35$, where the 2SL phases are metastable, although $t_f$ is varied for (c) and (d). These moments are a consequence of the unconventional $\vec{\Phi}$ order parameter and are absent in the 4SL and AFH$_{\Psi\perp|}$ phases. The  $m_u^{(z)}$ for 2SL$_\perp$ are suppressed by $D_f/D$ (equivalently $t_f/t_1$), while the $\vec{m}_u^{(\perp)}$ for 2SL$_z$ are suppressed by $D_fT_c/D^2$.\label{fig:Fig9}}
\vspace{-0.cm}
\end{figure}

The uniform magnetic moments can be calculated with an appropriate uniform conjugate magnetic field. All non-zero contributions come from the coupling to the $f$ and $c$ electrons, which are the first two terms in Eq. (\ref{eq:momcoupling}). In the 4SL and AFH$_{\Psi\perp}$ phases, these moments vanish for zero external field, but these are present in both 2SL$_z$ and 2SL$_\perp$ phases, where they are proportional to the $\vec{\Phi}$ order parameter and perpendicular to the staggered moments, in agreement with the Landau theory \cite{Kornjaca2020}.
Note, that these moments were also found in the cubic case \cite{Zhang2018}, were they were inappropriately discarded as gauge dependent moments. The temperature and $t_f$ dependence of these uniform moments are shown in Fig. \ref{fig:Fig9}.

The largest unconventional uniform moments are the $m_u^{(z)} \sim .02\mu_B$ moments found in 2SL$_\perp$ phases, which are still suppressed by a factor of $D_f/D$, as confirmed by the linear $D_f$ ($t_f$) scaling for small $t_f$, with the moments vanishing at $t_f=0$, where all AFH phases merge and $\vec{\Phi}$ vanishes. Detecting these moments experimentally may be the simplest signature for 2SL order. Some evidence of above $T_c$, $z$-directed, time-reversal breaking exists in sensitive Polar Kerr effect measurements \cite{Schemm2015} that might indicate a 2SL order with $\vec{\Phi}$ developing above $T_{HO}$, where $\vec{\Psi}$ develops. The in-plane unconventional moments in the 2SL$_z$ phases are suppressed by $D_fT_c/D^2$, as they arise completely from the conduction electron term. The temperature dependence of the unconventional moments close to $T_c$ is again  linear in $T_c-T$. The experimental picture is complicated by the existence of field-induced uniform moments in all AFH phases, as discussed further in Sec. \ref{sec:pBTphasediag}, meaning that 2SL phases are not necessarily favored in finite external field, although small training fields may be able to align domains without destabilizing the 2SL phase.

\subsection{Magnetic susceptibility and tetragonal symmetry breaking \label{sec:susc}}

\begin{figure}[!htb]
\includegraphics[width=1.0\columnwidth]{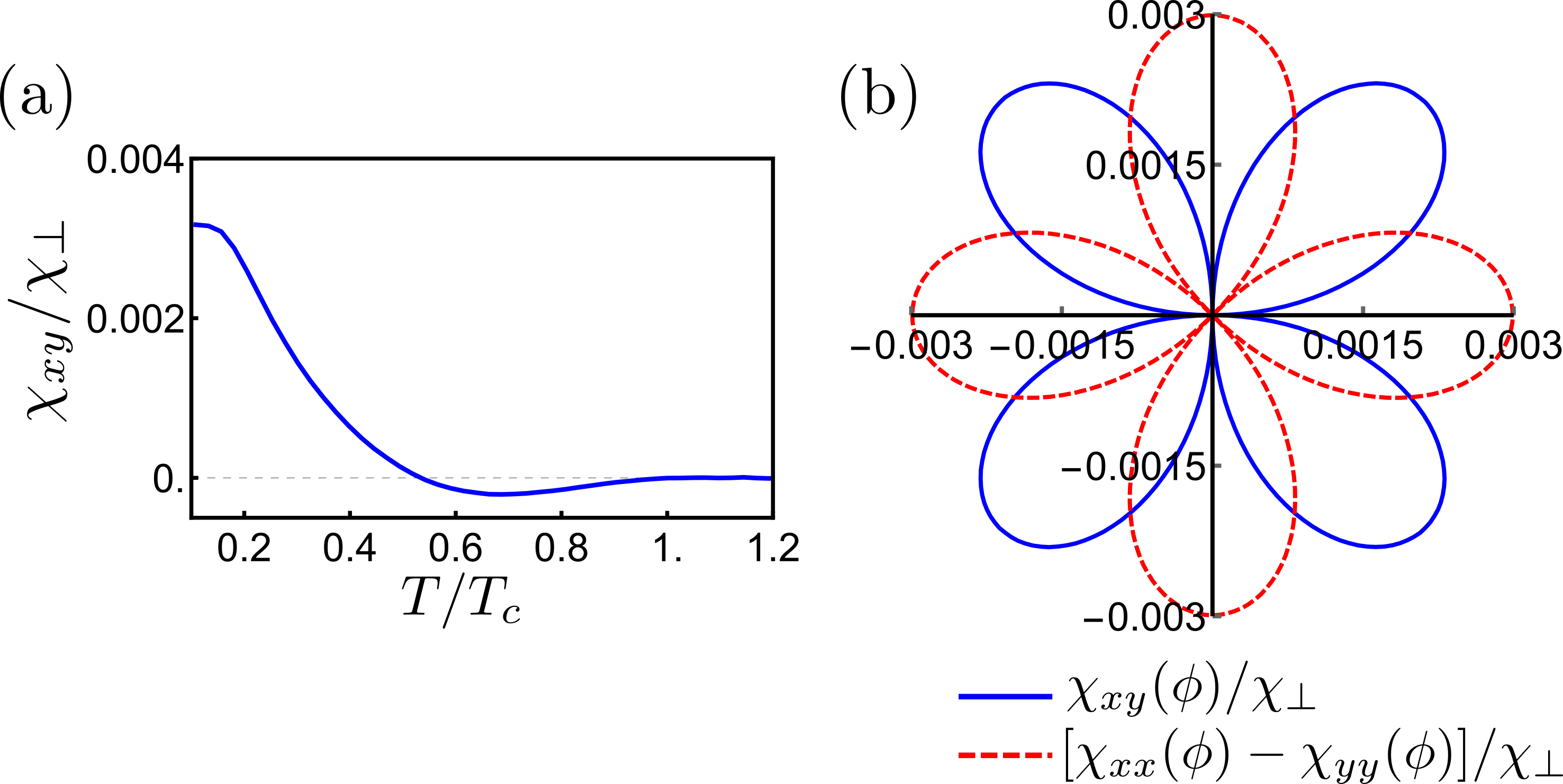}
\caption{Tetragonal symmetry breaking magnetic susceptibility matrix elements in the 4SL$_\perp$ phase.  (a) As a function of temperature for $\phi=\pi/4$.  The temperature dependence shows a prevalent quadratic component (positive) with a linear contribution in $T_c-T$ (negative) near the transition. (b) As a function of the hastatic spinor angle $\phi$ at zero temperature.  The in-plane angular dependence shows different types of tetragonal symmetry breaking for different hastatic order ($\vec{\Psi}$) orientations, mainly [110] type for $\phi=\pi/4$ (non-zero $\chi_{xy}$), [100] for $\phi=0$ (non-zero $\chi_{xx}-\chi_{yy}$) or both for the energetically unfavorable $0<\phi<\pi/4$. The plots were calculted self-consistently for the usual parameters with $n_c=1/2$, $V_7/V_6=0.35$. \label{fig:Fig10}}
\vspace{-0.cm}
\end{figure}

As all AFH phases have tetragonal symmetry breaking moments, we expect to observe a number of signatures of this tetragonal symmetry breaking in the absence of disorder.  In this section, we calculate the magnetic susceptibility anisotropy, while in the next two sections we examine the anisotropy in the elastic and electronic responses. We focus on the in-plane orders, and calculate the tetragonal symmetry breaking components of magnetic susceptibility using,
\begin{equation}\label{eq:mgsusc}
    \chi_{ij}=-\frac{\partial^2\mathcal{F}}{\partial B_i\partial B_j}\Bigg|_{B\rightarrow 0}.
\end{equation}
We again present results for the 4SL$_\perp$ phase found for $n_c = 1/2$, and $V_7/V_6 = 0.35$, in Fig. \ref{fig:Fig10}, although our results are generic to all AFH$_\perp$ phases. As contributions to tetragonal symmetry breaking susceptibility components ($\chi_{xy}$ or $\chi_{xx}-\chi_{yy}$) come from the excited state and conduction electron response, they are suppressed by $(T_c/D)^2$, as argued in \cite{Chandra2015}. This suppression is found both for mean-field $T_c-T$ term that is dominant close to $T_c$ and the quadratic $(T_c-T)^2$ that makes up the majority of the low temperature response. Comparing to torque anisotropy experiment \cite{Okazaki2011}, our linear term is comparable in magnitude to the experimentally found one, while the quadratic term is an order of magnitude smaller. For this particular parameter choice,  the sign of the linear term is opposite to the quadratic term, but this is not generic and both signs may be found for different parameter regimes.

Turning to the in-plane angular dependence of the tetragonal symmetry breaking susceptibility anisotropy, we find that $\chi_{xy}$ and $\chi_{xx}-\chi_{yy}$ follow the typical quadrupolar dependence. This dependence is expected from Landau theory, where these components stem from $\Psi_x\Psi_y$ and $\Psi_x^2-\Psi_y^2$ secondary order parameters \cite{Kornjaca2020}. The size of both components is similar along the two high symmetry directions, $\phi=0$ and $\phi=\pi/4$, as well as between 2SL$_\perp$, 4SL$_\perp$ and AFH$_{\Psi\perp}$ orders. It is worth noting that an extremely small, but otherwise similar, anisotropic magnetic susceptibility is seen in the $z$ phase as well, as a consequence of the $\vec{\Phi}$ order parameter, but it is suppressed by $T_c^2D_f/D^3$.

\subsection{Quadrupolar moments and tetragonal symmetry breaking \label{sec:quad}}

As a proxy for the elastic/lattice response, we can calculate the possible tetragonal symmetry breaking quadrupolar moments associated with different phases.  It is natural to expect some quadrupolar response, as the $\Gamma_5$ doublet has an Ising magnetic response along $\hat z$, while the in-plane $\Gamma_5$ moments are quadrupolar:
$Q_{xy}\sim\langle\overline{J_x J_y} \rangle$ (with $\overline{J_x J_y}$ acting like $\tau^{(y)}$ in pseudospin space)  and $Q_{x^2-y^2}\sim \langle J_x^2-J_y^2 \rangle$ (with $J_x^2-J_y^2$ acting like $\tau^{(x)}$ in pseudospin space). We can calculate the quadrupolar moments by introducing conjugate strains ($\epsilon_{xy},\, \epsilon_{x^2-y^2}$):
\begin{align}\label{eq:quadcoupling}
    H \rightarrow H &-\sum_{\vect{k}\alpha\alpha'}f\dg_{\vect{k}\alpha}\epsilon_{xy}\tau^{(y)}_{\alpha,\alpha'}f_{\vect{k}\alpha'}\\
    &-\sum_{\vect{k}\alpha\alpha'}f\dg_{\vect{k}\alpha}\epsilon_{x^2-y^2}\tau^{(x)}_{\alpha,\alpha'}f_{\vect{k}\alpha'}\notag,
\end{align}
and taking the appropriate free energy derivatives, similarly to the moment calculations. Here, we are setting the coupling between the strain and the quadrupolar moments to be one, as this quantity is not known. In principle, the conduction electrons will also couple to strain, and the hybridizations will be modified, but these couplings are expected to be smaller \cite{nakamura94,hazama00}
and we neglect them here. All of the quadrupolar moments are suppressed by $(T_{c}/D)^2$, similarly to the tetragonal symmetry breaking magnetic susceptibilities, which may have important consequences for resolving conflicting experimental measurements of elastic tetragonal symmetry breaking signatures \cite{Tonegawa2014, Choi2018, Ghosh2020}, as explored within Landau-Ginzburg theory framework in ref. (\onlinecite{Kornjaca2020}). For $\Psi_z$ phases, the quadrupolar moments are an order of magnitude smaller, due to the additional $D_f/D$ factor from their $\vec{\Phi}$ origin. 

\begin{figure}[!htb]
\includegraphics[width=1.0\columnwidth]{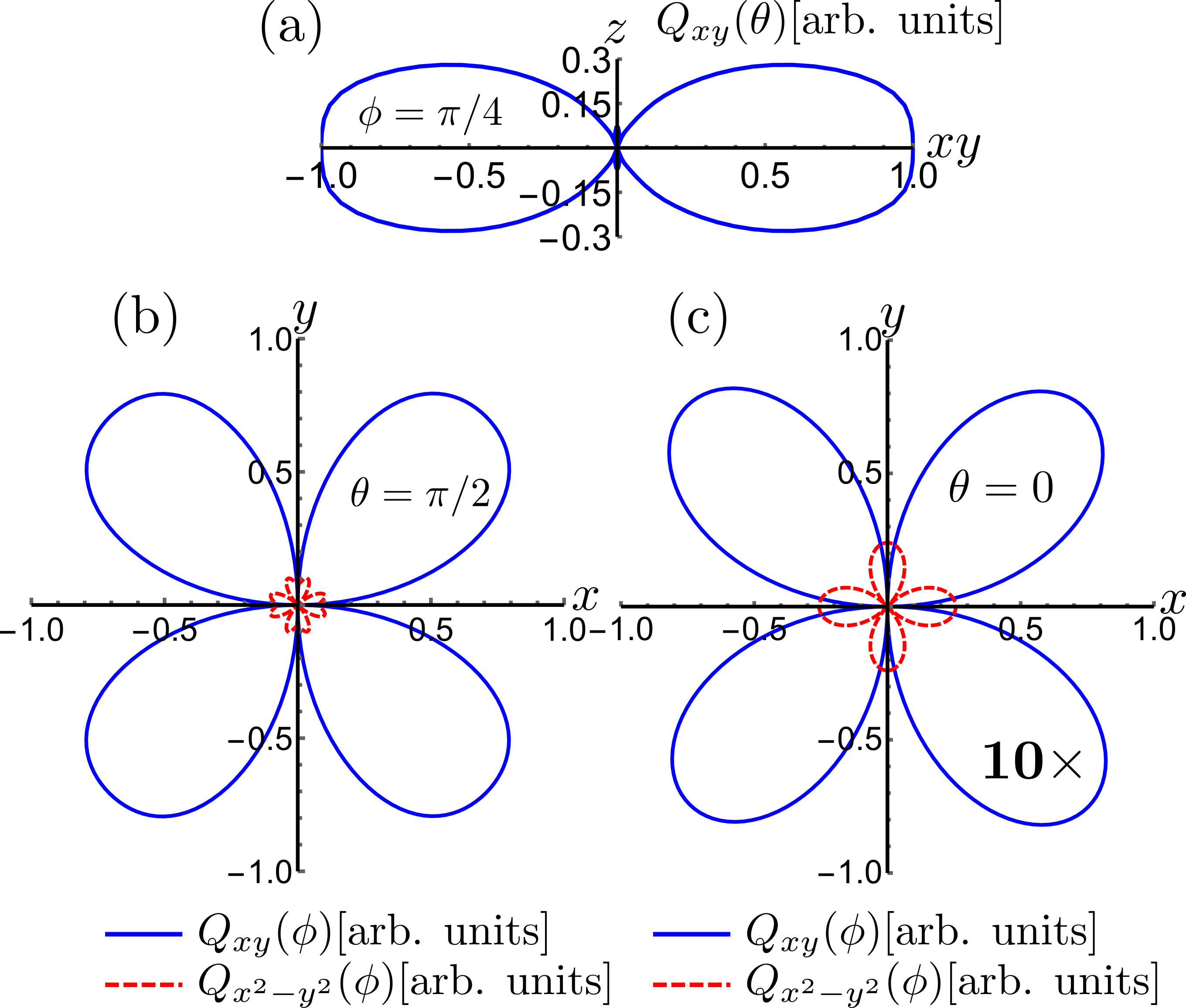}
\caption{Uniform quadrupolar moments ($Q_{xy}$, $Q_{x^2-y^2}$) in AFH phases at zero temperature as a function of: (a) $\theta$ with $\phi=\pi/4$; (b) $\phi$ for $\theta=\pi/2$ (in-plane phase); (c) $\phi$ for $\theta=0$ ($z$ phase) rescaled by a factor of 10. The quadrupolar moments are similar between 2SL, 4SL and AFH$_{\Psi\perp}$ phases. The plots are obtained self-consistently, for the usual set of parameters at $n_c=1/2$, $V_7/V_6=0.35$. All moments are rescaled by the maximum $Q_{xy}$ found for $\phi = \pi/4$ and an additional factor of 10 for (c). \label{fig:Fig11}}
\vspace{-0.cm}
\end{figure}

In Fig. \ref{fig:Fig11}, we show the $Q_{xy}$ and $Q_{x^2-y^2}$ moments found at zero temperature in the 4SL phases at $n_c = 1/2$, $V_7/V_6$.  Note that the in-plane, $\phi =\pi/4$ phase is the energetically favored state, although we show the moments as a function of $\theta$ and $\phi$ for a range of metastable states.  We find both $Q_{xy}$ and $Q_{x^2-y^2}$ moments, with the expected $\theta,\phi$ dependence, although surprisingly the $Q_{x^2-y^2}$ are an order of magnitude smaller for $\phi = 0$ than the $Q_{xy}$ phase is for $\phi = \pi/4$. The angular dependence both in $\theta$ and $\phi$ takes expected quadrupolar form, with an exception of small admixture of opposite sign higher order terms for $Q_{x^2-y^2}$ seen as dips around $\phi=m\pi/2$. We also checked that staggered tetragonal breaking quadrupolar moments vanish for all $AFH_\perp$ phases, which is expected as there is no secondary order parameter with appropriate symmetry to induce them \cite{Kornjaca2020}.

\subsection{Energy dependent tetragonal distortion and electronic nematicity\label{sec:FSnematicity}}

In the two previous sections, both magnetic and elastic tetragonal symmetry breaking signatures scale as $(T_c/D)^2$; here, we will show that the electronic nematicity signatures, as might be measured by elastoresistivity \cite{Riggs2015} are of order one. The electronic nematicity can be most straightforwardly seen in the Fermi surface shape, where we focus on the tetragonal distortion within heavy bands. This distortion can be made quantitative by defining the energy dependent nematicity, $\eta_{xy}(V)$ as average sign of  $k_x k_y$ normalized by the density of states:
\begin{equation}\label{eq:FStetdist}
    \eta_{xy}(V)=\frac{\sum_{\gamma}\int \mathrm{d}^3k \,\text{sgn}(k_x k_y) \delta\left(E_{\mathbf{k} \gamma}-V\right)}{\sum_{\gamma}\int \mathrm{d}^3k \, \delta\left(E_{\mathbf{k\gamma} \beta}-V\right)},
\end{equation}
where $\gamma$ denotes the hybridized band index and $V$ is the gate voltage/Fermi level, allowing for non-zero doping of the hybridized bands. This quantity is only non-zero if [110] tetragonal symmetry breaking is present. 

\begin{figure}[!htb]
\includegraphics[width=0.95\columnwidth]{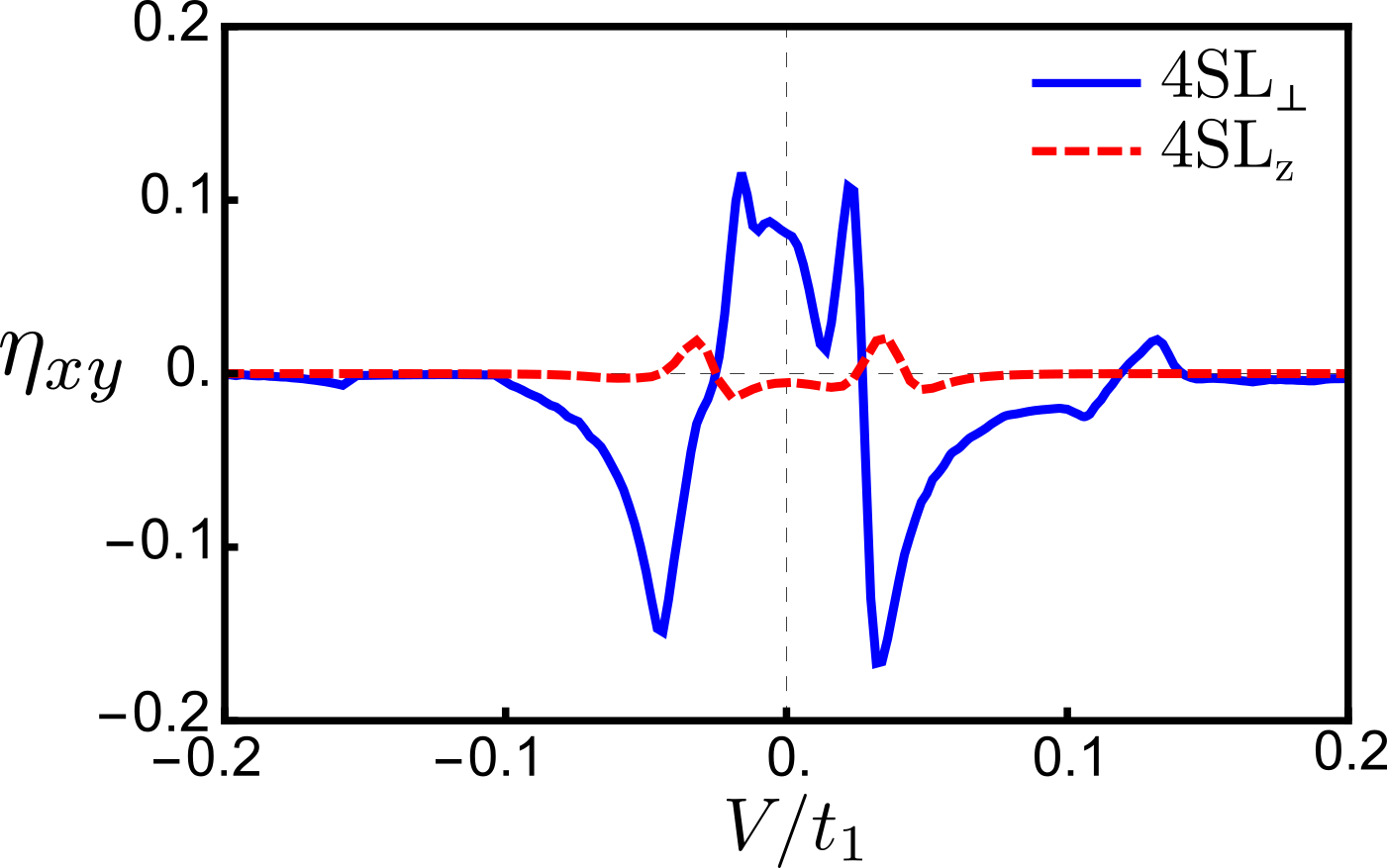}
\caption{Energy dependent nematicity, as defined by Eq. (\ref{eq:FStetdist}), for $\phi=\pi/4$ 4SL$_\perp$ (blue) and 4SL$_z$ (red) phases as a function of gate voltage or doping. The distortion is significant within the heavy bands for 4SL$_\perp$,  signifying a large tetragonal symmetry breaking electronic nematic response. The distortion is an order of magnitude smaller in the 4SL$_z$ phase, where it is suppressed by $D_f/D$, as it stems from the $\vec{\Phi}$ order parameter. The plots are obtained from the bandstructure in Fig. \ref{fig:Fig3b}.\label{fig:Fig12}}
\vspace{-0.cm}
\end{figure}

The energy dependent nematicity within the heavy bands is shown in Fig. \ref{fig:Fig12} for 4SL $\phi=\pi/4$ phases. The results are similar for all in-plane AFH phases, although $\phi = 0$ phases lead to an $\eta_{x^2-y^2}$ nematicity.  The nematic distortion is large ($\sim 0.1$) within the heavy bands, which can be seen in Fig. \ref{fig:Fig3b}; it is still present, but significantly smaller within the 4SL$_z$ phase. The large distortion is clearly the result of the heavy ($f$) bands around the Fermi surface. Even small tetragonal symmetry breaking hybridization gap magnitudes lead to significant distortions at the Fermi surface for relatively flat $f$ bands, and indeed the $f$-bandwidth is comparable to the hybridization gaps ($|V|^2/D$). Thus, the large Fermi surface distortion is expected to be a generic feature of hastatic order, and should lead to large transport anisotropies, even as other tetragonal symmetry breaking signatures are quite small.  In the presence of sufficiently strong random strain disorder, all tetragonal symmetry breaking is expected to be washed out, although the elastoresistivity is still expected to show signatures above the transition.

\section{Behavior in longitudinal magnetic field \label{sec:magfield}}

Longitudinal magnetic fields couple strongly to the $\Gamma_5$ doublet and so have a much more significant effect than in-plane fields.  In these sections, we explore the effect of $B_z$ on AFH phases, including how they evolve in both field and ``pressure'' ($V_7/V_6$).  We particularly focus on the competition between 2SL and 4SL phases, and how the magnetic moments evolve in field.

\subsection{Hastatic order and p-B-T phase diagram of \texorpdfstring{URu$_2$Si$_2$}{URu2Si2} \label{sec:pBTphasediag}}

\begin{figure}[!htb]
\includegraphics[width=0.95\columnwidth]{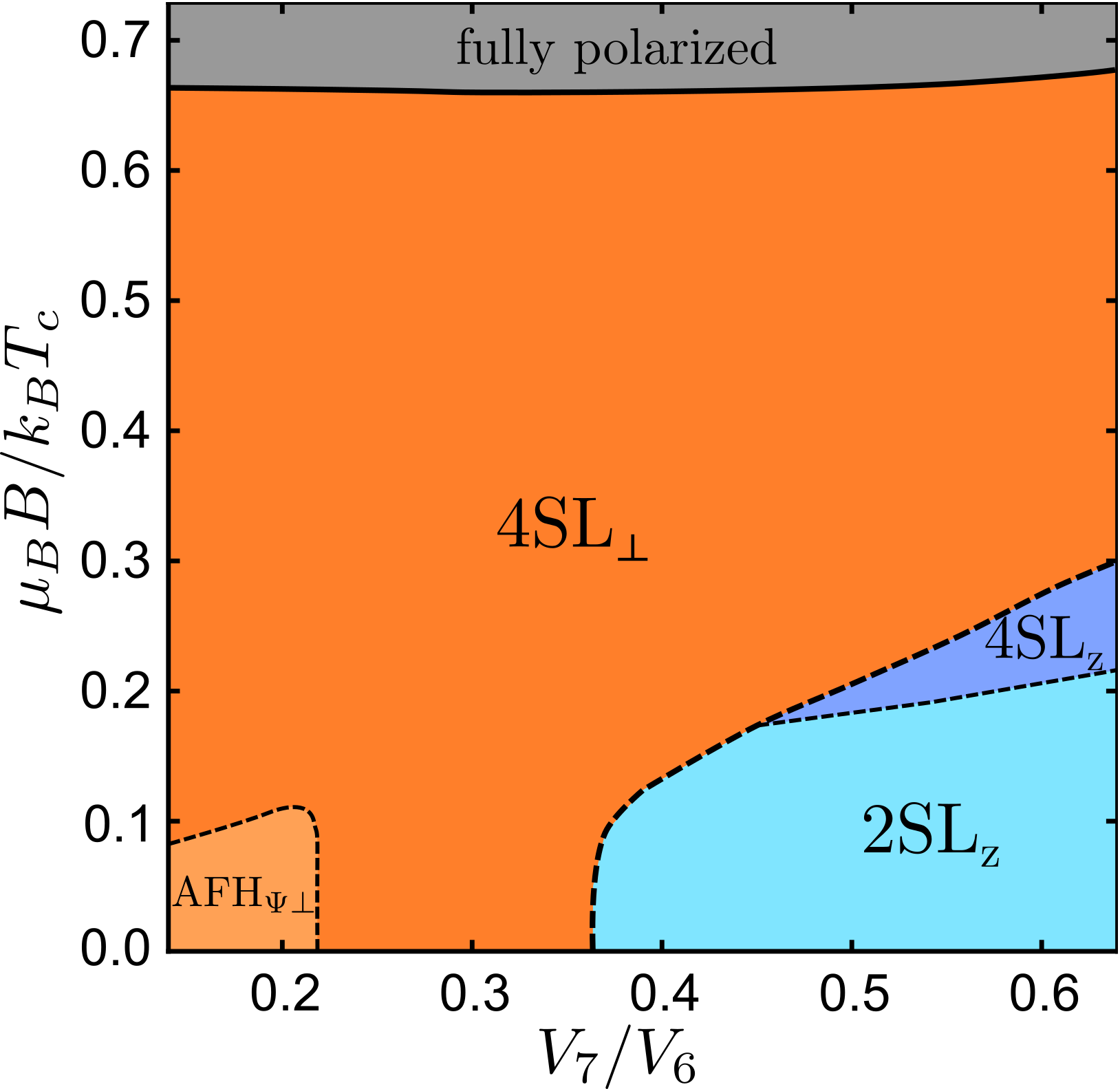}
\caption{Phase diagram of AFH order in magnetic field for $n_c=1/2$ at zero temperature, as the hybridization ratio, $V_7/V_6$ is varied. The usual set of parameters was used. Solid lines denote the second order transitions, while dashed lines denote the first order ones. Phases that differ only by $\vec{\Phi}$ are represented with different shadings of the same color: orange for AFH$_\perp$ phases and blue for AFH$_z$. Magnetic field suppresses the AFH$_z$ phases ($\vec{\Psi}$ parallel to field direction) in favor of the in-plane phases ($\vec{\Psi}$  orthogonal to field direction), similar to antiferromagnets in field.  We show  the uniform moments in Fig. \ref{fig:Fig14}. The phase diagram is qualitatively similar to the URu$_2$Si$_2$ p (Fe doping)-$B_z$ experimental phase diagrams. All AFH phases are suppressed in strong enough field ($0.65k_B T_c/\mu_B$, corresponding to $\sim$18T for $T_c = 17.5$K). \label{fig:Fig13}}
\vspace{-0.cm}
\end{figure}

We calculated the phase diagram self-consistently in $B_z$ as a function of $V_7/V_6$, for the same $n_c = 1/2$ used for the temperature versus $V_7/V_6$ phase diagram in Fig. \ref{fig:Fig7},  using the magnetic field couplings from Sec. \ref{sec:Bcoupling}. The result is shown in Fig. \ref{fig:Fig13}. Although we show one point in the parameter space, the obtained phase diagram is fairly generic for AFH regions of Fig. \ref{fig:Fig5}. 

For zero field, there are multiple AFH phases that evolve from an in-plane ``HO'' phase to a out of plane ``LMAFM'' phase as a function of $V_7/V_6$, similar to how URu$_2$Si$_2$ evolves with pressure.  In longitudinal field, we find that the AFH$_\perp$ phases are favored in magnetic field, similar to the material behavior \cite{Ran2017}.  Somewhat surprisingly, we do not find any FH or canted phases here, even as the AFH phases are suppressed. Both canted and uniform phases were found in strong magnetic fields for the cubic models \cite{Zhang2018,VanDyke2019}.  In the cubic model, the hastatic spinor is only weakly pinned to any high symmetry directions, and the ground state $\Gamma_3$ doublet in cubic symmetry only splits quadratically in field for all directions.  Therefore, we expect that the in-plane magnetic field response in tetragonal symmetry will resemble the cubic case, but the longitudinal field response is expected to be different, as we indeed find.  As long as we are sufficiently far from FH/AFH first order transition boundaries, the FH phase does not appear in longitudinal field. 

Within the AFH phases, the in-plane orientations are favoured in $B_z$ field over the out-of-plane.  This preference can be explained by an analogy to collinear antiferromagnets, where moments oriented perpendicular to the field are favoured over the parallel ones due to the canting of the spins \cite{SachdevBook}. While the hastatic spinor, and thus 5$f^1$ moments, do not cant, the conduction and 5$f^2$ moments do develop uniform components, as shown in Fig. \ref{fig:Fig14}, leading to an overall canting. The in-plane HO candidate phase is therefore re-entrant in field in the p (Fe doping)-$B_z$ phase diagram, giving qualitative agreement with experiment \cite{Aoki2009,Ran2017,Knafo2020}. We additionally find that magnetic field leads to a 2SL-4SL transition within the $z$ phase and AFH$_{\Psi\perp}$ to 4SL within the $\perp$; as discussed before, these phase transitions are likely difficult to observe due to the $D_f/D$ suppression of all signatures.

Finally, all AFH phases are suppressed for sufficiently large fields, leading to a fully polarized 5f$^2$ unhybridized state beyond the critical field, $\sim 0.65 k_B T_c/\mu_B$, around 18T for $T_c = 17.5$K. The critical field shows slight upturn with increasing $V_7/V_6$, mimicking $T_c(V_7/V_6)$ as seen in Fig. \ref{fig:Fig7}, the feature consistent with the experimental increase of the critical field with pressure or  Fe doping. This critical field is almost a factor of two smaller than the experimentally observed critical field \cite{Aoki2009,Ran2017,Knafo2020}.  We stress that while the qualitative features of the ``p''-B$_z$-T phase diagram are expected to be generic, quantitative features like the critical field depend strongly on the specifics of the microscopic model used and we have not optimized our parameters to attempt to reproduce quantitative features of URu$_2$Si$_2$, preferring to explore the generic possibilities of hastatic order here. 

\begin{figure}[!htb]
\includegraphics[width=1.0\columnwidth]{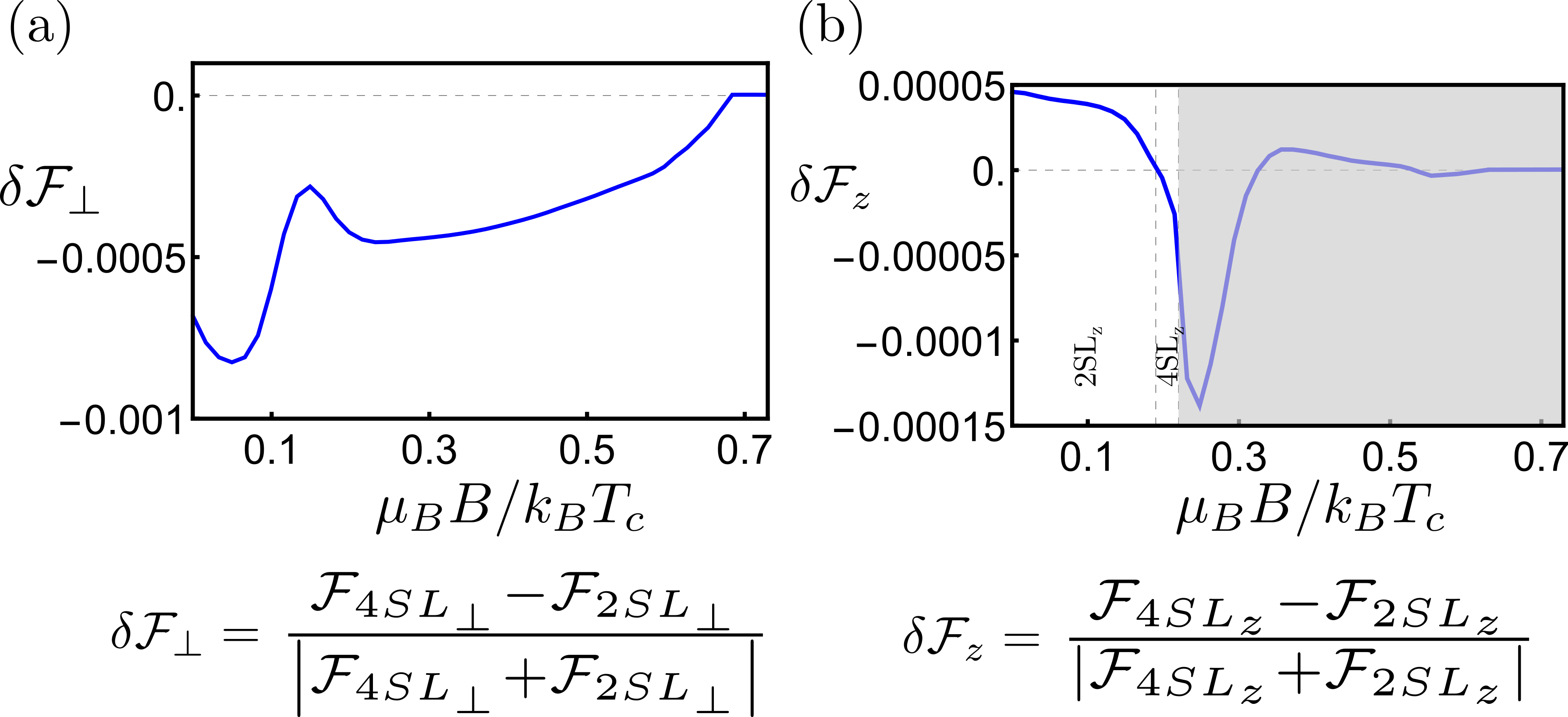}
\caption{Field ($B_z$) dependent relative free energy differences between 2SL and 4SL antiferrohastatic phases for (a) in-plane order with $\phi=\pi/4$; (b) $z$ order with $\phi=\pi/4$ at zero temperature. The greyed out part in (b) corresponds to the part of the phase diagram (see Fig. \ref{fig:Fig13}) where the $z$ phases are metastable. These relative energy differences stem from $\vec{\Phi}$ and are suppressed by $(D_f/D)^2$ compared to the $\theta$ pinning barriers. The magnetic field coupling to 2SL and 4SL phases is largely similar, with the difference proportional to $D_fT_c/D^2$, which allows field tuning. The plots are obtained self-consistently for the usual set of parameters with $n_c=1/2$ with $V_7/V_6=0.35$ for (a) and $V_7/V_6=0.5$ for (b).\label{fig:Fig14b}}
\vspace{-0.cm}
\end{figure}

\begin{figure}[!htb]
\includegraphics[width=1.0\columnwidth]{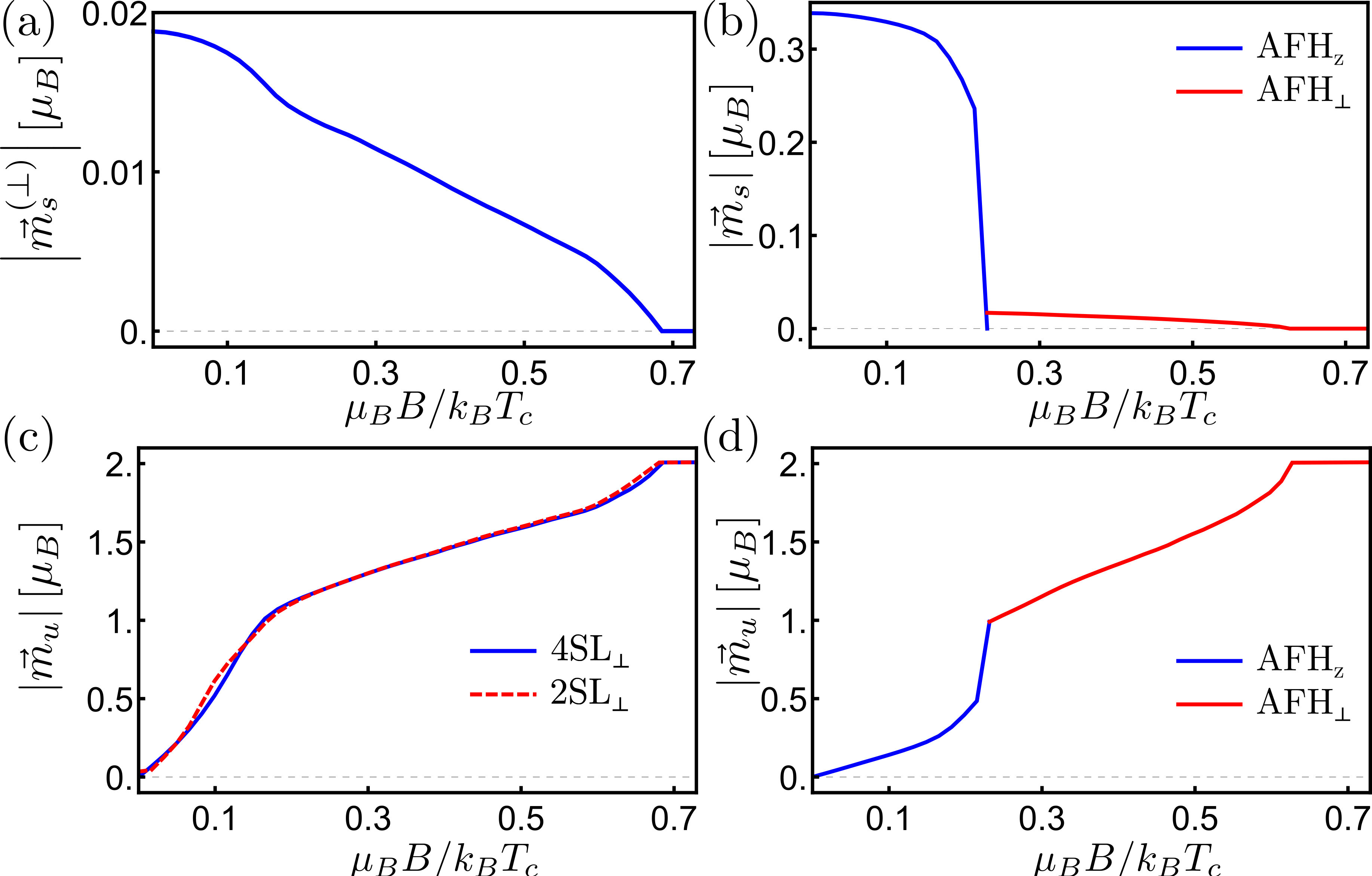}
\caption{Magnetic dipole moments for AFH phases at zero temperature. The moments are calculated self-consistently as a function of $B_z$ across cuts through the Fig. \ref{fig:Fig13} phase diagram: (a) staggered moments for the $V_7/V_6=0.35$ cut with 4SL$_\perp$ phase; (b) staggered moments for $V_7/V_6=0.5$ cut with the 2SL$_z$, 4SL$_z$ and 4SL$_\perp$ phases appearing; (c) uniform moments for $V_7/V_6=0.35$, where both 4SL$_\perp$ and 2SL$_\perp$ (metastable here) phases are shown for comparison; (d) uniform moments for $V_7/V_6=0.5$ with 2SL$_z$, 4SL$_z$ and 4SL$_\perp$ phases appearing. The staggered moments are suppressed as the hastatic order itself is suppressed in field. Uniform moments, primarily originating from 5$f^2$, develop with increasing field, leading eventually to full polarization of the $\Gamma_5$ doublet and complete destruction of hastatic order. The in-plane phases allow for larger 5f$^2$ polarization, making them more energetically favourable in field.\label{fig:Fig14}}
\vspace{-0.cm}
\end{figure}

The energy barriers between the 2SL and 4SL phases are small in all fields, as shown in  Fig. \ref{fig:Fig14b}.  We concentrate on the 2SL/4SL competition, but the AFH$_{\Psi\perp}$ barriers to 2SL$_\perp$ and 4SL$_\perp$ have similar orders of magnitude. The overall magnitude of the difference is again suppressed by $(D_f/D)^2$ and magnetic field can tune between the two phases. In general, the small energy differences mean extrinsic effects, like sample dependent strain, might lead to different $\vec{\Phi}$ phases in different URu$_2$Si$_2$ samples.

The evolution of the AFH phases in field can be understood by examining the evolution of staggered and uniform magnetic dipole moments in field. The moments were calculated self-consistently for several phase-diagram cuts and are shown in Fig. \ref{fig:Fig14}. In all phases, the staggered moments decrease in field and vanish with the hastatic order. The suppression is much stronger for the predominantly 5$f^2$ staggered moments of the $z$ phases than for in-plane $5f^1$ and conduction moments, reflecting their relative instability in field.  While the 2SL phase has small zero-field uniform moments along $z$ , this difference is quickly overcome in finite field, with all AFH$_\perp$ phases showing similar uniform moments across the field range. These uniform moments arise predominantly from partial polarization of the 5$f^2$ $\Gamma_5$ doublet along $\hat z$. As hastatic order is the consequence of two-channel Kondo physics, splitting of either the ground state or excited doublet is expected to suppress hastatic order, albeit relatively slowly. Once the polarization of the doublet reaches its bare value (given by $g_f/2$), the hastatic order is fully suppressed. The in-plane AFH phases allow for higher doublet polarization in-field (see Fig. \ref{fig:Fig14} (d)), effectively increasing the canting and explaining the instability of the AFH$_z$ phases.

\section{Fermi surfaces and effective \texorpdfstring{$g$}{g}-factors\label{sec:spinzeros}}

Now that our model can treat both the HO (AFH$_\perp$) and LMAFM (AFH$_z$) phases on equal footing, we can address two of the key original arguments for hastatic order in more detail.  First, there is the evolution of the Fermi surface across the first order transition from HO to LMAFM, as measured by dHvA \cite{Hassinger2010} and ARPES \cite{Frantzeskakis2021}, which was used to argue that both the HO and LMAFM share a $\bf{Q}$-vector and are related via a kind of spin-flop transition of the hidden order parameter\cite{Haule2009,Chandra2013}.  Second, there is the Ising anisotropy of the Fermi surface magnetization detected by the observation of spin-zeros in dHvA measurements in the HO phase\cite{Ohkuni1999,Altarawneh2011,Bastien2019}, which was used to argue that the heavy electrons acquired their anisotropy via hybridization with Ising 5f$^2$ $\Gamma_5$ moments\cite{Chandra2013}.  The continued existence of spin zero signatures is of particular concern for the 2SL and 4SL phases, where the bands are generically not doubly-degenerate.  We will show that, while the Fermi surfaces of the AFH$_\perp$ and AFH$_z$ are generically different, there are several moderately heavy Fermi surfaces whose volumes generically change little across the first order transition.  The spin zeroes do survive for sufficiently high magnetic fields, which we will show using a low-temperature effective model of the hastatic heavy bands.

\subsection{Fermi surface continuity \label{sec:FScontinuity}}

\begin{figure}[!htb]
\includegraphics[width=0.76\columnwidth]{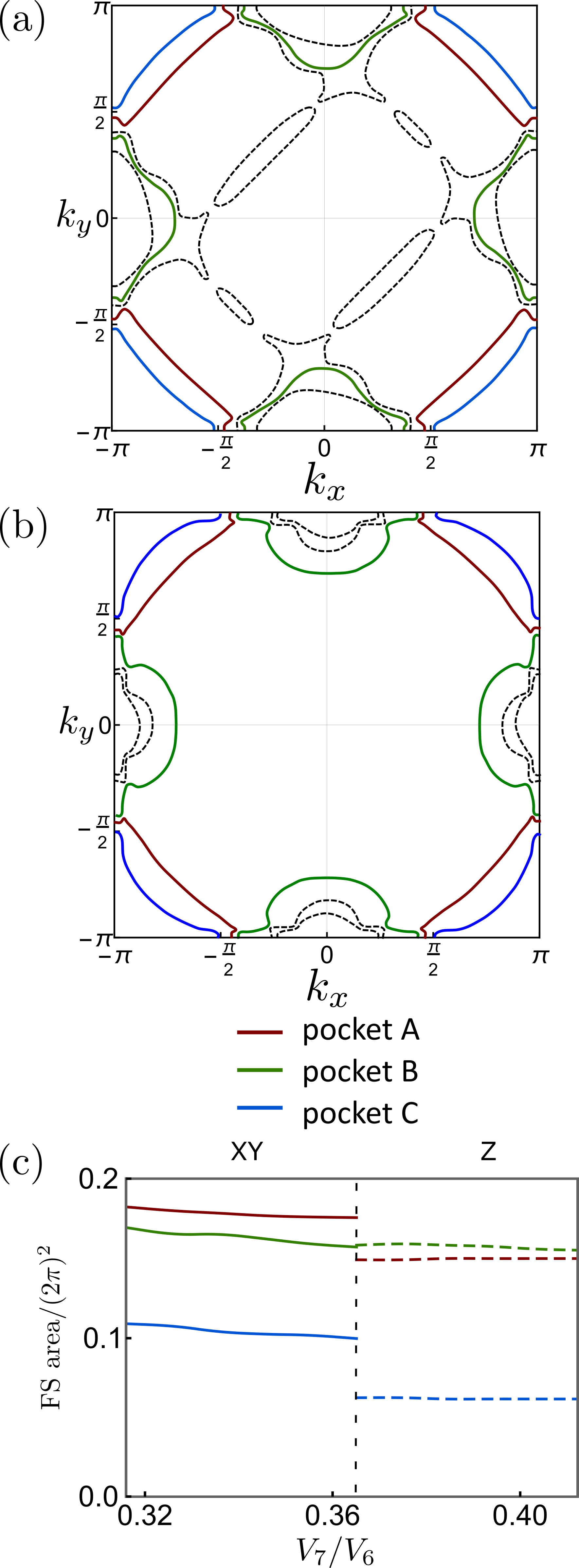}
\vspace{-0.2cm}
\caption{Fermi surface $k_z=0$ cuts for the 4SL$_\perp$ phase, (a), 4SL$_z$ phase, (b), and their surface areas as a function of $V_7/V_6$ (proxy for pressure), (c). The plots are found self-consistently using the usual parameters and $n_c = 1/2$.  Three ``moderately heavy'' pockets are emphasized in color; it is clear that they only weakly break tetragonal symmetry. The areas of these pockets have relatively small changes across AFH$_\perp$ to AFH$_z$ first-order transition, comparing favourably with experiments in URu$_2$Si$_2$. The ``very heavy'' pockets shown with black, dashed lines result in large tetragonal symmetry breaking and show more abrupt changes across the transition.  These pockets are likely difficult to observe with quantum oscillations, both due to the heavy mass and the effect of random strain on tetragonal symmetry breaking quantities.\label{fig:Fig18}}
\vspace{-0.7cm}
\end{figure}

One of the well-known features of URu$_2$Si$_2$ is the somewhat surprising Fermi surface continuity seen with quantum oscillation measurements across the first order HO-LMAFM transition \cite{Hassinger2010, Frantzeskakis2021}. This measurement strongly suggests that the LMAFM and HO and microscopically similar in origin.  To address this question, we present $k_z=0$ Fermi surface cuts in Fig. \ref{fig:Fig18}(a) for the 4SL$_\perp$ $\phi=\pi/4$ phase with usual parameter choices presented. The specific Fermi surfaces depend strongly on the exact parameter choices, however, there are several generic features shared between all AFH phases for similar filling.  We generically find two types of Fermi surface pockets formed from our heavy bands: one that can be described as ``moderately heavy'' and one that can be described as ``very heavy''. The moderately heavy pockets have somewhat lower masses and show only slight tetragonal symmetry breaking, while the very heavy ones have large tetragonal symmetry breaking, and are responsible for most of the electronic nematicity discussed in Sec. \ref{sec:FSnematicity}. The very heavy pockets change drastically across the first order transition between 4SL$_\perp$ and 4SL$_z$ phases, which is required by the significantly smaller tetragonal symmetry breaking in the 4SL$_z$ phase. As the tetragonal symmetry breaking is likely washed out by random strain disorder, these Fermi surfaces may be exceedingly difficult to observe with quantum oscillations. The moderately heavy pockets, on the other hand, are significantly less sensitive to such disorder, and generically have similar areas across the first-order line.  

An example of these Fermi surfaces is shown in Fig. \ref{fig:Fig18}, where the moderately heavy pockets are emphasized in color and appear around M (corner) and X (center of the side) points in the simple tetragonal BZ. Their areas across the AFH$_\perp$ to AFH$_z$ transition are shown, with resulting jumps qualitatively comparable to those measured experimentally \cite{Hassinger2010}. In particular, the larger pockets (denoted by A and B) show extremely small jumps, similarly to the experimental $\alpha, \beta$ pockets, while the smallest moderate pocket C shows a larger jump, resembling the experimental $\gamma$ pocket. We have chosen our initial band-structure parameters such that the positions of these pockets (one close to X and two close to M point) somewhat resemble electronic band structure calculations \cite{Oppeneer2010} and pockets measured by ARPES \cite{Meng2013, Frantzeskakis2021}, although this feature is of course not generic across parameter space. 

By contrast, the very heavy pockets shown with black dashed lines all exhibit a high degree of tetragonal symmetry breaking, and their areas change drastically across the transition. It is well-known that 
the pockets found in quantum oscillations or ARPES measurements \cite{Hassinger2010,Meng2013} account for approximately half of the Sommerfeld coefficient $\gamma$ measured by specific heat. One plausible scenario suggested by \cite{Hassinger2010} is that the missing pockets are have very heavy effective masses in excess of 70$m_e$, compared to $\lesssim 20 m_e$ for the measured pockets, thus being beyond the experimental sensitivity. This picture is very consistent with our very heavy pockets, even without invoking the effects of random strain disorder, which may make it difficult to observe these pockets with probes like quantum oscillations or ARPES that average over large areas.  It is also possible that more local measurements like quasiparticle interference measurements in scanning tunneling microscopy (STM-QPI) are more sensitive to these features and could track the varying nematicity across the material.

\subsection{Simplified model of hastatic order Fermi surfaces\label{sec:4bandmodel}}

The real Fermi surfaces in our microscopic models are complicated and difficult to follow for arbitrary magnetic fields.  In order to examine how the quantum oscillations evolve in field, and show that spin zeros are still expected even if the zero temperature bands are not degenerate, we construct a simple low temperature model of the AFH bands.  As we are interested in the extremal orbits, and the $k_z$ dispersion is weak, we construct a two-dimensional effective model that allows for easy calculation of the Fermi surface areas in field. The model captures the key features of the AFH Fermi surfaces found microscopically: there are heavy bands around the Fermi energy, there is a symmetry-breaking spin-orbit coupling hybridization, and the bands split similarly in magnetic field; the model can also treat different AFH phases by using different parameters, although it cannot accurately capture phase transitions.  We focus on 4SL$_\perp$ here, although the model can be modified to treat any of the AFH phases.

Our model has four bands, labeled by $\Gamma_5$ pseudospin ($\alpha$) and an additional degree of freedom ($\tau$) that is effectively a sublattice label. The effective Hamiltonian is:
\begin{align}\label{eq:4band}
H_{4}=&(\frac{k_x^2+k_y^2}{2m}-\mu)\mathds{1}_\alpha \mathds{1}_\tau\\
+&\left(\frac{\delta}{2}+\alpha' \left[k_xk_y \sin{2\phi}+\frac{k_x^2-k_y^2}{2}\cos{2\phi}\right]\right)\mathds{1}_\alpha \tau_z\notag\\
+ &\gamma \mathds{1}_\alpha\hat{\tau}_x+g_f B_z \alpha_z\mathds{1}_\tau+\beta B_z \left(k_y\cos{\phi}-k_x\sin{\phi}\right)\alpha_z\mathds{1}_\tau\notag\\
+&\eta\left(k_x^2+k_y^2\right)\tau_z\alpha_z\notag,
\end{align}
where the $\vec{\alpha}$ Pauli matrices act in the pseudospin space and $\vec{\tau}$ acts in the effective sublattice space. 

\begin{figure}[!htb]
\includegraphics[width=1.0\columnwidth]{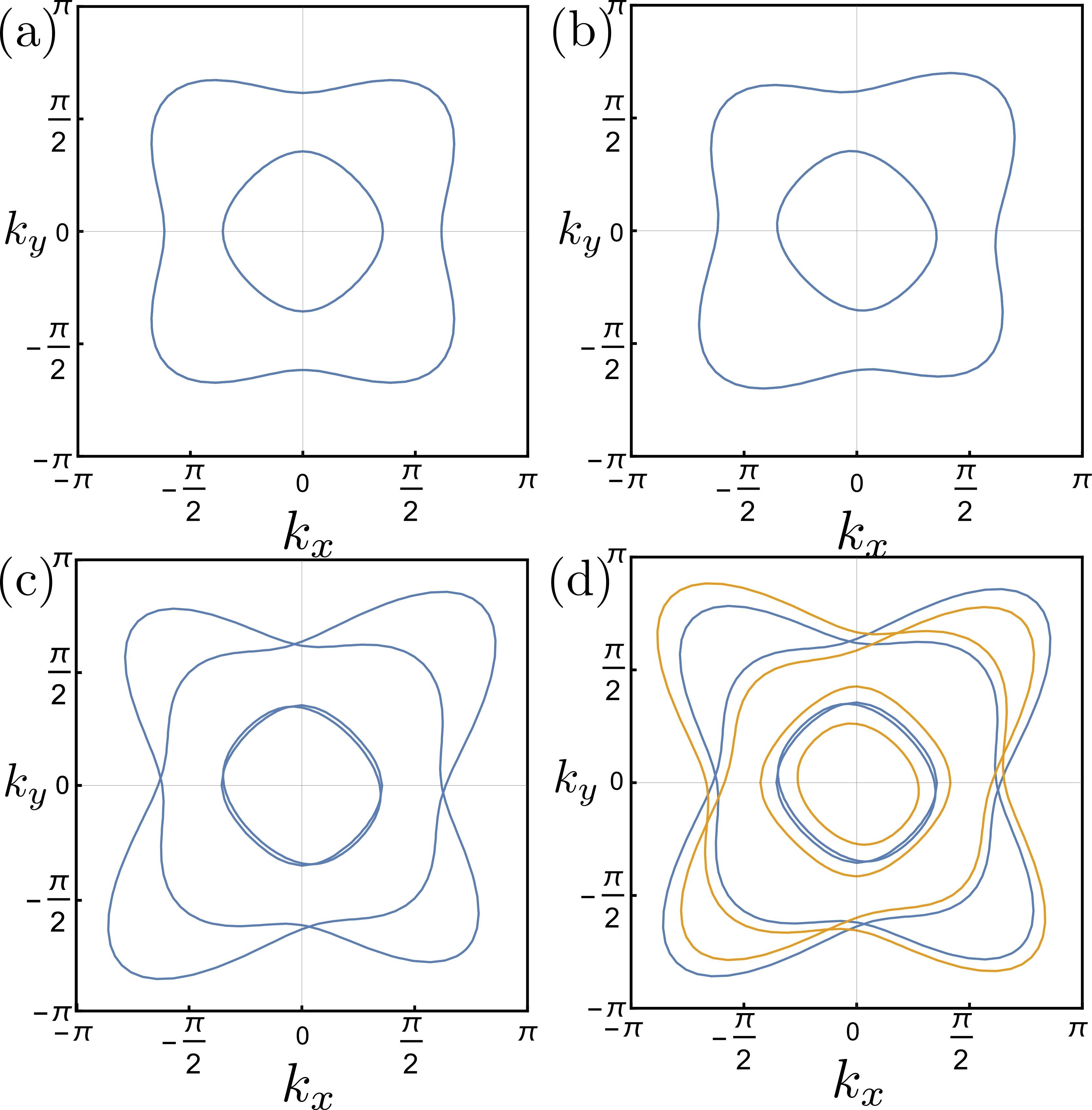}
\caption{Fermi surfaces in the effective four-band model of the AFH heavy bands for different sets of parameters.  All subfigures have $m=2.5$, $\mu=0.5$, $\alpha'=0.25$, $\phi=\pi/4$, $\gamma=0.25$, $g_f=1$ and $\beta=0.1$ in common.  In addition,  (a) has $\delta=B_z=\eta=0$, showing the bands with spin degeneracy in absence of tetragonal symmetry breaking; (b) has $\delta=0.1$, $B_z=\eta=0$ and introduces the characteristic tetragonal symmetry breaking, here of an $xy$ type; (c) has $\delta=0.1$, $\eta=0.03$, $B_z=0$ and introduces the zero-field spin-splitting seen generically in AFH phases; (d) has $\delta=0.1$, $\eta=0.03$, $B_z=0.1$, where the blue bands are the zero-field bands of (c) and the orange bands are for finite $B_z$. In (d), we see that strong magnetic field introduces further spin splitting of the bands as well as a shift of the Fermi pocket centers in field, a characteristic of the 4SL phase with broken inversion symmetry. Note that of the two larger, strongly tetragonal symmetry breaking bands in (c), the Fermi surface area of one grows while the other shrinks. \label{fig:Fig15}}
\vspace{-0.cm}
\end{figure}

This minimal model includes the following:
\begin{itemize}
    \item Unhybridized heavy bands diagonal in pseudospin and sublattice, given in the first line.
    \item A sublattice splitting shown in the second line, with both constant ($\delta$) and tetragonal symmetry breaking $k$-dependent ($\alpha'$) contributions, where $\phi$ is the in-plane angle of the hastatic spinor and tunes the nature of the tetragonal symmetry breaking. For $\delta=0$, the $k$-dependent term leads to additional modulations of the Fermi surfaces that are consistent with tetragonal symmetry, but when both terms are present, tetragonal symmetry is broken proportional to $\delta \alpha'$, as shown in  Fig. \ref{fig:Fig15}(b).
    \item $\gamma$ mixes the sublattice bands, leading to two sets of Fermi surfaces.
    \item $g_f$ captures the Ising splitting of the $\Gamma_5$ doublet.
    \item $\beta$ is only present in 4SL phases, and is necessary to reproduce the microscopic bandstructure; it introduces inversion symmetry breaking in the presence of magnetic field, and originates from the Landau-Ginzburg allowed triple product of magnetic field, $B_z$, $\vec{k}_\perp$ and the in-plane hastatic order parameter ($\vec{\Psi}_{\perp}$ or $\vec{\Phi}_{\perp}$). This term shifts the Fermi surface pocket centers  perpendicular to hastatic order and magnetic field directions.
    \item $\eta$ models the zero-field pseudospin splitting present in both 2SL and 4SL AFH phases, but absent in AFH$_{\Psi\perp}$. It makes the spin-zero analysis more complicated than for initially degenerate bands. There are two possibilities; here we have taken $\eta(k_x^2+k_y^2)$, which corresponds to the 4SL phase, which preserves pseudospin degeneracy at the $\Gamma$ point (see Sec. \ref{sec:4slsub}). In the 2SL phase, a constant $\eta$ term is more plausible, as pseudospin degeneracy is broken everywhere. 
\end{itemize}

The effect of these parameters are shown in Fig. \ref{fig:Fig15}. Now that we have a minimal model capturing the key microscopic features of the heavy bands in magnetic field, we turn to the effect of the zero-field pseudospin splitting. Indeed, the effect of the symmetry breaking hastatic order can make the ``split'' bands look extremely different, as seen in Fig. \ref{fig:Fig15}(c), where the two larger Fermi surfaces are singly degenerate, with one elongated along $x = y$ and the other along $x = -y$, and where these have different Fermi surface areas at zero field due to the tetragonal symmetry breaking.  In magnetic field, the bands further ``split'', with one shrinking and one growing with field, as shown in Fig. \ref{fig:Fig15}(d). These bands look different from the usual picture of quantum oscillations, where magnetic field introduces small splittings in degenerate bands that lead to a beating of similar oscillations and thus to spin-zeros.  Here, there will in fact be a similar beating and spin zeros, for sufficiently large fields, but we need to show this carefully, as we do in the next section.

\subsection{Recovering spin zeros with non-degenerate bands \label{sec:reczeros}}

To understand whether spin-zeros will be present in dHvA measurements of AFH order, let us first review what spin zeros are and where they come from in the case of degenerate bands.  In the usual Lifshitz-Kosevich formula for the magnetization measured in dHvA \cite{Lifshitz1955}, there is a pre-factor of the oscillatory term associated with a Fermi surface orbit of area $S$, $\cos \frac{\pi g(\theta) m_c^*}{2m_e}$, where $m_c^* = \frac{1}{2\pi}\frac{\partial S}{\partial E}$ is the cyclotron mass of this Fermi surface orbit, and we have allowed the effective g-factor to depend upon the angle from the c-axis, $\theta$.  This term arises from interference between bands that are degenerate at zero field and split linearly in field, leading to a beating between dHvA frequencies associated with Fermi surfaces of area $S_\sigma = S_0 + \sigma g(\theta) B \frac{\partial S_0}{\partial E} = S_0 + \sigma 2\pi g(\theta) B m_c^*$.  When $g(\theta)$ is angle dependent, it is possible to tune this prefactor through zero, for $\frac{\pi g(\theta) m_c^*}{2m_e} = (n+1/2)\pi$ and $n \in \mathds{Z}$.  The resulting ``spin zeros'' provide a very sensitive measurement of $g(\theta)$ that was used to identify the Ising-like g-factor in the HO phase of URu$_2$Si$_2$\cite{Altarawneh2011}.

In AFH order, there may already be a nontrivial splitting of the Fermi surfaces at zero field.  Here, we show how the Fermi surface areas ($S$) evolve in field (see Fig. \ref{fig:Fig15}). We treat the Fermi surfaces as pairs labeled by a pseudospin, $\sigma = \pm 1$, and our four band model has two such sets of Fermi surfaces.  For each pair, we can write the Fermi surface area as a function of field as,
\begin{equation}\label{eq:Sdhva}
    S^{dHvA}_{\sigma}(B)=S_0+S_1\sigma B + \delta S_1 \sigma+\delta S_0 B.
\end{equation}
Here, we neglect $\mathcal{O}(B^2)$ terms, but otherwise have four distinct terms:
\begin{itemize}
    \item $S_0$ is the zero-field average Fermi surface area for the pseudospin band pair, which is the primary contribution to the dHvA frequency.
    \item $\delta S_1\sigma$ describes the zero-field pseudospin splitting of the Fermi surface areas, which is generically present for both 2SL and 4SL phases. 
    \item $S_1\sigma B$ is the usual in-field splitting of the Fermi surface areas that results in beating of the dHvA signal and the appearance of spin-zeros.
    \item $\delta S_0 B$ describes the uniform shift of both pseudospin Fermi surface areas in field, which  introduces a weak field-dependence to the dHvA frequencies.
\end{itemize}

If there is no zero-field splitting, the dHvA frequency will be found from $S_0 + \delta S_0 B$, while $S_1 B$ will lead to a field-independent beating of the frequencies resulting in a $\cos \hbar  S_1 B/(2eB)$ coefficient.  Our model has only included a longitudinal field $B_z$, but as $S_1$ arises primarily from the coupling to the $\Gamma_5$ doublet, $g_f$, it effectively has an Ising $g_f B \cos \theta$ dependence consistent with the observed spin-zeros in URu$_2$Si$_2$ \cite{Ohkuni1999,Altarawneh2011}.

Once the zero-field splitting is included, the argument of the cosine becomes $\hbar \Delta S(B)/(2eB)$, where $\Delta S(B)=S^{dHvA}_{\uparrow}(B)-S^{dHvA}_{\downarrow}(B) = 2 S_1 B + 2 \delta S_1$. If $\delta S_1$ is sufficiently large, this prefactor is no longer field independent and the two Fermi surfaces should really be treated as having two independent dHvA frequencies.  However, as long as the in-field splitting dominates the zero-field splitting in the field range used for measurements, the coefficient is approximately field independent and should exhibit spin zeros.

In this four-band model, we can estimate how large the field must be to have an effectively field independent prefactor. 
We can calculate the effective $g$-factor as a function of field ($g_{\mathrm{eff}}(B)$) that would result from the Fermi surface splitting as \cite{AshcroftMerminBook}:
\begin{equation}\label{eq:geff}
    \frac{g_{\mathrm{eff}}}{g_{f}}=\frac{\Delta S(B)}{4\pi B m_c^*},
\end{equation}
where $g_f$ is the bare (Ising) $g$-factor and $m_c^*$ is the cyclotron mass of the orbit \cite{AshcroftMerminBook}. 

\begin{figure}[!htb]
\includegraphics[width=0.95\columnwidth]{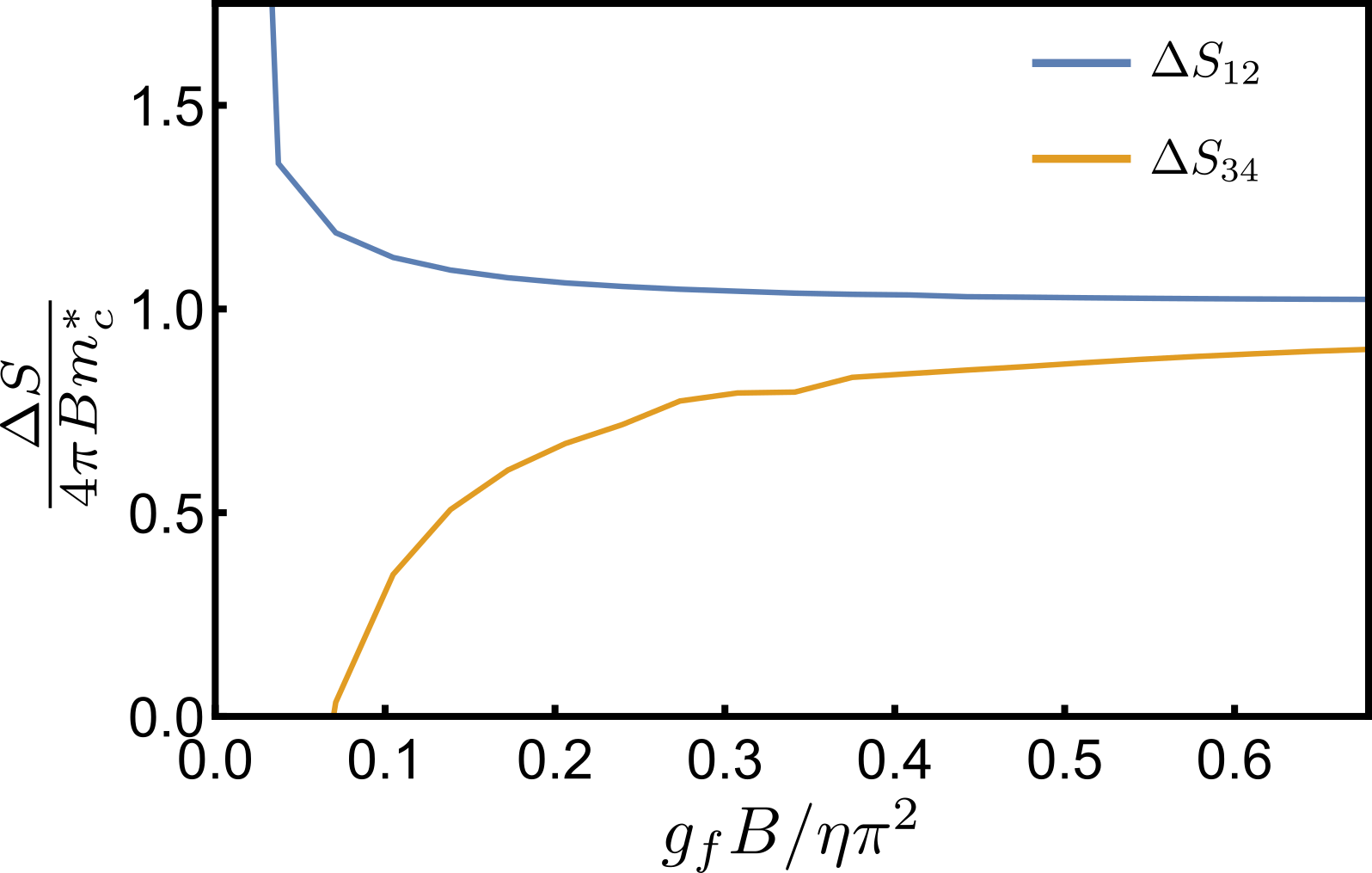}
\caption{The effective $g$-factor, $g_{\mathrm{eff}}/g_f = \frac{\Delta S}{4\pi B m_c^*}$ is plotted as a function of magnetic field strength for the effective four band model.  There are two sets of bands, seen in Fig. \ref{fig:Fig15}(c), which are 12 = inner and 34 = outer. Although clear zero-field spin splitting is seen in AFH phases, it is overcome by strong enough magnetic fields, where the ratio approaches unity. Thus, spin zeros appear largely as predicted by $g_f$, the Ising $g$-factor, for sufficiently large fields. In these units, the critical field for destroying the AFH order is on the order of .5-1. \label{fig:Fig16}}
\vspace{-0.cm}
\end{figure}

We plot this quantity as a function of field in Fig. \ref{fig:Fig16}, for the inner (12) and outer bands (34). As the field is increased, the effective $g$-factor approaches the bare value for both pairs of bands. This approach happens faster for bands with smaller area (12) as they have lower zero-field splitting in the 4SL-like case. Thus, in large enough fields, as determined by the ratio of the in-field splitting ($g_fB_z$) to the zero-field splitting ($\sim \eta \pi^2$), the original prediction of Ising anisotropic spin zeros holds, where ``large enough'' means $g_fB_z/\eta \pi^2 \sim .1-.4$, depending on the bands. Fitting the microscopic 4SL band-structures used throughout the paper in the vicinity of $\Gamma$ point, we find the band-dependent $\eta \approx 0.01-0.02$, meaning that for the critical field, $B_z^{(c)}$, where the AFH order is fully suppressed, the ratio $g_fB_z^{(c)}/\eta \pi^2\approx 0.5-1$ and thus spin zeros are expected for the typical fields used in dHvA measurements in URu$_2$Si$_2$.
 
Note that the Ising anisotropy arises from the $g_f$ in the four band model, which is not the bare 
$\Gamma_5$ $g$-factor, but rather the effective $g$-factor for the hybridized bands.  This effective $g_f$ is slightly renormalized by the admixture of some conduction electron states at the Fermi energy, which can lead to small deviations from a pure Ising response.  In addition, the degree of anisotropy, as well as the values for $g$-factor measured on different Fermi surface pockets will generically be different due to different degrees of hybridization, as was also found experimentally \cite{Bastien2019}.  However, this theory predicts similarly anisotropic $g$-factors in both the HO and LMAFM phases.

\section{Experimental prediction \label{sec:exppredictions}}

Hastatic order is a channel symmetry breaking heavy Fermi liquid with an Ising-like magnetic response, as well as additional symmetry breaking signatures. Many of these experimental signatures were discussed in the original proposal \cite{Chandra2013, Chandra2015}, or in our Landau-Ginzburg analysis \cite{Kornjaca2020}.  Here, we summarize the key features coming out of this more realistic microscopic theory.

\begin{itemize}
    \item \emph{Competition between HO and LMAFM}: The inclusion of two conduction bands and generic hybridization form-factors allows us to tune between HO and LMAFM candidates: AFH$_\perp$ and  AFH$_z$. These candidate phases fill large parts of parameter space, and first order transitions between the two can be induced by varying a proxy for pressure (the hybridization strength ratio) and longitudinal magnetic field. The resulting phase diagrams agree well with the experimental p-B-T phase diagram of URu$_2$Si$_2$ \cite{MydoshReview,Ran2017}. Both HO and LMAFM would then be hastatic phases with heavy Ising bands and hybridization gaps, but with very different symmetry breaking signatures. Most importantly, the LMAFM = AFH$_z$ phase has large ($.4\mu_B$) 5f$^2$ U moments along the $c$ axis, while the AFH$_\perp$ has only small $< .02\mu_B$ 5f$^3$ U moments in the plane, which are likely washed out by strain disorder.
    
    \item \emph{Separation of energy scales and tetragonal symmetry breaking:} We find a distinct separation of energy scales between the out of plane ($\theta$) and in-plane ($\phi$) energy barriers between hastatic phases, with smaller barriers between the different types of AFH order (2SL, 4SL, AFH$_{\Psi\perp}$) and even smaller $\mathds{Z}_4$ pinning of the in-plane moments. In addition, the in-plane moments, and the elastic and magnetic responses are heavily suppressed by $T_c/D$. This suppression, together with the smallness of in-plane energy barriers suggests that the in-plane staggered moments and other tetragonal symmetry breaking signatures (elastic and magnetic) are strongly affected by disorder \cite{Kornjaca2020}. Thus we can reconcile the absence of in-plane staggered moments \cite{Metoki2013,Das2013,Ross2014} in experiments, and the contradicting tetragonal symmetry breaking signatures \cite{Okazaki2011,Tonegawa2014,Choi2018,Bridges2020,Ghosh2020}. We predict that the elastoresistivity nematicity \cite{Riggs2015,Wang2020} should be the most reliable experiment for detecting tetragonal symmetry breaking effects in the hastatic phases, as the underlying microscopic electronic nematicity is quite large within the heavy bands around Fermi energy.
    
    \item \emph{Spinorial signatures of hastatic order:} Hastatic order is fundamentally spinorial, and generically does break double-time-reversal symmetry (2SL, 4SL), although it is not required (AFH$_{\Psi\perp}$). The spinor itself, however, is not an observable, leading the broken double time-reversal symmetry to be captured by an additional vectorial order parameter, $\vec{\Phi}$ \cite{Kornjaca2020} with additional symmetry breaking signatures. The main signatures we expect to be robust to disorder are uniform $c$-axis moments in the 2SL$_\perp$ HO candidate phase and inversion symmetry breaking in all of the 4SL phases. The small uniform moments are best detected in a sensitive experiment like the nonlinear Kerr effect, where some signatures of the $c$-axis moments might already have been seen \cite{Schemm2015}, while the inversion breaking might be best explored with second harmonic generation. Given the smallness of the 2SL/4SL energy barriers, there may be tuning between different $\vec{\Phi}$ phases by field, strain or sample quality.
    
    \item \emph{Fermi surfaces and quantum oscillations:} As HO and LMAFM stem from the same microscopic order parameter in the hastatic theory, our results naturally explain the continuity between the Fermi surface areas seen in the quantum oscillation experiments \cite{Hassinger2010, Frantzeskakis2021}. The hastatic theory also predicts that missing heat capacity contributions stem from the very heavy Fermi surface pockets that also potentially show large tetragonal symmetry breaking on short length scales. Our updated hastatic theory still generically predicts Ising anisotropic spin-zeros \cite{Ohkuni1999,Altarawneh2011,Bastien2019}, for sufficiently large fields. In addition, the unified theory we present predicts Ising anisotropic spin zeros in both HO and LMAFM phases, which could be tested in quantum oscillation experiments on URu$_2$Si$_2$ under pressure or Fe doping.
\end{itemize}

\section{Conclusions\label{sec:conclusions}}

To summarize, we have investigated a realistic microscopic model of tetragonal hastatic order as a candidate for explaining the phase diagram of URu$_2$Si$_2$ Compared to the original hastatic theory of hidden order \cite{Chandra2013}, our more generic model was able to treat both hidden order and antiferromagnetism on the same footing, and can tune between the two using either a pressure analogue ($V_7/V_6$ hybridization channel strength ratio) or longitudinal magnetic field. We explored the phase diagram of the model in detail, considering different ferrohastatic, two-sublattice and four-sublattice antiferrohastatic Ans\" atze. The calculated symmetry breaking signatures of the order (magnetic and quadrupolar moments, susceptibilities, nematicity) were found to fit well with the predictions of the phenomenological Landau theory \cite{Kornjaca2020}. In particular, we confirmed the theoretical presence of the spinorial order signatures, including uniform moments orthogonal to hastatic spinor and tetragonal symmetry breaking in $z$ phases. These signatures were notably absent in the initial hastatic proposal due to particularly simple band choices.  If observed experimentally, these would be the first signatures of spinorial order in a material.  We also replicated the successes of the original theory in our more complex model, in particularly spin-zeros associated with the $g$-factor anisotropy are still present, despite non-degenerate bands. Finally, we were able to capture the similarities between the electronic properties of LMAFM and HO candidate phases.

While we concentrated on the hidden order in URu$_2$Si$_2$, these conclusions are quite general for tetragonal two-channel Kondo models, and hexagonal and other symmetries with Ising doublets should behave similarly.  Future work could examine the nature of the two-channel Kondo insulator, expected near quarter filling, which has been proposed to host Majorana zero modes at the center of defects \cite{Majorana2021}, or study the effect of the hastatic bands on any superconducting orders arising out of hastatic order. 

\vspace{12pt}
\begin{acknowledgments}
We acknowledge stimulating discussions with Premala Chandra, Piers Coleman, John van Dyke, Victor L. Quito and Brad Ramshaw. This work was supported by the U.S. Department of Energy, Office of Science, Basic Energy Sciences, under Award No. DE-SC0015891. R.F. thanks the Aspen Center for Physics, supported by the NSF Grant PHY-1607611, for their hospitality.
\end{acknowledgments}

\appendix
\section{Symmetry allowed hybridization and hoppings\label{app:SC}}

Here we derive the symmetry allowed hybridization form-factors and hoppings quoted in main text using the Slater-Koster method \cite{Slater1954}. For hybridization form factors, we start from Eq. (\ref{eq:Valfluct}) and construct the Wannier states of the necessary symmetries ($\Gamma_6$ and $\Gamma_7^-$) from the $d_{z^2}$ Ru electrons. The Ru orbitals can be written in the $|l,m_l,s,m_s\rangle$ basis as,
\begin{equation}\label{eq:Rustates}
|d_{z^2},\sigma\rangle=|2, 0, \frac{1}{2}, \frac{\sigma}{2}\rangle,
\end{equation}
while the resulting $\Gamma_6$ and $\Gamma_7^-$ Wannier states are equivalent on-site to the $J=5/2$ states:
\begin{align}\label{eq:Gstates}
|\Gamma_6,\alpha\rangle&=|\frac{5}{2}, \frac{\alpha}{2}\rangle,\\
|\Gamma_7^-,\alpha\rangle&=\cos{\eta}|\frac{5}{2}, \frac{5\alpha}{2}\rangle+\sin{\eta}|\frac{5}{2}, -\frac{3\alpha}{2}\rangle\notag,
\end{align}
in the $|j,m_j\rangle$ basis.

To construct the $\Gamma_6$ and $\Gamma_7^-$ Wannier states, we expand the $J=5/2$ states in the Ru basis: 
\begin{align}\label{eq:Coverlap}
    c_{\Gamma_7^-\alpha}(j)&=\sum_{j',\sigma,\beta}\langle\Gamma_7^-,\alpha,j|d_{z^2},\sigma,j',\beta\rangle c_{j'\beta\sigma},\\
    c_{\Gamma_6\alpha}(j)&=\sum_{j',\sigma,\beta}\langle\Gamma_6,\alpha,j|d_{z^2},\sigma,j',\beta\rangle c_{j'\beta\sigma}\notag,
\end{align}
where the Ru $d_{z^2}$ electrons are on neighboring ($j'$) sites to the Wannier state ($j$). The overlaps can be calculated by integrating the corresponding wave-functions in real-space:
\begin{align}\label{eq:Intoverlap}
    \langle\Gamma_7^-,\alpha,j|d_{z^2},\sigma,j',\beta\rangle&=\int
\mathrm{d}\vect{r} \langle\Gamma_7^-,\alpha,j|\vect{r}\rangle\langle\vect{r}|d_{z^2},\sigma,j',\beta\rangle,\cr
\langle\Gamma_6,\alpha,j|d_{z^2},\sigma,j',\beta\rangle&=\int
\mathrm{d}\vect{r} \langle\Gamma_6,\alpha,j|\vect{r}\rangle\langle\vect{r}|d_{z^2},\sigma,j',\beta\rangle.
\end{align}
To do this, the angular dependence of the wave-functions is needed, and so we rewrite them using spherical harmonics:
\begin{align}\label{eq:Yoverlap}
   \langle\vect{r}|d_{z^2},\sigma,j',\beta\rangle&=Y_2^0\left(\vect{r}-\vect{R}_{j'}\right)|\frac{1}{2},\frac{\sigma}{2}\rangle,\\
   \langle\vect{r}|\Gamma_6,\alpha,j|\rangle&=\sum_{m,\sigma}\langle 3,m,\frac{1}{2},\frac{\sigma}{2}|\frac{5}{2},\frac{\alpha}{2}\rangle\cr
   &\times Y_3^m\left(\vect{r}-\vect{R}_{j}\right)|\frac{1}{2},\frac{\sigma}{2}\rangle,\cr
   \langle\vect{r}|\Gamma_7^-,\alpha,j\rangle&=\sum_{m, \sigma}\Bigg(\cos{\eta}\langle 3,m,\frac{1}{2},\frac{\sigma}{2}|\frac{5}{2},\frac{5\alpha}{2}\rangle\cr
   &\times Y_3^m\left(\vect{r}-\vect{R}_{j}\right)|\frac{1}{2},\frac{\sigma}{2}\rangle\cr
   &+\sin{\eta} \langle 3,m,\frac{1}{2},\frac{\sigma}{2}|\frac{5}{2},-\frac{3\alpha}{2}\rangle\cr
   &\times Y_3^m\left(\vect{r}-\vect{R}_{j}\right)|\frac{1}{2},\frac{\sigma}{2}\rangle\Bigg)\notag,
\end{align}
where $\langle 3,m,\frac{1}{2},\frac{\sigma}{2}|\frac{5}{2},\frac{\alpha}{2}\rangle$, $\langle 3,m,\frac{1}{2},\frac{\sigma}{2}|\frac{5}{2},\frac{5\alpha}{2}\rangle$, and $\langle 3,m,\frac{1}{2},\frac{\sigma}{2}|\frac{5}{2},-\frac{3\alpha}{2}\rangle$ are the respective Clebsch-Gordan coefficients.

We numerically calculate the overlaps between $\Gamma_6$ and $\Gamma_7^-$ taken at the origin ($\vect{R}_j=0$) and one of the neighboring Ru sites, mainly Ru$_A$ site at $\vect{R}_{j'}=(\delta_x,0,\delta_z)$ and find four independent overlaps:
\begin{align}\label{eq:nnoverlaps}
    \langle d_{z^2},\uparrow|\Gamma_6,+\rangle&=-\langle d_{z^2},\downarrow|\Gamma_6,-\rangle=\tilde{V}_6^{(1)},\\
     \langle d_{z^2},\uparrow|\Gamma_6,-\rangle&=\langle d_{z^2},\downarrow|\Gamma_6,+\rangle=\tilde{V}_6^{(2)},\cr
     \langle d_{z^2},\uparrow|\Gamma_7^-,+\rangle&=-\langle d_{z^2},\downarrow|\Gamma_7^-,-\rangle=-\tilde{V}_7^{(1)},\cr
     \langle d_{z^2},\uparrow|\Gamma_7^-,-\rangle&=\langle d_{z^2},\downarrow|\Gamma_7^-,+\rangle=\tilde{V}_7^{(2)}.\notag
\end{align}

Fourier transforming these using Eq. (\ref{eq:v67def}), we obtain the hybridization matrices for the $(\delta_x,0,\delta_z)$ Ru:
\begin{align}\label{eq:nnmatrix}
    \hat{V}_{6,(\beta\sigma,\alpha)}^{(\delta_x,\delta_z)}(\vect{k})&=\mathrm{e}^{i \left(k_z/4+k_x/2\right)}
\begin{pmatrix}
\tilde{V}_{6}^{(1)} & \tilde{V}_{6}^{(2)}\\
\tilde{V}_{6}^{(2)} & -\tilde{V}_{6}^{(1)}\\
0 &  0\\
0 & 0
\end{pmatrix}_{(\beta\sigma,\alpha)},\\
\hat{V}_{7,(\beta\sigma,\alpha)}^{(\delta_x,\delta_z)}(\vect{k})&=\mathrm{e}^{i \left(k_z/4+k_x/2\right)}
\begin{pmatrix}
-\tilde{V}_{7}^{(1)} & \tilde{V}_{7}^{(2)}\\
\tilde{V}_{7}^{(2)} & \tilde{V}_{7}^{(1)}\\
0 &  0\\
0 & 0
\end{pmatrix}_{(\beta\sigma,\alpha)},\notag
\end{align}
with $\beta\sigma$ denoting the $(\mathrm{Ru}_A\uparrow, \mathrm{Ru}_A\downarrow,\mathrm{Ru}_B\uparrow, \mathrm{Ru}_B\downarrow)$ basis and $\alpha$ denoting the two states in the $\Gamma_6$ and $\Gamma_7^-$ multiplets, respectively.

To obtain the full hybridization matrix, we employ the symmetry operations that relate $(\delta_x,0,\delta_z)$ Ru to other neighbors \cite{Slater1954}. One possibility is provided by combining multiple $\pi/2$ rotations around the $z$ axis and $\pi$ rotation around the $y$ axis, with the full hybridization matrices being:
\begin{align}\label{eq:rotatingforV}
    \hat{V}_{6/7}(\vect{k})&=\hat{V}_{6}^{(\delta_x,\delta_z)}(\vect{k})\\
    &+\mathcal{R}_{z,\pi/2}^{(Ru)} \hat{V}_{6}^{(\delta_x,\delta_z)}(R_{z,\pi/2}^{-1}\vect{k})\left[\mathcal{R}_{z,\pi/2}^{(6/7)}\right]^{-1}\cr
    &+\left[\mathcal{R}_{z,\pi/2}^{(Ru)}\right]^2 \hat{V}_{6}^{(\delta_x,\delta_z)}(R_{z,\pi/2}^{-2}\vect{k})\left[\mathcal{R}_{z,\pi/2}^{(6/7)}\right]^{-2}\cr
    &+\left[\mathcal{R}_{z,\pi/2}^{(Ru)}\right]^3 \hat{V}_{6}^{(\delta_x,\delta_z)}(R_{z,\pi/2}^{-3}\vect{k})\left[\mathcal{R}_{z,\pi/2}^{(6/7)}\right]^{-3}\cr
    &+\mathcal{R}_{y,\pi}^{(Ru)}\hat{V}_{6}^{(\delta_x,\delta_z)}(R_{y,\pi}^{-1}\vect{k})\left[\mathcal{R}_{y,\pi}^{(Ru)}\right]^{-1}\cr
    &+\mathcal{R}_{y,\pi}^{(Ru)}\mathcal{R}_{z,\pi/2}^{(Ru)} \hat{V}_{6}^{(\delta_x,\delta_z)}(R_{y,\pi}^{-1}R_{z,\pi/2}^{-1}\vect{k})\cr
    &\times\left[\mathcal{R}_{z,\pi/2}^{(6/7)}\right]^{-1}\left[\mathcal{R}_{y,\pi}^{(Ru)}\right]^{-1}\cr
    &+\mathcal{R}_{y,\pi}^{(Ru)}\left[\mathcal{R}_{z,\pi/2}^{(Ru)}\right]^2 \hat{V}_{6}^{(\delta_x,\delta_z)}(R_{y,\pi}^{-1}R_{z,\pi/2}^{-2}\vect{k})\cr
    &\times\left[\mathcal{R}_{z,\pi/2}^{(6/7)}\right]^{-2}\left[\mathcal{R}_{y,\pi}^{(Ru)}\right]^{-1}\cr
    &+\mathcal{R}_{y,\pi}^{(Ru)}\left[\mathcal{R}_{z,\pi/2}^{(Ru)}\right]^3 \hat{V}_{6}^{(\delta_x,\delta_z)}(R_{y,\pi}^{-1}R_{z,\pi/2}^{-3}\vect{k})\cr
    &\times\left[\mathcal{R}_{z,\pi/2}^{(6/7)}\right]^{-3}\left[\mathcal{R}_{y,\pi}^{(Ru)}\right]^{-1}.\notag
\end{align}
Rotation matrices acting to the left and to the right in the equation above differ, as they act in different Hilbert spaces. The rotation matrices themselves can be determined from the angular momentum representation of the states given in Eq. (\ref{eq:Rustates})-(\ref{eq:Gstates}) using Wigner functions. In the  $(\mathrm{Ru}_A\uparrow, \mathrm{Ru}_A\downarrow,\mathrm{Ru}_B\uparrow, \mathrm{Ru}_B\downarrow)$ basis, one also has to keep track of switching between Ru sites, with the rotation matrices reducing to:
\begin{align}\label{eq:rotmatricesRu}
    \mathcal{R}_{z,\pi/2}^{(Ru)}&=\begin{pmatrix}
0 & 0 &\mathrm{e}^{-i\pi/4}&0\\
0 & 0 & 0 & \mathrm{e}^{i\pi/4}\\
\mathrm{e}^{i\pi/4} & 0 & 0 & 0\\
0 & \mathrm{e}^{-i\pi/4} & 0 & 0
\end{pmatrix},\\
    \mathcal{R}_{y,\pi}^{(Ru)}&=\begin{pmatrix}
0 & 0 & 0 & -1\\
0 & 0 & 1 & 0\\
0 & 1 & 0 & 0\\
-1 & 0 & 0 & 0
\end{pmatrix},\notag
\end{align}
while the direct application of Wigner functions in $\Gamma_6$ and $\Gamma_7^-$ bases, respectively, gives:
\begin{align}\label{eq:rotmatrices67}
    \mathcal{R}_{z,\pi/2}^{(6)}&=- \mathcal{R}_{z,\pi/2}^{(7)}=\begin{pmatrix}
\mathrm{e}^{-i\pi/4}&0\\
0 & \mathrm{e}^{i\pi/4}
\end{pmatrix},\\
    \mathcal{R}_{y,\pi}^{(6)}&=\mathcal{R}_{y,\pi}^{(7)}=\begin{pmatrix}
0 & -1\\
1 & 0
\end{pmatrix}.\notag
\end{align}

The evaluation of Eq. (\ref{eq:rotatingforV}) leads to the form factors presented in the main text. The analytical construction of the full hybridization matrices was also checked numerically by calculating the overlaps to all nearest neighbor Ru sites and parametrizing the results.

The same method was applied to obtain the conduction and $f$-electron overlaps. The results follow from simple geometry and symmetry considerations. For the conduction electron overlaps, the nearest neighbor in-plane (within constant $z$ plane), Ru$_A$-Ru$_B$ hopping can be parametrized by a single parameter, $t_1$, as the overlaps between $d_{z^2}$ orbitals to all four neighbors are equivalent due to four-fold rotational symmetry symmetry. The same is true for the Ru$_A$-Ru$_A$ and Ru$_B$-Ru$_B$ next-nearest in-plane neighbor hopping, which is the same for both Ru sublattices, and can be parametrized by $t_2$. The Ru$_A$-Ru$_B$ $z$-directed hopping is also easily shown to be equivalent for both neighbors by the application of $\pi$ rotations around $x$ and $y$. We therefore arrive at the conduction electron dispersion quoted in Eq. (\ref{eq:elbands}). As for the $f$-overlaps, the Slater-Koster method shows that all nearest neighbors to the central U $\Gamma_5$ site are equivalent, leading to the simple dispersion of Eq. (\ref{eq:fbands}).

\section{Crystal field model for U and magnetic field couplings \label{app:CEF}}

To obtain the relevant $g$ factors for the 5$f^2$ $\Gamma_5$ ground state and 5$f^1$ $\Gamma_7$ excited state doublets, we use a crystal field model for U in tetragonal environment with parameters fit to the thermodynamic measurements. The most general form of crystal-field Hamiltonian for an $f$-ion in tetragonal symmetry is \cite{Bleaney1953}:
\begin{equation}\label{eq:hcef}
    H_{CEF}(J)=aO_2^0[J]+bO_4^0[J]+cO_4^4[J]+dO_6^0[J]+eO_6^4[J],
\end{equation}
where the Stevens operators are \cite{Lea1962}:
\begin{align}\label{eq:Oops}
    O_2^0[J]&=3J_z^2-J(J+1),\\
    O_4^0[J]&=35J_z^4-30J(J+1)J_z^2+3J^2(J+1)^2\cr
    &+25J_z^2-6J(J+1),\cr
    O_4^4[J]&=\frac{1}{2}\left(J_+^4+J_-^4\right),\cr
    O_6^0[J]&=231 J_z^6-315J(J+1)J_z^4+735J_z^4\cr
    &105J^2(J+1)^2J_z^2-525J(J+1)J_z^2+294J_z^2\cr
    &-5J^3(J+1)^3+40J^2(J+1)^2-60J(J+1),\cr
    O_6^4[J]&=\frac{1}{4}\left(11J_z^2-J(J+1)-38\right)\left(J_+^4+J_-^4\right)\cr
    &+\frac{1}{4}\left(J_+^4+J_-^4\right)\left(11J_z^2-J(J+1)-38\right)\notag,
\end{align}
in each $J$-multiplet. 

The free parameters are then obtained by a numerical search with aim to reproduce the presence of a ground state doublet, the $3.5\mu_B$ Curie-Weiss moment and the Van Vleck susceptibility from high-temperature measurements. The values of Stevens parameters thus obtained are $a=1.4$ K, $b=0.58$ K, $c=3.9$ K, $d=0.0125$ K and $e=0.15$ K. While the search does not guarantee uniqueness of the solution, it allows us to obtain reasonable ground and excited state wave-functions that are self consistent.

To calculate the $g$-factors, we introduce magnetic field as:
\begin{equation}\label{eq:hcefB}
    \delta H(J,L,S,\vec{B})=g_{JLS}\mu_B\vec{J}\cdot\vec{B},
\end{equation}
where  $g_{JLS}$ is the Lande factor for the corresponding multiplet ($J=4$, $L=5$, $S=1$ for 5$f^2$ $\Gamma_5$; $J=7/2$, $L=3$, $S=1/2$ for 5$f^1$ $\Gamma_7$). By analyzing the field splittings of the doublet states for different field directions, we then obtain the $g$-factors used in the main text.

\begin{widetext}
\section{Details of AFH mean-field Ans\" atze\label{app:AFH}}

In this appendix, we show the fermionic parts of AFH Hamiltonians whose mean-field Ans\" atze were described in Sec. \ref{sec:AFHph}. In all cases, $\vect{Q} = [001]$. Using the spatial dependence of the hastatic spinors, the 2SL fermionic part reduces to:

\begin{align}\label{eq:2SLham}
    \mathcal{H}^{2SL}(\vect{k})&= \sum_{\vect{k}\beta\beta'\sigma}\left[\epsilon_c\left(\vect{k}\right)\mathds{1}_\sigma-\frac{1}{2}g_c\mu_B B_z\mathds{1}_\beta \sigma_z\right]_{(\beta\sigma,\beta'\sigma')}c\dg_{\vect{k}\beta\sigma}c_{\vect{k}\beta'\sigma}+\sum_{\vect{k}\alpha}\left[\left(\epsilon_f\left(\vect{k}\right)+\lambda\right)\mathds{1}_\alpha-\frac{1}{2}g_f\mu_B B_z\tau_z\right]_{(\alpha)} f\dg_{\vect{k}\alpha}f_{\vect{k}\alpha}\cr
    &+\sum_{\vect{k}\alpha\beta\sigma}\Bigg[\left(\hat{V}_{6,(\beta\sigma,\alpha)}(\vect{k})a_2(\theta,\phi)+\hat{V}_{7,(\beta\sigma,\alpha)}(\vect{k})a_2(\theta,\phi)\tau_x\right)c\dg_{\vect{k}\beta\sigma}f_{\vect{k}\alpha}\cr
    &+\left(\hat{V}_{6,(\beta\sigma,\alpha)}(\vect{k})a_3(\theta,\phi)+\hat{V}_{7,(\beta\sigma,\alpha)}(\vect{k})a_3(\theta,\phi)\tau_x\right)c\dg_{\vect{k}\beta\sigma}f_{\vect{k}+\vect{Q}\alpha}\cr
    &+\left(\hat{V}_{6,(\beta\sigma,\alpha)}(\vect{k}+\vect{Q})a_3(\theta,\phi)+\hat{V}_{7,(\beta\sigma,\alpha)}(\vect{k}+\vect{Q})a_3(\theta,\phi)\tau_x\right)c\dg_{\vect{k}+\vect{Q}\beta\sigma}f_{\vect{k}\alpha}\cr
    &+\left(\hat{V}_{6,(\beta\sigma,\alpha)}(\vect{k}+\vect{Q})a_2(\theta,\phi)+\hat{V}_{7,(\beta\sigma,\alpha)}(\vect{k}+\vect{Q})a_2(\theta,\phi)\tau_x\right)c\dg_{\vect{k}+\vect{Q}\beta\sigma}f_{\vect{k}+\vect{Q}\alpha}+h.c.\Bigg],
\end{align}
where the spinors are rewritten in terms of matrices:
\begin{align}\label{eq:a2}
    a_2(\theta,\phi)=\frac{|b|}{2}\begin{pmatrix}
    \left(\cos{\frac{\theta}{2}}-\sin{\frac{\theta}{2}}\right)\mathrm{e}^{i\phi/2}&0\\
    0& \left(\cos{\frac{\theta}{2}}+\sin{\frac{\theta}{2}}\right)\mathrm{e}^{-i\phi/2}
    \end{pmatrix},\quad
    a_3(\theta,\phi)&=\left[a_2(\theta,\phi)\tau_x\right].\dg
\end{align}

In the 4SL case we have an additional band folding with the resulting fermionic Hamiltonian:
\begin{align}\label{eq:4SLham}
    \mathcal{H}^{4SL}(\vect{k})&= \sum_{\vect{k}\beta\beta'\sigma}\left[\epsilon_c\left(\vect{k}\right)\mathds{1}_\sigma-\frac{1}{2}g_c\mu_B B_z\mathds{1}_\beta \sigma_z\right]_{(\beta\sigma,\beta'\sigma')}c\dg_{\vect{k}\beta\sigma}c_{\vect{k}\beta'\sigma}+\sum_{\vect{k}\alpha}\left[\left(\epsilon_f\left(\vect{k}\right)+\lambda\right)\mathds{1}_\alpha-\frac{1}{2}g_f\mu_B B_z\tau_z\right]_{(\alpha)} f\dg_{\vect{k}\alpha}f_{\vect{k}\alpha}\cr
    &+\sum_{\vect{k}\alpha\beta\sigma}\Bigg[\left(\hat{V}_{6,(\beta\sigma,\alpha)}(\vect{k})a_4(\theta,\phi)+\hat{V}_{7,(\beta\sigma,\alpha)}(\vect{k})a_4(\theta,\phi)\tau_x\right)c\dg_{\vect{k}\beta\sigma}f_{\vect{k}+\frac{\vect{Q}}{2}\alpha}\cr
    &+\left(\hat{V}_{6,(\beta\sigma,\alpha)}(\vect{k})a_5(\theta,\phi)+\hat{V}_{7,(\beta\sigma,\alpha)}(\vect{k})a_5(\theta,\phi)\tau_x\right)c\dg_{\vect{k}\beta\sigma}f_{\vect{k}-\frac{\vect{Q}}{2}\alpha}\cr
    &+\left(\hat{V}_{6,(\beta\sigma,\alpha)}(\vect{k}+\frac{\vect{Q}}{2})a_5(\theta,\phi)+\hat{V}_{7,(\beta\sigma,\alpha)}(\vect{k}+\frac{\vect{Q}}{2})a_5(\theta,\phi)\tau_x\right)c\dg_{\vect{k}+\frac{\vect{Q}}{2}\beta\sigma}f_{\vect{k}\alpha}\cr
    &+\left(\hat{V}_{6,(\beta\sigma,\alpha)}(\vect{k}+\frac{\vect{Q}}{2})a_4(\theta,\phi)+\hat{V}_{7,(\beta\sigma,\alpha)}(\vect{k}+\frac{\vect{Q}}{2})a_4(\theta,\phi)\tau_x\right)c\dg_{\vect{k}+\frac{\vect{Q}}{2}\beta\sigma}f_{\vect{k}+\vect{Q}\alpha}\cr
    &+\left(\hat{V}_{6,(\beta\sigma,\alpha)}(\vect{k}-\frac{\vect{Q}}{2})a_4(\theta,\phi)+\hat{V}_{7,(\beta\sigma,\alpha)}(\vect{k}-\frac{\vect{Q}}{2})a_4(\theta,\phi)\tau_x\right)c\dg_{\vect{k}-\frac{\vect{Q}}{2}\beta\sigma}f_{\vect{k}\alpha}\cr
    &+\left(\hat{V}_{6,(\beta\sigma,\alpha)}(\vect{k}-\frac{\vect{Q}}{2})a_5(\theta,\phi)+\hat{V}_{7,(\beta\sigma,\alpha)}(\vect{k}-\frac{\vect{Q}}{2})a_5(\theta,\phi)\tau_x\right)c\dg_{\vect{k}-\frac{\vect{Q}}{2}\beta\sigma}f_{\vect{k}+\vect{Q}\alpha}\cr
    &+\left(\hat{V}_{6,(\beta\sigma,\alpha)}(\vect{k}+\vect{Q})a_5(\theta,\phi)+\hat{V}_{7,(\beta\sigma,\alpha)}(\vect{k}+\vect{Q})a_5(\theta,\phi)\tau_x\right)c\dg_{\vect{k}+\vect{Q}\beta\sigma}f_{\vect{k}+\frac{\vect{Q}}{2}\alpha}\cr
    &+\left(\hat{V}_{6,(\beta\sigma,\alpha)}(\vect{k}+\vect{Q})a_4(\theta,\phi)+\hat{V}_{7,(\beta\sigma,\alpha)}(\vect{k}+\vect{Q})a_4(\theta,\phi)\tau_x\right)c\dg_{\vect{k}+\vect{Q}\beta\sigma}f_{\vect{k}+\frac{\vect{Q}}{2}\alpha}+h.c.\Bigg],
\end{align}
where the hastatic spinor matrices are defined as:
\begin{align}\label{eq:a4}
    a_4(\theta,\phi)=\frac{|b|}{2}\mathrm{e}^{i\theta/2}\begin{pmatrix}
    \mathrm{e}^{i\phi/2}&0\\
    0& -i\mathrm{e}^{-i\phi/2}
    \end{pmatrix}, \quad
    a_5(\theta,\phi)&=\left[a_4(\theta,-\phi)\right]^*.
\end{align}

Finally, we turn to AHF$_\Psi$ phase only exists for in-plane directed spinors. After the unitary transformation described in Sec. \ref{sec:afhperp} we obtain the following fermionic Hamiltonian:
\begin{align}\label{eq:AFHpham}
    \mathcal{H}^{AFH_\Psi}(\vect{k})&= \sum_{\vect{k}\beta\beta'\sigma}\left[\epsilon_c\left(\vect{k}\right)\mathds{1}_\sigma-\frac{1}{2}g_c\mu_B B_z\mathds{1}_\beta \sigma_z\right]_{(\beta\sigma,\beta'\sigma')}c\dg_{\vect{k}\beta\sigma}c_{\vect{k}\beta'\sigma}+\sum_{\vect{k}\alpha}\left[\left(\epsilon_f\left(\vect{k}\right)+\lambda\right)\mathds{1}_\alpha-\frac{1}{2}g_f\mu_B B_z\tau_z\right]_{(\alpha)} \chi\dg_{\vect{k}\alpha}\chi_{\vect{k}\alpha}\cr
    &+\sum_{\vect{k}\alpha\beta\sigma}\Bigg[\hat{V}_{6,(\beta\sigma,\alpha)}(\vect{k})a_6(\phi)c\dg_{\vect{k}\beta\sigma}\chi_{\vect{k}\alpha}+\hat{V}_{7,(\beta\sigma,\alpha)}(\vect{k})a_7(\phi)\tau_x c\dg_{\vect{k}\beta\sigma}\chi_{\vect{k}+\vect{Q}\alpha}\cr
    &+\hat{V}_{7,(\beta\sigma,\alpha)}(\vect{k}+\vect{Q})a_7(\phi)\tau_x c\dg_{\vect{k}+\vect{Q}\beta\sigma}\chi_{\vect{k}\alpha}+\hat{V}_{6,(\beta\sigma,\alpha)}(\vect{k}+\vect{Q})a_6(\phi)c\dg_{\vect{k}+\vect{Q}\beta\sigma}\chi_{\vect{k}+\vect{Q}\alpha}+h.c.\Bigg],
\end{align}
with the corresponding spinor matrices being:
\begin{align}\label{eq:a6}
    a_6(\phi)=\frac{|b|}{\sqrt{2}}\begin{pmatrix}
    1&0\\
    0& 1
    \end{pmatrix}, \quad
    a_7(\phi)&=\frac{|b|}{\sqrt{2}}\begin{pmatrix}
    \mathrm{e}^{i\phi}&0\\
    0& \mathrm{e}^{-i\phi}
    \end{pmatrix}.
\end{align}
The form of the AHF$_\Psi$  Hamiltonian is analogous to the one derived in the original hastatic model \cite{Chandra2013}, although we now have two conduction bands.
\end{widetext}


%


\end{document}